\documentclass{article}

\usepackage{amsmath,amstext,amssymb,epsf}

 \newtheorem{lemma}{Lemma}[section]
 \newtheorem{theorem}[lemma]{Theorem}
 \newtheorem{proposition}[lemma]{Proposition}

 \newtheorem{corollary}[lemma]{Corollary}
 
\newtheorem{notation}[lemma]{Notation}
 \newtheorem{definition}[lemma]{Definition}
\newtheorem{rem}[lemma]{Remark}
                                                                                
\numberwithin{equation}{section} 

\newenvironment{proof}{\par \sf Proof.\rm}{\hspace*{\fill}$\Box$\vspace{1ex}}

\newenvironment{remark}{\begin{rem}}{\hspace*{\fill}$\diamondsuit$\end{rem}}
 \newtheorem{ex}[lemma]{Example}
\newenvironment{example}{\begin{ex}}{\hspace*{\fill}$\diamondsuit$\end{ex}}

\newcommand{\len}[2]{l_{#1}(#2)}

\newcommand{\m}{{\bf m}}
\newcommand{\lea}{\stackrel{{}_+}{<}}
\newcommand{\gea}{\stackrel{{}_+}{>}}
\newcommand{\eqa}{\stackrel{{}_+}{=}}

\newcommand{\Lint}{L_{{\mathcal N}}}
\newcommand{\eps}{\epsilon}

\newcommand{\commentout}[1]{}

\begin{document}
\title{Shannon Information and Kolmogorov Complexity}
\author{Peter Gr\"unwald and
Paul Vit\'anyi\thanks{
Manuscript received xxx, 2004;
revised yyy 200?. 
This work supported in part
by the EU fifth framework project QAIP, IST--1999--11234,
the NoE QUIPROCONE IST--1999--29064,
the ESF QiT Programmme, and the EU Fourth Framework BRA
NeuroCOLT II Working Group
EP 27150, the EU NoE PASCAL, and by the Netherlands Organization for
Scientific Research (NWO) under Grant 612.052.004.
Address: CWI, Kruislaan 413,
1098 SJ Amsterdam, The Netherlands.
Email: {\tt Peter.Grunwald@cwi.nl, Paul.Vitanyi@cwi.nl}.}
}
                                                                                

\maketitle
\begin{abstract}
We compare the 
elementary theories of Shannon information and Kolmogorov 
complexity, the extent to which they have a common purpose, and where
they are fundamentally different. We discuss and relate the basic
notions  of both theories: 
Shannon entropy versus Kolmogorov complexity, the relation of both
to universal coding, Shannon mutual information
versus Kolmogorov (`algorithmic') mutual information, 
probabilistic sufficient statistic versus algorithmic sufficient
statistic (related to lossy compression in
the Shannon theory versus
meaningful 
information in the Kolmogorov theory),
and
rate distortion theory versus Kolmogorov's structure function.
Part of the material has appeared in print before, scattered
through various publications, but this is the first comprehensive
systematic comparison. The last mentioned relations are new.

\end{abstract}
\tableofcontents
\section{Introduction}
%
{\em Shannon information} theory, usually called just `information'
theory was introduced in 1948, \cite{Sh48}, by C.E. Shannon (1916--2001). {\em
  Kolmogorov complexity} theory, also known as `algorithmic
information' theory,
was introduced with different
motivations (among which Shannon's probabilistic notion
of information), independently by R.J. Solomonoff
(born 1926), A.N. Kolmogorov (1903--1987) and G. Chaitin (born 1943)
in 1960/1964, \cite{So64}, 1965, \cite{Ko65}, and 1969 \cite{Ch69},
 respectively. Both theories 
aim at providing a means for measuring `information'.  They
use the same unit to do this: the {\em bit}. In both cases, the amount
of information in an object may be interpreted as the length of a
description of the object.  In the Shannon approach, however, the
method of encoding objects is based on the presupposition that the
objects to be encoded are outcomes of a known random source---it is
only the characteristics of that random source that determine the
encoding, not the characteristics of the objects that are its
outcomes.  In the Kolmogorov complexity approach we consider the
individual objects themselves, in isolation so-to-speak, and the
encoding of an object is a short computer program 
(compressed version of the object) that
generates it and then halts.  In the Shannon approach we are
interested in the minimum expected number of bits to transmit a
message from a random source of known characteristics 
through an error-free channel.  Says Shannon \cite{Sh48}:
\begin{quote}
 ``The fundamental problem 
of communication is that of reproducing at one point
either exactly or approximately a message selected at another point.
Frequently the messages have {\em meaning}; that is they refer to or are
correlated according to some system with certain physical or conceptual
entities. These semantic aspects of communication are irrelevant to the
engineering problem. The significant aspect is that the actual message
is one {\em selected from a set} of possible messages. The system must
be designed to operate for each possible selection, not just the one which
will actually be chosen since this is unknown at the time of design.'' 
\end{quote}
In
Kolmogorov complexity we are interested in the minimum number of bits
from which a particular message or file 
can effectively be reconstructed: the minimum
number of bits that suffice to store the file in reproducible format.
This is the basic question 
of the ultimate compression
of given individual files.  A
little reflection reveals that this is a great difference: for {\em
  every} source emitting but two messages the Shannon information (entropy) is
at most 1 bit, but we can choose both messages concerned of
arbitrarily high Kolmogorov complexity. Shannon stresses in his
founding article that his notion is only concerned with {\em
  communication}, while Kolmogorov stresses in his founding article
that his notion aims at supplementing the gap left by Shannon theory
concerning the information in individual objects.
Kolmogorov 
\commentout{\begin{quote}
``The probabilistic approach is natural in
the theory of information transmission over communication channels
carrying `bulk' information consisting of a large number of unrelated or
weakly related messages obeying definite probabilistic laws. $\dots$
But what real meaning is there, for example, in asking how much information
is contained in `War and Peace'? Is it reasonable to include this
novel in the set of `possible novels,' or even to postulate
some probability distribution for this set? Or, on the other hand, must
we assume that the individual scenes in this book form a random
sequence with `stochastic relations' that damp out quite rapidly over a 
distance of several pages?''
\end{quote}
And in 
}
\cite{Ko83}: 
\begin{quote}
``Our definition of the 
quantity of information has the advantage that it refers to individual
objects and not to objects treated as members of a set of objects
with a probability distribution given on it. The probabilistic
definition can be convincingly applied to the information contained,
for example, in a stream of congratulatory telegrams. But it would
not be clear how to apply it, for example, to an estimate of the quantity
of information contained in a novel or in the translation of a novel
into another language relative to the original. I think that the
new definition is capable of introducing in similar applications
of the theory at least clarity of principle.'' 
\end{quote}
To be sure, both notions are natural: Shannon ignores the object itself
but considers only the characteristics of the random source of which the
object is one of the possible outcomes, while Kolmogorov considers
only the object itself to determine the number of bits in the ultimate
compressed version irrespective of the manner in which the object arose.
In this paper, we introduce, compare and contrast the Shannon and Kolmogorov
approaches. 
An early comparison between Shannon entropy and Kolmogorov
complexity is \cite{ChCo78}. 
\paragraph{How to read this paper:}
We switch back and forth between the two
theories concerned according to the following pattern: we first discuss a
concept of Shannon's theory, discuss its properties as well as some
questions it leaves open.  We then provide Kolmogorov's analogue of
the concept and show how it answers the question left open by
Shannon's theory. 
To ease understanding of the two theories and
how they relate, we supplied the overview below 
and then Sections~\ref{sec:coding} and
Section~\ref{sec:basic}, which discuss preliminaries, fix
notation and introduce the basic notions. The other sections are
largely independent from one another.
Throughout the text, 
we assume some basic familiarity with elementary notions of
probability theory and computation, but we have kept the treatment
elementary. This may provoke scorn in the information theorist, who sees
an elementary treatment of basic matters in his discipline, and likewise
from the computation theorist concerning the treatment
of aspects of the elementary theory of computation. But experience has shown
that what one expert views as child's play is an insurmountable
mountain for his opposite number. Thus, we decided to 
ignore background knowledge and
cover both areas from first principles onwards, so that
the opposite expert can easily access the unknown discipline, possibly
helped along by the familiar analogues in his own ken of knowledge.
\subsection{Overview and Summary}
A summary of the basic ideas is given
below. In the paper, these notions are discussed in the same order.
\begin{description}
\item[1. Coding: Prefix codes, Kraft inequality]
(Section~\ref{sec:coding}) Since descriptions or {\em encodings\/} of objects are
fundamental to both theories, we first review some elementary facts
about coding. The most important of these is the {\em Kraft
  inequality}. This inequality gives the 
fundamental relationship between {\em probability density functions and
  prefix codes}, which are the type of codes we are interested in. 
Prefix codes and the Kraft inequality underly most of Shannon's, and a
large part of Kolmogorov's theory.
\item[2. Shannon's Fundamental Concept: Entropy]
(Section~\ref{sec:shannon}) Entropy is defined by a functional that maps 
{\em probability distributions\/} or, 
equivalently, {\em random variables},
to {\em real numbers}. This notion is derived from first
principles as the only `reasonable' way to measure the 
`average amount of information conveyed when an outcome of the random
variable is observed'. The notion is then related to 
encoding and communicating messages by Shannon's famous `coding theorem'.
\item[3. Kolmogorov's Fundamental Concept: Kolmogorov Complexity]
(Section~\ref{sec:kolmogorov})
Kolmogorov complexity is defined by a function that maps {\em
  objects\/} (to be thought of as natural numbers or sequences of
symbols, for example outcomes of the random variables 
figuring in the Shannon theory) to the {\em natural numbers\/}. Intuitively, the Kolmogorov
complexity of a sequence is the length (in bits) of the shortest computer
program that prints the sequence and then halts.  
\item[4. Relating entropy and
  Kolmogorov complexity ]
(Section~\ref{sec:KCSE} and Appendix~\ref{sec:universal})
Although their primary aim is quite different, and they are functions
defined on different spaces,  there are close relations
between entropy and Kolmogorov complexity. The formal relation
``entropy = expected Kolmogorov complexity'' is discussed in
Section~\ref{sec:KCSE}. The relation is further illustrated
by explaining `universal coding' (also introduced by Kolmogorov in 1965)
which combines elements from both
Shannon's and Kolmogorov's theory, and which lies at the basis of most
practical data compression methods. While related to the main theme
of this paper, universal coding plays no direct role in the later
sections, and therefore we delegated it to Appendix~\ref{sec:universal}. 
\end{description}
Entropy and Kolmogorov Complexity are the basic
notions of the two theories. They serve as building blocks for all
other important notions in the respective theories. Arguably the most
important of these notions is {\em mutual information\/}:
\begin{description}
\item[5. Mutual Information---Shannon and Kolmogorov Style]
(Section~\ref{sec:mutual})
Entropy and Kolmogorov complexity are 
concerned with information in a single object: a random variable 
(Shannon)
or an individual sequence (Kolmogorov). Both theories provide
a (distinct) notion of  {\em mutual information\/} that 
measures the information that {\em one
object gives about another object}. In Shannon's theory, this is the
information that one random variable carries about another; in
Kolmogorov's theory (`algorithmic mutual information'), 
it is the information one sequence gives about another.
In an appropriate setting the former notion can be shown to
be the expectation of the latter notion.
\item[6. Mutual Information Non-Increase]
(Section~\ref{sect.mini})
In the probabilistic setting the mutual information between two 
random variables cannot be increased by processing the outcomes.
That stands to reason, since the mutual information is expressed
in probabilities of the random variables involved. But in the algorithmic
setting, where we talk about mutual information between two
strings this is not evident at all. Nonetheless, up to some precision,
the same non-increase law holds. This result was used recently to
refine and extend the celebrated G\"odel's incompleteness theorem.
\item[7. Sufficient Statistic] (Section~\ref{sect.sufstat}) Although
  its roots are in the statistical literature, the notion of
  probabilistic ``sufficient statistic'' has a natural formalization
  in terms of mutual Shannon information, and can thus also be
  considered a part of Shannon theory. The probabilistic sufficient
  statistic extracts the information in the data about a model class.
  In the algorithmic setting, a sufficient statistic extracts the
  meaningful information from the data, leaving the remainder as
  accidental random ``noise''.  In a certain sense the probabilistic version of
  sufficient statistic is the expectation of the algorithmic version.
  These ideas are generalized significantly in the next item.
\item[8. Rate Distortion Theory versus Structure Function]
  (Section~\ref{sect.rdsf}) Entropy, Kolmogorov complexity and mutual
  information are concerned with {\em lossless\/} description or
  compression: messages must be described in such a way that from the
  description, the original message can be completely reconstructed.
  Extending the theories to {\em lossy\/} description or compression
  leads to rate-distortion theory in the Shannon setting, and the
  Kolmogorov structure function in the Kolmogorov section. The basic
  ingredients of the lossless theory (entropy and Kolmogorov
  complexity) remain the building blocks for such extensions.  The
  Kolmogorov structure function significantly extends the idea of
  ``meaningful information'' related to the algorithmic sufficient
  statistic, and can be used to provide a foundation for inductive
  inference principles such as Minimum Description Length (MDL). Once again, the Kolmogorov
  structure function can be related to Shannon's rate-distortion
  function by taking expectations in an appropriate manner.
\end{description}

\subsection{Preliminaries}
\label{sec:preliminaries}
\paragraph{Strings:}
Let ${\cal B}$ be some finite or countable set. We use the notation
${\cal B}^*$ to denote the set of finite 
{\em strings\/} or {\em sequences\/} over ${\cal X}$. For example,
$$\{0,1\}^* = \{ \epsilon,0,1,00,01,10,11,000,\ldots \},$$
with $\epsilon$ denoting the {\em empty word} `' with no letters.
Let
${\cal N}$ denotes the natural
numbers. We identify
${\cal N}$ and $\{0,1\}^*$ according to the
correspondence 
\begin{equation}
\label{eq:correspondence}
(0, \epsilon ), (1,0), (2,1), (3,00), (4,01), \ldots
\end{equation}
The {\em length} $l(x)$ of $x$ is the number of bits
in the binary string $x$. For example,
$l(010)=3$ and $l(\epsilon)=0$. 
If $x$ is interpreted as an integer, we get $ l(x) =  \lfloor \log
(x+1) \rfloor$ and, for $x \geq 2$,
\begin{equation}
\label{eq:intlength}
\lfloor \log x \rfloor
\leq l(x) \leq \lceil \log x \rceil.
\end{equation}
Here, as in the sequel, $\lceil x \rceil$ is the smallest integer larger than or equal to
$x$, $\lfloor x \rfloor$ is the largest integer smaller than or equal
to $x$ and $\log$ denotes logarithm  to base two.
We shall typically be concerned with
encoding finite-length binary strings by other finite-length binary strings.
The emphasis is on binary strings only for convenience;
observations in any alphabet can be so encoded in a way
that is `theory neutral'.

\paragraph{Precision and $\lea, \eqa$ notation:}
It is customary in the area of Kolmogorov complexity
 to use ``additive constant $c$'' or
equivalently ``additive $O(1)$ term'' to mean a constant,
accounting for the length of a fixed binary program,
independent from every variable or parameter in the expression
in which it occurs. In this paper we use the prefix complexity
variant of Kolmogorov complexity for convenience. Since 
(in)equalities in the Kolmogorov complexity setting
typically hold up to an additive constant, we use a special notation.

We will denote by $\lea$ an
inequality to within an additive constant. More precisely, let $f,g$
be functions from $\{0,1\}^*$ to ${\cal R}$, 
the {\em real numbers}. Then by `$f(x) \lea g(x)$'
we mean that there exists a $c$ such that for all $x \in \{0,1\}^*$,
$f(x) < g(x) + c$.  We denote by $\eqa$ the situation when both $\lea$
and $\gea$ hold.

\paragraph{Probabilistic Notions:}
Let ${\cal X}$ be a finite or countable set. A function $f: {\cal X}
\rightarrow [0,1]$ is a {\em probability mass function} if $\sum_{x
  \in {\cal X}} f(x) = 1$. We call $f$ a {\em sub-probability mass
  function} if $\sum_{x \in {\cal X}} f(x) \leq 1$. Such sub-probability
mass functions will sometimes be used for technical convenience. We
can think of them as ordinary probability mass functions by
considering the surplus probability to be concentrated on an undefined
element $u \not\in {\cal X}$.

In the context of (sub-) probability mass functions, ${\cal
  X}$ is called the {\em sample space}. Associated with mass function $f$ and
sample space ${\cal X}$ is the {\em random variable\/} $X$ and the
probability distribution $P$ such that $X$ takes value $x \in {\cal
  X}$ with probability $P(X=x) = f(x)$. A subset of ${\cal X}$ is
called an {\em event}. We extend the probability of individual
outcomes to events.  With this terminology, $P(X= x) = f(x)$ is the
probability that the singleton event $\{x\}$ occurs, and $P(X \in
{\cal A}) = \sum_{x \in {\cal A}} f(x)$. In some cases (where the use
of $f(x)$ would be confusing) we write $p_x$ as an abbreviation of
$P(X= x)$.  In the sequel, we often refer to probability distributions
in terms of their mass functions, i.e. we freely employ phrases like
`Let $X$ be distributed according to $f$'.

Whenever we refer to probability mass functions without explicitly
mentioning the sample space ${\cal X}$ is assumed to
be ${\cal N}$ or, equivalently, $\{ 0,1\}^*$.

For a given probability mass function $f(x,y)$ on sample space ${\cal
  X} \times {\cal Y}$ with random variable $(X,Y)$, we define the {\em
  conditional probability mass function\/} $f(y \mid x)$ of outcome
$Y=y$ given outcome $X=x$ as
$$
f (y|x) :=  {f(x,y)  \over \sum_{y}  f(x,y)}.
$$
Note that $X$ and $Y$ are not necessarily independent. 

In some cases (esp. Section~\ref{sec:relpa} and
Appendix~\ref{sec:universal}), the notion of {\em sequential
  information source\/} will be needed. This may be thought of as a
probability distribution over arbitrarily long binary sequences, of
which an observer gets to see longer and longer initial segments.
Formally, a sequential information source $P$ is a probability
distribution on the set $\{0,1\}^{\infty}$ of one-way infinite
sequences. It is characterized by a {\em sequence of probability mass
  functions\/} $(f^{(1)},f^{(2)}, \ldots)$ where
$f^{(n)}$ is a probability mass function on $\{0,1\}^n$ that 
denotes the {\em marginal\/} distribution of
$P$ on the first $n$-bit segments. By definition, the sequence 
$f \equiv (f^{(1)}, f^{(2)},
\ldots)$ represents a sequential information source if for all $n >
0$, $f^{(n)}$ is related to $f^{(n+1)}$ as follows: for all $x \in
\{0,1\}^n$, $\sum_{y \in \{0,1\}} f^{(n+1)}(xy) = f^{(n)}(x)$ and
$f^{(0)}(x)=1$. This is also called Kolmogorov's {\em compatibility
  condition\/} \cite{Ri89}.

Some (by no means all!) probability mass functions on $\{ 0,1\}^*$ can
be thought of as information sources. Namely, given a probability mass
function $g$ on $\{0,1 \}^*$, we can define $g^{(n)}$ as the
conditional distribution of $x$ given that the length of $x$ is $n$,
with domain restricted to $x$ of length $n$.  That is, $g^{(n)}:
\{0,1\}^n \rightarrow [0,1]$ is defined, for $x \in \{0,1\}^n$, as 
$g^{(n)}(x) = g(x) / \sum_{y \in \{0,1\}^n} g(y)$. Then $g$ can be
thought of as an information source if and only if the sequence 
$(g^{(1)}, g^{(2)},  \ldots)$ represents an information source.
\paragraph{Computable Functions:} 
Partial functions on the natural numbers ${\cal N}$ are 
functions $f$ such that $f(x)$ can be `undefined' for some $x$. We
abbreviate `undefined' to `$\uparrow$'. A
central notion in the theory of computation is that of the {\em
  partial recursive functions}. Formally, a function $f: {\cal N}
\rightarrow {\cal N} \cup \{ \uparrow \}$ is called {\em partial
  recursive\/} or {\em computable\/} if there exists a Turing Machine
$T$ that implements $f$. This means that for all $x$
\begin{enumerate}
\item 
If $f(x) \in {\cal N}$, then $T$, 
when run with input $x$ outputs $f(x)$ and then halts. 
\item
If $f(x) = \uparrow$ (`$f(x)$ is undefined'), then $T$ with input $x$ never halts.
\end{enumerate}
Readers not familiar with computation theory may think of a Turing
Machine as a computer program written in a general-purpose language such as
C or Java.

A function $f: {\cal N} \rightarrow {\cal N} \cup \{ \uparrow \}$ is
called {\em total\/} if it is defined for all $x$ (i.e. for all $x$,
$f(x) \in {\cal N}$). A {\em total recursive\/} function is thus a
function that is implementable on a Turing Machine that halts on all
inputs. These definitions are extended to several arguments as
follows: we fix, once and for all, some standard invertible pairing
function $\langle \cdot, \cdot \rangle: {\cal N} \times {\cal N}
\rightarrow {\cal N}$ and we say that $f: {\cal N} \times {\cal N}
\rightarrow {\cal N} \cup \{ \uparrow \}$ is computable if there
exists a Turing Machine $T$ such that for all $x_1, x_2$, $T$ with
input $\langle x_1, x_2 \rangle$ outputs $f(x_1,x_2)$ and halts if
$f(x_1,x_2) \in {\cal N}$ and otherwise $T$ does not halt. By
repeating this construction, functions with arbitrarily many arguments
can be considered.

{\em Real-valued Functions:} We call a
distribution $f: {\cal N} \rightarrow {\cal R}$ {\em recursive\/} or
{\em computable\/} if there exists a Turing machine that, when input
$\langle x, y\rangle$ with $x \in \{0,1\}^*$ and $y \in {\cal N}$,
outputs $f(x)$ to precision $1/y$; more precisely, it outputs a pair
$\langle p, q \rangle$ such that $| p/q - |f(x)| | < 1/y $ and an
additional bit to indicate whether $f(x)$ larger or smaller than $0$.
Here $\langle \cdot, \cdot \rangle$ is the standard pairing function.
In this paper all real-valued functions we consider are by definition
total. Therefore, in line with the above definitions, for a
real-valued function `computable' (equivalently, recursive), means
that there is a Turing Machine which for {\em all\/} $x$, computes
$f(x)$ to arbitrary accuracy; `partial' recursive real-valued
functions are not considered.

It is convenient to distinguish between {\em upper\/} and {\em lower
  semi-computability}.  For this purpose we consider both the argument
of an auxiliary function $\phi$ and the value of $\phi$ as a pair of
natural numbers according to the standard pairing function $\langle
\cdot \rangle$. We define a function from ${\cal N}$ to the reals
${\cal R}$ by a Turing machine $T$ computing a function $\phi$ as
follows. Interpret the computation $\phi(\langle x,t \rangle ) =
\langle p,q \rangle$ to mean that the quotient $p/q$ is the rational
valued $t$th approximation of $f(x)$.
\begin{definition}\label{def.enum.funct}
\rm
\label{def.semi}
A function $f: {\cal N} \rightarrow {\cal R}$ is
{\em lower semi-computable} if there is a Turing machine $T$ computing a
total function $\phi$ 
such that $\phi (x,t+1) \geq \phi (x,t)$ and
$\lim_{t \rightarrow \infty} \phi (x,t)=f(x)$. This means
that $f$ can be computably approximated from below.
A function $f$ is {\em upper semi-computable} if
$-f$ is lower semi-computable, 
Note that, if $f$ is both upper- and lower semi-computable, then
$f$ is computable.
\end{definition}

{\em (Sub-) Probability mass functions\/:} Probability mass
functions on $\{0,1\}^*$ may be thought of as real-valued functions on
${\cal N}$. Therefore, the definitions of `computable' and
`recursive' carry over unchanged from the real-valued function case.
\subsection{Codes}
\label{sec:coding}
We repeatedly consider the following scenario: a {\em
  sender\/} (say, A) wants to communicate or transmit some information
  to a {\em receiver\/} (say, B). The information to be transmitted is
  an element from some set ${\cal X}$ (This set may or may not consist
  of binary strings). 
It will be communicated by sending a 
binary string, called the {\em message}. 
When B receives the message, he can decode it again and (hopefully)
  reconstruct the element of ${\cal X}$ that was sent.
To achieve this, A and B need to agree
  on a {\em code\/} or {\em description method\/} before
  communicating. Intuitively, this is a binary relation between {\em
  source words} and associated {\em code words}. The relation is fully
  characterized by the {\em decoding function}. Such a decoding function
$D$ can be any function $D: \{ 0, 1 \}^* \rightarrow {\cal X}$.
The domain of $D$ is the set of %
\it code words
\rm and the range of $D$ is the set of %
\it source words. \rm $D(y) = x$ is interpreted as ``$y$ is a code
word for the source word $x$''.
The set of all code words
for source word $x$ is the set $D^{-1} (x) = \{ y: D(y) = x \}$.
Hence, $E=D^{-1}$ can be called the %
\it encoding %
\rm substitution
($E$ is not necessarily a function). With each code $D$ we can
associate a {\em length function\/} $L_D: {\cal X} \rightarrow {\cal N}$ 
such that, for each source
word $x$, $L(x)$ is the length of the shortest encoding of $x$:
$$
L_D(x) = \min \{ l(y): D(y) = x  \}.
$$
We denote by $x^*$ the shortest $y$ such that $D(y) = x$; if there is
more than one such $y$, then $x^*$ is defined to be the
first such $y$ in some agreed-upon order---for example,
the lexicographical order.

In coding theory attention is often restricted to
the case where the source word set is finite, say
${\cal X} =  \{  1, 2,  \ldots , N  \}  $. If there is a constant $l_0$
such that $l(y) = l_0$ for all code words $y$ (which implies, $L(x) =
l_0$ for all source words $x$),
then we call $D$ a %
\it fixed-length
\rm code. It is
easy to see that $l_0   \geq   \log N$.
For instance, in teletype transmissions the source
has an alphabet of $N = 32$ letters, consisting
of the 26 letters in the Latin alphabet plus
6 special characters. Hence, we need $l_0 = 5$
binary digits per source letter. In electronic computers
we often use the fixed-length ASCII code\index{code!ASCII}
with $l_0=8$.
\paragraph{Prefix code:}
It is immediately clear that in general
we cannot uniquely recover $x$ and $y$ from $E(xy)$.
Let $E$ be
the identity mapping.
Then we have $E(00)E(00) = 0000 = E(0)E(000)$.
We now introduce {\em prefix codes}, which do not suffer from this defect. 
A binary string $x$
is a {\em proper prefix} of a binary string $y$
if we can write $y=xz$ for $z \neq \epsilon$.
 A set $\{x,y, \ldots \} \subseteq \{0,1\}^*$
is {\em prefix-free} if for any pair of distinct
elements in the set neither is a proper prefix of the other.
A function $D: \{ 0, 1 \}^*  \rightarrow  {\cal N}$
defines a {\it prefix-code}\index{code!prefix-}
if its domain is prefix-free.
In order to decode a code sequence of a prefix-code,
we simply start at the beginning and decode one
code word at a time. When we come to the end of
a code word, we know it is the end, since no
code word is the prefix of any other code word
in a prefix-code.

Suppose we encode each binary string $x=x_1 x_2 \ldots x_n$ as
\[ \bar x = \underbrace{11 \ldots 1}
_{n \mbox{{\scriptsize  \ times}}}0x_1x_2 \ldots x_n .\]
The resulting code is prefix because we can determine where the
code word $\bar x$ ends by reading it from left to right without
backing up. Note $l(\bar{x}) = 2n+1$; thus, we have encoded strings in
$\{0,1\}^*$ in a prefix  manner at the price of doubling their
length. We can  get a much more efficient code by applying the
construction above to the length $l(x)$ of $x$ rather than $x$ itself:
define $x'=\overline{l(x)}x$, where $l(x)$ is interpreted as a binary
string according to the correspondence (\ref{eq:correspondence}). Then the code $D'$ with
$D'(x') = x$ is a prefix  code satisfying, for all $x \in
\{0,1\}^*$, $l(x') = n+2 \log n+1$ (here we ignore the `rounding error'
in \eqref{eq:intlength}). $D'$ is used throughout this paper as
a standard code to encode natural numbers in a prefix free-manner; we call it
the {\em standard prefix-code for the natural numbers}. We use
$\Lint(x)$ as notation for $l(x')$. When $x$ is
interpreted as an integer (using the correspondence
(\ref{eq:correspondence}) and (\ref{eq:intlength})), we see that,
up to rounding,
$\Lint(x) = \log x +
2 \log \log x+1$. 

\paragraph{Prefix codes and the Kraft inequality:}
Let ${\cal X}$ be the set of natural numbers and
consider the straightforward non-prefix representation 
(\ref{eq:correspondence}).
There are two elements of ${\cal X}$ with
a description of length $1$, four with a description of
length $2$ and so on. However, for a prefix code $D$ for the natural numbers
there are less binary prefix code words of each length: 
if $x$ is a prefix code word
then no $y = xz$ with $z \neq \epsilon$ is a prefix code word. 
Asymptotically there are less prefix code words of length $n$
than the $2^n$ source words of length $n$.
Quantification of this intuition for countable ${\cal X}$ and
arbitrary prefix-codes leads to
a precise constraint on the number of code-words of given lengths.
This important relation is known as the
{\em Kraft Inequality\index{Kraft Inequality|bold}}
and is due to L.G. Kraft\index{Kraft, L.G.} \cite{Kr49}.
\begin{theorem}
\label{kraft}
Let
$l_1 , l_2 , \ldots $
be a finite or infinite sequence
of natural numbers.
There is a prefix-code with this sequence as
lengths of its binary code words iff
$$
\sum_n  2^{{-}  l_n }  \leq 1.
$$
\end{theorem}
\paragraph{Uniquely Decodable Codes:}
We want to code elements of ${\cal X}$ in a way that they can be
uniquely reconstructed from the encoding. Such codes are called
`uniquely decodable'.
Every prefix-code is a uniquely decodable code. For example, if
$E(1) = 0$, $E(2) = 10$, $E(3) = 110$, $E(4) = 111$
%
then
$1421$ is encoded as $0111100$, which can be
easily decoded from left to right in 
a unique way.

On the other hand, not every uniquely decodable code satisfies the prefix
condition. 
Prefix-codes are
distinguished from other uniquely decodable codes
by the property that the end of a code word is always
recognizable as such. This means that decoding
can be accomplished without the delay of observing
subsequent code words, which is why prefix-codes
are also called instantaneous codes.

There is 
good reason for our emphasis on prefix-codes.  
Namely, it turns out that
Theorem~\ref{kraft} stays valid if we replace
``prefix-code'' by ``uniquely decodable code.'' 
This important fact means that every
uniquely decodable code can be replaced
by a prefix-code without changing the set of
code-word lengths. 
In Shannon's and Kolmogorov's theories, we are only interested in code
word {\em lengths\/} of uniquely decodable codes rather than actual
encodings. By the previous
argument, we may restrict the set of codes we work with to prefix
codes, which are much easier to handle.
\paragraph{Probability distributions and complete prefix codes:}
A uniquely decodable code is %
\it complete\index{code!uniquely decodable} \rm if the addition of any
new code word to its code word set results in a non-uniquely decodable
code.  It is easy to see that a code is complete iff equality holds in
the associated Kraft Inequality.  Let $l_1, l_2, \ldots$ be the
code words of some complete uniquely decodable code. Let us define
$q_x = 2^{- l_x}$.  By definition of completeness, we have $\sum_x q_x
= 1$. Thus, the $q_x$ can be thought of as {\em probability mass
  functions\/} corresponding to some probability distribution $Q$. We
say $Q$ is the distribution {\em corresponding\/} to $l_1,l_2,\ldots$.
In this way, each complete uniquely decodable code is mapped to a
unique probability distribution. Of course, this is nothing more than
a formal correspondence: we may choose to encode outcomes of $X$ using
a code corresponding to a distribution $q$, whereas the outcomes are
actually distributed according to some $p \neq q$. But, as we show
below, if $X$ is distributed according to $p$, then the code to which
$p$ corresponds is, in an average sense, the code that achieves
optimal compression of $X$.
\section{Shannon Entropy versus Kolmogorov Complexity}
\label{sec:basic}
\subsection{Shannon Entropy} 
\label{sec:shannon}
It seldom happens that a detailed mathematical theory springs forth in
essentially final form from a single publication. Such was the case
with Shannon information theory, which properly started only with the
appearance of C.E. Shannon's paper ``The mathematical theory of
communication'' \cite{Sh48}.
In this paper, Shannon proposed a measure of 
information in a distribution, which he
called the `entropy'. The
entropy $H(P)$ of a distribution $P$ measures the
`the inherent  uncertainty in $P$', or (in fact
equivalently), `how much information is gained when an outcome of $P$
is observed'. To make this a bit more precise, let us imagine an
observer  who knows that $X$ is distributed
according to $P$. The observer then observes $X=x$. The entropy of $P$
stands for the `uncertainty of the observer about the outcome $x$
{\em before\/} he observes it'. Now think of the observer as a
`receiver' who receives the message conveying the value of $X$. From this dual point of
view, the entropy stands for  
\begin{quote}
the average amount of information that the observer has gained {\em after\/}
receiving a realized outcome $x$ of the random variable $X$. $(*)$
\end{quote}
Below, we first give Shannon's mathematical definition of entropy, and
we then connect it to its intuitive meaning $(*)$.
\begin{definition} \rm Let ${\cal X}$ be a finite or countable
\label{def.entropy}
set, let $X$ be a random variable taking values in ${\cal X}$ with
  distribution $P(X=x)=p_x$. Then
the  (Shannon-) 
\it entropy\index{entropy|bold}\index{$H$: entropy stochastic source} %
\rm of random variable $X$
is given by
\begin{equation}
\label{eq:entropy}
H(X)  =   \sum_{x \in {\cal X}} p_x \log 1/p_x ,
\end{equation}
Entropy is defined here as a functional mapping random
variables to real numbers. In many texts, entropy is, essentially
equivalently, defined as a map from {\em distributions\/} of random variables to 
the real numbers. Thus, by definition:
$
H(P) := H(X) =  \sum_{x \in {\cal X}} p_x \log 1/ p_x
$.
\end{definition}
\paragraph{Motivation:} The entropy function \eqref{eq:entropy}
can be motivated in different ways. The two most
important ones are the {\em axiomatic\/} approach and the {\em coding
  interpretation}.  In this paper we concentrate on the latter, but we
first briefly sketch the former. The idea of the axiomatic approach is
to postulate a
small set of self-evident axioms that
any measure of information relative to a distribution should
satisfy. One then shows that the only measure satisfying all the
postulates is the Shannon entropy. We
outline this approach for
finite sources ${\cal X} = \{1,\ldots, N\}$. We look for a function
$H$ that maps probability distributions on ${\cal X}$ to real
numbers. For given distribution $P$, $H(P)$ should measure
`how much information is gained on average  when an outcome is made
available'. We can write $H(P) = H(p_1,\ldots,p_N)$ where
$p_i$ stands for the
  probability of $i$. 
Suppose we require that
\begin{enumerate}
\item $H(p_1,\ldots,p_N)$ is continuous in $p_1,\ldots,p_N$.
\item If all the $p_i$ are equal, $p_i = 1/N$, then $H$ should be a
  monotonic increasing function of $N$. With equally likely events  
there is more choice, or uncertainty, when there are more possible
events.
\item If a choice is broken down into two successive choices, the
  original $H$ should be the weighted sum of the individual values of
  $H$. Rather than formalizing this condition, we will give a specific
  example. Suppose that ${\cal X} = \{ 1,2,3\}$, and $p_1 = \frac{1}{2}, p_2 =
  1/3, p_3 = 1/6$. We can think of $x \in {\cal X}$ as being generated
  in a two-stage process. First, an outcome in ${\cal X'} =\{0,1\}$ is
  generated according to a distribution $P'$ with
 $p'_0 = p'_1 = \frac{1}{2}$. If $x'=1$, we set $x=1$ and the process
 stops. If $x'= 0$, then  outcome `$2$' is generated with probability
 $2/3$ and outcome `$3$' with probability $1/3$, and the process
 stops. The final results
 have the same probabilities as before. In this particular case we
 require that 
$$H(\frac{1}{2},\frac{1}{3},\frac{1}{6}) = H(\frac{1}{2},\frac{1}{2}) + \frac{1}{2} H(\frac{2}{3},\frac{1}{3}) + \frac{1}{2} H(1).$$ 
Thus, the entropy of $P$ must be equal to entropy of the first
 step in the generation process, plus the weighted sum (weighted
 according to the probabilities in the first step) of the entropies of the
 second step in the generation process. 

As a special case, if ${\cal X}$
 is the $n$-fold product space of another space ${\cal Y}$, $X =
 (Y_1,\ldots, Y_n)$ and the $Y_i$ are all independently distributed
 according to $P_Y$, then $H(P_X) = n H(P_Y)$. For example, the total
 entropy of $n$ independent tosses of a coin with bias $p$ is $n
 H(p,1-p)$.
\end{enumerate}
\begin{theorem}
\label{thm:axiomatic}
The only $H$ satisfying the three above assumptions is of the form
$$
H =  K \sum_{i=1}^N p_i \log 1/p_i,
$$
with $K$ a constant.
\end{theorem}
Thus, requirements (1)--(3) lead us to the definition of entropy
(\ref{eq:entropy}) given above up to an (unimportant) scaling
factor. We shall give a concrete interpretation of this factor later
on. Besides  the defining characteristics (1)--(3), the function $H$ has a few other
properties that make it attractive as a measure of information. 
We mention:
\begin{description}
\item[\rm 4.]  $H(p_1,\ldots,p_N)$ is a concave function of the $p_i$.
\item[\rm 5.] For each $N$, $H$ achieves its unique maximum for the uniform distribution $p_i =
  1/N$.
\item[\rm 6.] $H(p_1,\ldots,p_N)$ is zero iff one of the $p_i$ has value $1$.
  Thus, $H$ is zero if and only if we do not gain any information at
  all if we are told that the outcome is $i$ (since we already knew
  $i$ would take place with certainty).
\end{description}
\paragraph{The Coding Interpretation:}
Immediately after 
stating Theorem~\ref{thm:axiomatic}, Shannon \cite{Sh48} continues, ``this theorem, and the
assumptions required for its proof, are in no way necessary for the
present theory. It is given chiefly to provide a certain plausibility
to some of our later definitions. The {\em real justification\/} of these
definitions, however, will reside in their implications''.
Following this injunction, we emphasize the main practical
interpretation of entropy as the length
(number of bits) needed to encode outcomes in ${\cal X}$. This
provides much clearer intuitions,
it lies at the root of the many practical applications
of information theory, and, most importantly for us, 
it simplifies the comparison to Kolmogorov complexity.
 
\begin{example}
\rm
The entropy
of a random variable $X$ with equally likely outcomes
in a finite sample space ${\cal X}$ is given by
$H(X) = \log |{\cal X}|$. 
By choosing a particular message $x$ from ${\cal X}$,
we remove the entropy from $X$ by the
assignment $X := x$ and produce
or transmit {\em information}\index{information}
$I = \log |{\cal X}|$ by our selection of $x$. We  show below
that $I = \log |{\cal X}|$ (or, to be more precise, the integer 
$I' =  \lceil
\log |{\cal X}|  \rceil $) can be interpreted as the number of bits
needed to be transmitted from an (imagined) sender 
to an (imagined) receiver.
\end{example}
We now connect entropy to minimum average code lengths. These are
defined as follows:
\begin{definition}
\rm
Let source words $x \in \{0,1\}^*$ be
produced by a random variable $X$ with probability
$P(X=x)=p_x$ for the event $X=x$. The characteristics of 
$X$ are fixed. Now consider prefix codes
$D: \{0,1\}^* \rightarrow {\cal N}$ 
with one code word per source word,
and denote the length of the code word for $x$ by $l_x$.
We want to minimize the expected number of bits 
we have to transmit for the given
source $X$ and choose a prefix code $D$ that achieves this.
In order to do so, we must minimize the
\it average code-word length\index{code!average word length|bold}
\rm 
$\bar{L}_{D} = \sum_x p_x l_x$%
\rm .
We define the %
\it minimal average code word
length
\rm as $\bar{L} = \min   \{  \bar{L}_{D}: D \mbox{ is a prefix-code}\}  $.
A prefix-code $D$ such that $\bar{L}_{D} = \bar{L}$ is called
an %
\it optimal prefix-code\index{code!optimal prefix-}
\rm with respect to prior
probability $P$ of the source words.
\end{definition}
The (minimal) average code length of an
(optimal) code does not depend on the details of the set
of code words, but only on the set of code-word lengths.
It is just the expected code-word length
with respect to the given distribution.
Shannon\index{Shannon, C.E.} discovered that the
minimal average code word 
length is about equal to the entropy of
the source word set. This is known as the
{\it Noiseless Coding Theorem}.\index{Theorem!Noiseless Coding|bold}
The adjective ``noiseless'' emphasizes that we ignore the possibility
of errors.
\begin{theorem}
\label{thm:noiseless} 
Let $\bar{L}$ and $P$ be as above.
If $H(P) =  \sum_x p_x  \log 1/ p_x$
is the entropy\index{entropy}, then
\begin{equation}
\label{eq:entopt}
H(P)  \leq \bar{L}  \leq H(P) + 1.
\end{equation}
\end{theorem}
We are typically interested in encoding a binary string
of length $n$ with  entropy proportional to $n$
(Example~\ref{ex:universal}). The essence of
(\ref{eq:entopt})  is that, 
for all but the smallest $n$, the difference between 
entropy and minimal expected
code length is completely negligible.

It turns out that the optimum $\bar{L}$ in (\ref{eq:entopt}) is relatively easy to achieve,
with the Shannon-Fano code.
Let there be $N$ symbols
(also called basic messages or source words).
Order these symbols
according to decreasing probability,
say ${\cal X} = \{ 1,2, \ldots ,N \}$ with probabilities $p_1 ,p_2 , \ldots ,p_N$.
Let $P_r = \sum_{i=1}^{r-1} p_i$, for $r = 1, \ldots ,N$.
The binary code $E: {\cal X} \rightarrow \{0,1\}^*$ is obtained
by coding $r$ as a binary number $E(r)$, obtained by
truncating the binary expansion of $P_r$ at length
$l(E(r))$ such that
$$
 \log 1/ p_r   \leq l(E(r))  <  1 + \log 1/ p_r .
$$
This code is the {\em Shannon-Fano code}.
It has the property that highly probable symbols
are mapped to short code words and symbols with low
probability are mapped to longer code words (just like in a less optimal,
non-prefix-free, setting is done in the Morse code). 
Moreover,
$$
2^{-l(E(r))}  \leq p_r  <  2^{-l(E(r))+1} .
$$
Note that the code for symbol $r$ differs from all
codes of symbols $r+1$ through $N$ in one or more
bit positions, since for all $i$  with $ r+1  \leq i  \leq N$,
\[ P_i \geq P_r + 2^{-l(E(r))}.\]
Therefore the binary
expansions of $P_r$ and $P_i$ differ in the first $l(E(r))$
positions.  This means that $E$ is one-to-one,
and it has an inverse: the decoding mapping $E^{-1}$.
Even better, 
since
no value of $E$ is a prefix of any other value of $E$,
the set of code words is
a prefix-code\index{code!prefix-}. This means we
can recover the source message
from the code message
by scanning it from left to right
without look-ahead.
If $H_1$ is the average
number of bits used per symbol of an original
message, then $H_1 = \sum_r  p_r l(E(r))$.
Combining this with the previous inequality we obtain (\ref{eq:entopt}):
$$
 \sum_r  p_r \log 1/ p_r    \leq
H_1  <  
\sum_r  (1+ \log 1/ p_r )p_r  = 1 + \sum_r  p_r \log 1/ p_r .
$$
\commentout{
\begin{example}
\label{ex:00}
\rm
Assuming that $x$ is emitted by a random source $X$
with probability $P(X=x)$, we can transmit $x$ using the Shannon-Fano
code. This uses (up to rounding) $ \log 1/ P(X=x)$ bits.
By Shannon's noiseless coding theorem this is optimal {\em on average},
the average taken over the probability distribution of outcomes
from the source. Thus, if $x = 00 \ldots 0$ ($n$ zeros), and the
random source emits $n$-bit messages with equal probability $1/2^n$
each, then we require $n$ bits to transmit $x$ (the same as 
transmitting $x$ literally). However, we can transmit $x$ 
in about $\log n$ bits if we ignore probabilities and
just describe $x$ individually. Thus, the optimality with
respect to the average may be very sub-optimal in individual cases. 
\end{example}
}

\ \\
{\bf Problem and Lacuna:}
Shannon observes, ``Messages have %
\index{Shannon, C.E.}
\it meaning %
\rm [ $\ldots$ however $\ldots$ ]
the semantic aspects of communication are irrelevant
to the engineering problem.'' In other words, can we answer a
question like ``what is the information in this book''
by viewing it as an element of a set of possible books
with a probability distribution on it? Or that the
individual sections in this book form
a random sequence with stochastic relations that
damp out rapidly over a distance of several pages?
And how to measure the quantity of hereditary information in
biological organisms, as encoded in DNA? Again there is the
possibility of seeing a particular form of animal as one of a set of
possible forms with a probability distribution on it. This seems
to be contradicted by the fact that the calculation of
all possible lifeforms in existence at any one time on earth
would give a ridiculously low figure like 
$2^{100}$.

Shannon's classical
information theory\index{information theory}\index{Shannon, C.E.}
assigns a quantity of information to an ensemble of
possible messages. All messages in the ensemble being equally probable,
this quantity is the number of bits needed to
count all possibilities. 
This expresses the fact that
each message in the ensemble can be communicated
using this number of bits.
However, it does not say
anything about the number of bits needed to convey any
individual message in the ensemble. To illustrate this,
consider the ensemble consisting of all binary strings
of length 9999999999999999.

By Shannon's measure, we require
9999999999999999 bits
on the average to encode a string in such an ensemble. However, the
string consisting of 9999999999999999 1's can be encoded in about
55 bits by expressing 9999999999999999 in binary and adding the
repeated pattern ``1.'' A requirement for this to work is
that we have agreed on an algorithm that decodes the encoded
string. We can compress the string still further when we note that
9999999999999999 equals $3^2 \times 1111111111111111$, and that
1111111111111111 consists of $2^4$ 1's.

Thus, we
have discovered an interesting phenomenon: the description of
some strings can be compressed considerably,
provided they exhibit enough regularity. 
However, if regularity is lacking, it becomes more cumbersome
to express large numbers. For instance, it seems easier to 
compress the number ``one billion,'' than the number
``one billion seven hundred thirty-five million two hundred 
sixty-eight thousand and three hundred ninety-four,'' even though they
are of the same order of magnitude.

We are interested in a measure of information that, unlike Shannon's,
does not rely on (often untenable) probabilistic assumptions, 
and that takes into account the phenomenon that
`regular' strings are compressible. Thus, we aim for a measure of information
content of an %
\it individual finite object%
\rm ,
and in the information conveyed about an individual finite
object by another individual finite object. Here, we want
the information content of an object $x$ to be
an attribute of $x$ %
\it alone%
\rm , and not to depend
on, for instance, the means chosen to describe this information
content.  Surprisingly, this turns
out to be possible, at least to a large extent. The resulting theory
of information is based on Kolmogorov complexity, a
notion independently proposed by Solomonoff (1964), Kolmogorov (1965)
and Chaitin (1969); Li and Vit\'anyi (1997) describe the history of the
subject.
\subsection{Kolmogorov Complexity}
\label{sec:kolmogorov}
Suppose we want to describe a given object by a
finite binary string. We do not care whether the object
has many descriptions; however, each description
should describe but one object.
From among all descriptions
of an object we
can take the length of the shortest description as
a measure of the object's complexity.
It is natural to call an object ``simple'' if it has
at least one short description, and to call it ``complex''
if all of its descriptions are long.

As  in Section~\ref{sec:coding}, consider a description method
$D$, to be used to transmit messages from a sender to a receiver. 
%
%
If $D$ is known
to both a sender and receiver, then a message $x$ can be transmitted
from sender to receiver by transmitting the description $y$ with
$D(y)=x$. The cost of this transmission is measured by $l(y)$,
the length of $y$. The least cost of transmission of $x$ is determined
by the length function $L(x)$: recall that $L(x)$ is the length of 
the shortest $y$ such that $D(y)=x$. 
We choose this length function
as the descriptional complexity of $x$ under specification
method $D$.
 
Obviously, this descriptional complexity of 
$x$ depends crucially
on $D$.
The general principle involved is that the syntactic
framework of the description language
determines the succinctness of description.

In order to objectively compare descriptional complexities
of objects, to be able to say ``$x$ is more complex than $z$,''
the descriptional complexity of $x$
should depend on $x$ alone. This complexity can be viewed as related to
a universal description method that is a priori
assumed by all senders and receivers.
This complexity is optimal if no other description method
assigns a lower complexity to any object.

We are not really interested in optimality with respect to
all description methods.
For specifications to be useful at all it is
necessary that the mapping from $y$ to $D(y)$
can be executed in an effective manner. That is,
it can at least in principle be performed by humans or machines.
This notion has been 
formalized as that of ``partial recursive functions'', 
also known simply as ``computable functions'', which are
formally defined later. 
According to
generally accepted mathematical viewpoints it coincides
with the intuitive notion of effective computation.
 
The set of partial recursive functions
contains an optimal function that minimizes
description length of every other such function. We denote
this function by $D_0$.
Namely, for any other recursive function $D$,
for all objects $x$,
there is a description $y$ of $x$ under $D_0$ that is
shorter than any description $z$ of $x$ under $D$. (That is,
shorter up to an
additive constant that is independent of $x$.)
Complexity with respect to $D_0$ minorizes 
the complexities with respect
to all partial recursive functions.

We identify the
length of the description of $x$ with respect
to a fixed specification function $D_0$ with
the ``algorithmic (descriptional) complexity'' of $x$.
The optimality of $D_0$ in the sense above
means that the complexity of an object $x$
is invariant (up to an additive constant
independent of $x$) under transition
from one optimal specification function to another.
Its complexity is an objective attribute
of the described object alone: it is an
intrinsic property of that object, and it does
not depend on the description formalism.
This complexity can be viewed as ``absolute information content'':
the amount of information that needs to be transmitted
between all senders and receivers when they communicate the
message in absence of any other a priori knowledge
that restricts the domain of the message.
%
%
%
Thus, we have outlined the program for
a general theory of algorithmic complexity.
The three 
major innovations are as follows:
\begin{enumerate}
\item
In restricting
ourselves to formally effective descriptions,
our definition covers every form of description
that is intuitively acceptable as being effective
according to general viewpoints in mathematics and logic.
\item
The restriction to effective descriptions
entails that there is a universal description
method that minorizes the description length or complexity
with respect to any other effective description
method.
Significantly, this implies Item 3.
\item
The description length or complexity of an object
is an intrinsic attribute of the object independent
of the particular description method or formalizations
thereof.
\end{enumerate}

\subsubsection{Formal Details}
The Kolmogorov complexity $K(x)$ of a finite object $x$
will be defined as the length of the
shortest effective binary description of $x$. Broadly speaking, $K(x)$
may be thought of as the length of the shortest computer program that
prints $x$ and then halts. This computer program may be written in
C, Java, LISP or any other universal language: we shall see that,
for any two universal languages, 
the resulting program lengths differ at most by a constant not
depending on $x$.

To make this precise, 
let $T_1 ,T_2 , \ldots$ be a standard enumeration \cite{LiVi97}
of all Turing machines, and let $\phi_1 , \phi_2 , \ldots$
be the enumeration of corresponding functions
which are computed by the respective Turing machines.
That is, $T_i$ computes $\phi_i$.
These functions are the {\em partial recursive} functions
or {\em computable} functions, Section~\ref{sec:preliminaries}. For technical reasons we are interested in the
so-called prefix complexity, which is associated with Turing machines
for which the set of programs (inputs) resulting in a halting computation
is prefix free\footnote{There exists a version of Kolmogorov
  complexity corresponding to programs that are not necessarily
  prefix-free, but we will not go into it here.}. We can realize this by equipping the Turing
machine with a one-way input tape, a separate work tape,
and a one-way output tape. Such Turing
machines are called prefix machines
since the halting programs for any one of them form a prefix free set.
%

We first define $K_{T_i}(x)$, the prefix Kolmogorov complexity of $x$ relative to a
given prefix machine $T_i$, where $T_i$ is the $i$-th prefix machine
in a standard enumeration of them. $K_{T_i}(x)$  is defined as the length of the shortest
input sequence $y$ such that $T_i(y) = \phi_i(y) = x$. If no such
input sequence exists, $K_{T_i}(x)$ remains undefined. Of course, this
preliminary definition is still highly sensitive to the particular
prefix machine $T_i$ that we use. But now the  `universal
prefix machine' comes to our rescue. Just as there exists universal ordinary
Turing machines, there also exist universal prefix machines. These
have the remarkable property that they can simulate every other prefix
machine. More specifically, there exists a prefix machine $U$ such
that, with as input the pair $\langle i, y\rangle$, it outputs $\phi_i(y)$
and then halts. We now fix, once and for all, 
a prefix machine $U$ with this property and call $U$ the {\em reference
  machine}. The Kolmogorov complexity $K(x)$ of $x$ is defined as $K_U(x)$. 

Let us formalize this definition.
Let $\langle \cdot \rangle$ be a standard invertible
effective one-one encoding from ${\cal N} \times {\cal N}$
to a prefix-free  subset of ${\cal N}$. $\langle \cdot \rangle$ may be
thought of as the encoding function of a prefix code. 
For example, we can set $\langle x,y \rangle = x'y'$.
Comparing to the definition of  in
Section~\ref{sec:preliminaries}, we note that from now on, we require
$\langle \cdot \rangle$ to map to a prefix-free set.
We insist on prefix-freeness and
effectiveness because we want a universal Turing
machine to be able to read an image under $\langle \cdot \rangle$ 
from left to right and
determine where it ends.
\begin{definition}\label{def.KolmK}
\rm
Let $U$ be our reference prefix machine satisfying for all $i \in {\cal N},
y \in \{0,1\}^*$, 
$U(\langle i,y \rangle) = \phi_i(y)$. The {\em prefix Kolmogorov complexity} of $x$ is
\begin{eqnarray}
K(x) & = &
\min_{z} \{ l(z) : U(z) = x , z \in \{0,1\}^*\} =  \nonumber \\
& = &  \min_{i,y}\{l(\langle i, y \rangle): \phi_i (y )=x , y \in \{0,1\}^*, i
\in {\cal N} \}.
\end{eqnarray}
\end{definition}
We can alternatively think of $z$ as a program that prints $x$ and
then halts, or as $z = \langle i,y \rangle$ where $y$ is a program such
that, when $T_i$ is input program $y$, it prints $x$ and then halts. 

Thus, by definition $K(x)=l(x^*)$, where $x^*$ is the
lexicographically first shortest
self-delimiting (prefix) program for $x$ with respect to the
reference prefix machine. Consider the mapping $E^*$ defined by $E^*(x)=x^*$.
This may be viewed as the encoding function of a prefix-code (decoding
function) $D^*$ with $D^*(x^*) = x$. By its definition, $D^*$ is  a
very parsimonious code. The reason for working with prefix rather than standard
Turing machines is that, for many of the subsequent developments, 
we need $D^*$ to be prefix.
%
%

Though defined in terms of a 
particular machine model, the Kolmogorov complexity
is machine-independent up to an additive
constant 
 and acquires an asymptotically universal and absolute character
through Church's thesis, from the ability of universal machines to
simulate one another and execute any effective process.
  The Kolmogorov complexity of an object can be viewed as an absolute
and objective quantification of the amount of information in it.
%

\subsubsection{Intuition}
To develop some intuitions, it is useful to think of $K(x)$ as
  the shortest program for $x$
in some standard programming language such as
  LISP or Java. Consider the lexicographical enumeration
  of all syntactically correct LISP programs $ \lambda_1, \lambda_2,
  \ldots$, and the lexicographical enumeration of all syntactically
  correct Java programs $ \pi_1, \pi_2, \ldots$. We assume that both
  these programs are encoded in some standard prefix-free manner. With
  proper definitions we can view the programs in both enumerations as
  computing partial recursive functions from their inputs to their
  outputs. Choosing reference machines in both enumerations we can
  define complexities $K_{\mbox{\scriptsize LISP}}(x)$ and
  $K_{\mbox{\scriptsize  Java}}(x)$
  completely analogous to $K(x)$.  All of these measures of the
  descriptional complexities of $x$ coincide up to a fixed additive
  constant. Let us show this directly for $K_{\mbox{\scriptsize LISP}}(x)$ and
  $K_{\mbox{\scriptsize Java}}(x)$. Since LISP is universal, there exists a LISP
  program $\lambda_P$ implementing a Java-to-LISP compiler.
  $\lambda_P$ translates each Java program to an equivalent LISP
  program. Consequently, for all $x$, $K_{\mbox{\scriptsize LISP}}(x) \leq
  K_{\mbox{\scriptsize Java}}(x) + 2l(P)$. Similarly, there is a Java program
  $\pi_L$ that is a LISP-to-Java compiler, so that for all $x$,
  $K_{\mbox{\scriptsize Java}}(x) \leq K_{\mbox{\scriptsize LISP}}(x) + 2l(L)$. It follows
  that $|K_{\mbox{\scriptsize Java}}(x) - K_{\mbox{\scriptsize LISP}}(x)| \leq 2l(P) + 2 l(L)$
  for all $x$!  

The programming  language view immediately tells us that $K(x)$ must be
small for `simple' or `regular' objects $x$. For example, 
there exists a fixed-size program that, when input
$n$, outputs the first $n$ bits of $
\pi$ and then halts. Specification of $n$ takes at most $L_{\cal N}(n)
= \log n + 2 \log \log n + 1 $ bits. Thus, if $x$
consists of the first $n$ binary digits of $\pi$, then $K(x) \lea \log
n + 2 \log \log n$. Similarly, if $0^n$ denotes the string
consisting of $n$ $0$'s, then $K(0^n) \lea \log n + 2 \log \log n$. 

On the other hand, for all $x$, there exists a program `print $x$;
 halt'. This shows that for all $K(x) \lea l(x)$. As was previously noted, for any prefix code,
 there are no more than $2^m$ strings $x$ which can be described by
 $m$ or less bits. In particular, this holds for the prefix code $E^*$
 whose length function is $K(x)$. Thus, the fraction of strings $x$ of
 length $n$ with $K(x) \leq m$ is at most  $2^{m-n}$: the overwhelming majority
 of sequences cannot be compressed by more than a
 constant. Specifically, if $x$ is determined by $n$ independent
 tosses of a fair coin, then with overwhelming probability,  $K(x) \approx
 l(x)$. Thus, while for very regular strings, the Kolmogorov complexity is
small (sub-linear in the length of the string), 
{\em most\/} strings are `random' and have Kolmogorov
complexity about equal to their own length.
\subsubsection{Kolmogorov complexity of sets, functions and
  probability distributions}
\paragraph{Finite sets:}
The  class of {\em finite sets} consists of the set
of finite subsets $S \subseteq \{0,1\}^*$. The  {\em complexity
of the finite set} $S$ is
$K(S)$---the length (number of bits) of the
shortest binary program $p$ from which the reference universal
prefix machine $U$
computes a listing of the elements of $S$ and then
halts.
That is, if $S=\{x_1 , \ldots , x_{n} \}$, then
$U(p)= \langle x_1,\langle x_2, \ldots, \langle x_{n-1},x_n\rangle \ldots\rangle \rangle $.
The {\em conditional complexity} $K(x \mid S)$ of $x$ given $S$,
is the length (number of bits) in the
shortest binary program $p$ from which the reference universal
prefix machine $U$, given $S$ literally as auxiliary information,
computes $x$.
\paragraph{Integer-valued functions:}
The (prefix-) complexity $K(f)$ of a
partial recursive function $f$ is defined by
$
K(f) = \min_i \{K(i): \mbox{\rm Turing machine } T_i
\; \; \mbox{\rm computes }
f \}.
$
If $f^*$ is a shortest program for computing the function $f$
(if there is more than one of them then $f^*$ is the first one in
enumeration order), then $K(f)=l(f^*)$.
\begin{remark}
\rm
In the above definition of $K(f)$, the objects being
described are functions instead of finite binary strings.
To unify the approaches, we can
consider a finite binary string $x$ as corresponding
to a function having value $x$ for argument 0.
Note that we can upper semi-compute (Section~\ref{sec:preliminaries}) 
$x^*$ given $x$,
but we cannot upper semi-compute $f^*$ given $f$ (as an oracle),
since we should be able to
verify agreement of a program for a function and an oracle for the
target function, on all infinitely many arguments.
\end{remark}
\paragraph{Probability Distributions:}
In this text we identify 
probability distributions on finite and countable sets ${\cal
  X}$ with their corresponding mass functions
(Section~\ref{sec:preliminaries}). Since any 
(sub-) probability mass function $f$ is a total real-valued function, $K(f)$
is defined in the same way as above. 
\subsubsection{Kolmogorov Complexity and the Universal Distribution}
\label{sec:m}
Following the definitions
above we now consider lower semi-computable and computable probability
mass functions (Section~\ref{sec:preliminaries}).   
By the fundamental
Kraft's inequality, Theorem~\ref{kraft}, we know that
if $l_1 , l_2 , \ldots$ are the code-word lengths of a  prefix code,
then $\sum_x 2^{-l_x} \leq 1$. Therefore,
since $K(x)$ is the length of
a prefix-free program for $x$, 
we can interpret $2^{-K(x)}$
as a sub-probability mass function, and 
 we define ${\bf m}(x)=2^{-K(x)}$.
This is the so-called
universal distribution---a rigorous form of Occam's razor.
The following two theorems are to be considered as major achievements
in the theory of Kolmogorov complexity, and will be used
again and again in the sequel. For the proofs we refer to
\cite{LiVi97}.

\begin{theorem}\label{PR1}
Let $f$ represent a
lower semi-computable (sub-) probability distribution on the
natural numbers (equivalently, finite binary strings).
(This implies $K(f) < \infty$.) 
Then, $2^{c_f} {\bf m}(x) > f(x)$ for all $x$, where $c_f =K(f)+O(1)$.
We call ${\bf m}$ a {\em universal distribution}.
\end{theorem}

The family of lower semi-computable sub-probability mass functions
contains all distributions with computable parameters which have a
name, or in which we could conceivably be interested, or which have
ever been considered\footnote{To be sure, in statistical applications,
  one often works with model classes containing distributions that are
  neither upper- nor lower semi-computable. An example is the
  Bernoulli model class, containing the distributions with $P(X=1) =
  \theta$ for all $\theta \in [0,1]$. However, every concrete {\em
    parameter estimate\/} or {\em predictive distribution\/} based on
  the Bernoulli model class that has ever been considered or in which we
  could be conceivably interested, is in fact computable; typically,
  $\theta$ is then rational-valued. See also Example~\ref{ex:appy} in
  Appendix~\ref{sec:universal}.}.  In particular, it contains the
computable distributions.  We call $\hbox{\bf m}$ ``universal'' since
it assigns at least as much probability to each object as any other
lower semi-computable distribution (up to a multiplicative factor),
and is itself lower semi-computable.

\begin{theorem}\label{PR2} 
\begin{equation}\label{eq.m}
  \log 1/\hbox{\bf m} (x)=K(x) \pm  O( 1). 
\end{equation}
\end{theorem}
That means that $\hbox{\bf m}$ assigns high probability to simple
objects
and low probability to complex or random objects.
For example, for $x=00 \ldots 0$ ($n$ 0's) we have
$K(x) = K(n) \pm O(1) \leq \log n + 2 \log \log n +O(1) $ since the program
\[ \mbox{\tt print } n \mbox{\tt \_times a ``0''} \]
prints $x$. (The additional $2 \log \log n$ term
is the penalty term for a prefix encoding.)
Then, $1/ (n \log^2 n ) = O( \hbox{\bf m}(x))$. 
But if we flip a coin to obtain a string $y$ of $n$ bits,
then with overwhelming probability $K(y) \geq n \pm O(1) $
(because $y$ does not contain effective regularities
which allow compression),
and hence $\hbox{\bf m}(y) = O( 1/2^n)$.

\paragraph*{Problem and Lacuna:} Unfortunately $K(x)$ is not a recursive
function: the Kolmogorov complexity is 
not computable in general. This means that
there exists no computer program that, when input an arbitrary string,
outputs the Kolmogorov complexity of that string and then halts. 
While Kolmogorov complexity is upper semi-computable
(Section~\ref{sec:preliminaries}), it cannot be approximated in
general in a
practically useful sense; and even though 
there
exist `feasible', resource-bounded forms of Kolmogorov
complexity (Li and Vit\'anyi 1997), these lack some of the elegant
properties of the original, uncomputable notion.

Now suppose we are interested in efficient storage and transmission of
long sequences of data. According to Kolmogorov, we can compress such
sequences in an essentially optimal way by storing or transmitting the
shortest program that generates them. Unfortunately, as we have just
seen, we cannot find such a program in general. According to Shannon,
we can compress such sequences optimally in an average sense (and
therefore, it turns out, also with high probability) if they are
distributed according to some $P$ and we know $P$. Unfortunately, in
practice, $P$ is often unknown, it may not be computable---bringing us
in the same conundrum as with the Kolmogorov complexity approach---or
worse, it may be nonexistent. In Appendix~\ref{sec:universal}, we
consider {\em universal coding}, which can be considered a sort of
middle ground between Shannon information and Kolmogorov complexity.
In contrast to both these approaches, universal codes can be directly
applied for practical data compression. Some basic knowledge of
universal codes will be very helpful in providing intuition for the
next section, in which we relate Kolmogorov complexity and Shannon
entropy. Nevertheless, universal codes are not directly needed in any
of the statements and proofs of the next section or, in fact, anywhere
else in the paper, which is why delegated their treatment to an
appendix.
\subsection{Expected Kolmogorov Complexity Equals Shannon Entropy}
\label{sec:KCSE}

Suppose the source words $x$ are distributed as a random variable
$X$ with probability $P(X=x) = f(x)$.
%
While $K(x)$ is
fixed for each $x$ and gives the shortest code word length
(but only up to a fixed constant) and is {\em independent} of the
probability distribution $P$, we may wonder whether
$K$ is also universal in the following sense:
If we weigh each individual code word length for
$x$ with its probability $f(x)$, does the resulting $f$-expected
code word length $\sum_x f(x)K(x)$
achieve the minimal average code word
length $ H(X)=  \sum_x f(x) \log 1/ f(x)$?
Here we sum over the entire support of $f$; restricting summation
to a small set, for example the singleton set $\{x\}$, can give
a different result.
The reasoning above implies that, under some mild restrictions on the
distributions $f$,
the answer is yes.
This is expressed in the following theorem, where, instead of the quotient 
we look at the difference of
$\sum_x f(x) K(x)$ and $ H(X)$. 
This allows
us to express really small distinctions.
%

\begin{theorem}\label{theo.eq.entropy}
Let $f$ be a computable probability mass function (Section~\ref{sec:preliminaries}) $f(x)=P(X=x)$ on
sample space ${\cal X} = \{0,1\}^*$
associated with a random source $X$ and
entropy
$H(X)=\sum_x f(x) \log 1/f(x)$. Then,
\[ 0 \leq \left( \sum_x f(x) K(x) - H(X) \right) \leq  K(f) + O(1). \]
\end{theorem}

\begin{proof}
Since $K(x)$ is the code word length of a prefix-code for $x$,
the first inequality of the Noiseless Coding Theorem~\ref{thm:noiseless}
states that
\[H(X) \leq \sum_x f(x) K(x).\]
Since $f(x) \leq 2^{K(f)+O(1)} {\bf m}(x)$ (Theorem~\ref{PR1})
and $\log {\bf m} (x) = K(x)+O(1)$ (Theorem~\ref{PR2}), we have
$ \log 1/ f(x) \geq K(x) - K(f) - O(1)$.
It follows that
\[\sum_x  f(x) K(x) \leq H(X)  + K(f)+ O(1).\]
Set the constant $c_f$ to
\[c_f := K(f)+O(1), \]
and the theorem is proved.
As an aside, the constant implied in the $O(1)$ term
depends on the lengths of the programs occurring in the proof of the cited
Theorems~\ref{PR1}, \ref{PR2} (Theorems 4.3.1 and 4.3.2 in \cite{LiVi97}). 
These depend only
on the reference universal prefix machine.
\end{proof}

The theorem shows that for simple (low complexity)
distributions the expected Kolmogorov complexity is close to
the entropy, but these two quantities may be wide apart for distributions
of high complexity. This explains the apparent problem arising
in considering a distribution $f$ that concentrates all probability
on an element $x$ of length $n$.  Suppose we choose $K(x)>n$.
Then $f(x)=1$ and hence the entropy $H(f)=0$. On the other hand
the term $ \sum_{x \in \{0,1\}^* } f(x) K(x) = K(x)$. Therefore,
the discrepancy between the expected Kolmogorov complexity and the entropy
exceeds the length $n$ of $x$. One may think this contradicts
the theorem, but that is not the case: The complexity of the distribution 
is at least that of $x$, since we can reconstruct $x$ given $f$
(just compute $f(y)$ for all $y$ of length $n$ in lexicographical
 order until we meet one that has probability 1). Thus, $c_f = K(f)+O(1)
\geq K(x)+O(1) \geq n+O(1)$. Thus, if we pick a probability distribution
with a complex support, or a trickily skewed probability distribution,
than this is reflected in the complexity of that distribution, and
as consequence in the closeness between the entropy and the expected
Kolmogorov complexity. 

For example, bringing the discussion in line with the universal coding
counterpart of Appendix~\ref{sec:universal} by considering $f$'s that
can be interpreted as sequential information sources and denoting the
conditional version of $f$ restricted to strings of length $n$ by
$f^{(n)}$ as in Section~\ref{sec:preliminaries}, we find by the same
proof as the theorem that for all $n$,
\[ 0 \leq \sum_{x \in \{0,1\}^n } f^{(n)}(x) K(x) - H(f^{(n)}) \leq
c_{f^{(n)}}, \]
where $c_{f^{(n)}} = K(f^{(n)})+O(1) \leq  K(f) + K(n)+O(1)$ is now a constant
depending on both $f$ and $n$.
On the other hand, we can eliminate
the complexity of the distribution, or its recursivity for that matter,
and / or restrictions to a conditional version of $f$
restricted to a finite support $A$
(for example $A = \{0,1\}^n$), denoted by $f^A$,
in the following conditional formulation (this involves
a peek in the future since
the precise meaning of the ``$K(\cdot \mid \cdot)$'' notation
is only provided in Definition~\ref{def.KolmKb}):
\begin{equation}\label{eq.condentropy}
 0 \leq \sum_{x \in A } f^A (x) K(x \mid f,A) - H(f^A) = O(1) .
\end{equation}

The Shannon-Fano code for a computable distribution is
itself computable. Therefore, for every computable 
distribution $f$, the universal code $D^*$
whose length function is the Kolmogorov complexity compresses
on average at least as much as the Shannon-Fano code for $f$. 
This is the intuitive reason
why, no matter what computable distribution $f$ we take, its expected
Kolmogorov complexity is close to its entropy.

%


\section{Mutual Information}
\label{sec:mutual}
\subsection{Probabilistic Mutual Information}
\label{sec:probmutual}
How much information can a random variable $X$ convey about a
random variable $Y$?
Taking a purely combinatorial approach,
this notion is captured as follows:
If $X$ ranges over ${\cal X}$
and $Y$ ranges over ${\cal Y}$, then we look at the set $U$
of possible events $(X=x,Y=y)$ consisting of
joint occurrences of event $X=x$ and event $Y=y$.
If $U$ does not equal the Cartesian product ${\cal X} \times {\cal Y}$,
then this means there is some dependency between $X$ and $Y$.
Considering the set $U_x =  \{ (x,u):(x,u) \in U \}$ for $x \in {\cal X} $,
it is natural to define the 
\it conditional entropy 
\rm of $Y$ 
\rm given $X = x$ as $H(Y|X=x) = \log d(U_x )$. This suggests
immediately that the information given by $X=x$ about $Y$ is
$$
I(X=x: Y) = H(Y) - H(Y| X=x).  
$$
For example, if $U =  \{ (1,1), (1,2),(2,3) \} $, $U  \subseteq {\cal X} 
\times {\cal Y}$
with ${\cal X} =  \{ 1,2 \} $ and ${\cal Y} =  \{ 1,2,3,4 \} $,
then $I(X=1: Y) = 1$ and $I(X=2: Y) = 2$.

In this formulation it is obvious that $H(X|X=x) = 0$,
and that $I(X=x : X) = H(X)$. 
This approach amounts
to the assumption of a 
{\em uniform distribution}\index{distribution!uniform}
of the probabilities concerned.   

We can generalize this approach,
taking into account
the frequencies or probabilities of the occurrences of the different
values $X$ and $Y$ can assume. 
Let the {\em joint probability}\index{probability!joint}
$f(x,y)$ be the ``probability of
the joint occurrence of event $X=x$ and event $Y=y$.''
The {\em marginal probabilities} $f_1 (x)$ and $f_2(y)$ are
defined by $f_1 (x)= \sum_y f(x,y)$ and $f_2 (y)= \sum_x f(x,y)$ and
are ``the probability of the occurrence of the event $X=x$''
and the ``probability of the occurrence of the event $Y=y$'',
respectively. 
This leads to the self-evident formulas for joint variables $X,Y$:
\begin{eqnarray*}
 && H(X,Y)  =   \sum_{x,y}  f(x,y) \log 1/ f(x,y), \\
 && H(X)  =   \sum_{x}  f(x) \log  1/f(x), \\
 && H(Y)  =   \sum_{y}  f(y) \log 1/ f(y) ,
\end{eqnarray*}
where summation over $x$ is taken over all outcomes of the random variable
$X$ and summation over $y$ is taken over all outcomes of random variable $Y$.
One can show that
\begin{equation}
H(X,Y)  \leq H(X) + H(Y) ,
\label{I4}
\end{equation}
with equality only in the case that $X$ and $Y$ are independent.
In all of these equations the
entropy quantity on the left-hand side increases if
we choose the probabilities on the right-hand side
more equally.

\paragraph{Conditional entropy:}
We start
the analysis of
the information in $X$ about $Y$ by first considering
the %
\it conditional
entropy \index{entropy!conditional|bold}%
\rm of $Y$ %
\rm given $X$ as the average of the
entropy for $Y$ for each value of $X$ %
\rm weighted
by the probability of getting that particular value:
\begin{eqnarray*}
H(Y| X)
 & = &   \sum_x  f_1(x) H(Y|X=x) \\
 & = &   \sum_x  f_1(x) \sum_{y} f(y|x) \log 1/ f(y|x) \\
 & = &  \sum_{x,y} f(x,y) \log 1/f(y| x) .
\end{eqnarray*}
Here $f(y|x)$ is the conditional probability mass function as defined
in Section~\ref{sec:preliminaries}.

The quantity on the left-hand side tells us
how uncertain we are on average about the outcome of $Y$
when we know an outcome of $X$. With
\begin{eqnarray*}
H(X) & = &  \sum_x  f_1(x) \log 1/ f_1(x) \\
&  = &  \sum_{x} \left(\sum_y f(x,y) \right)
\log  \sum_y 1/ f(x,y)  \\
&  = &  \sum_{x,y} f(x,y)
\log  \sum_y 1/ f(x,y) ,
\end{eqnarray*}
and substituting the formula for $f(y|x)$, we find
$H(Y| X)  =  H(X,Y) - H(X)$. Rewrite this expression as
the Entropy Equality
\begin{equation}
H(X,Y)  =  H (X) + H(Y| X).
\label{I5}  
\end{equation} 
This can be interpreted as, ``the uncertainty of 
the joint event $(X,Y)$ is the uncertainty of $X$
plus the uncertainty of $Y$ given $X$.''
Combining Equations~\ref{I4}, \ref{I5} gives 
$H(Y)  \geq H(Y| X)$, which can be taken to imply
that, on average, knowledge of $X$ can never increase uncertainty
of $Y$. In fact, uncertainty in $Y$ will be decreased
unless $X$ and $Y$ are independent.
\paragraph{Information:}
The
\it information %
\rm in the outcome $X=x$ %
\rm about $Y$ is defined as
\begin{equation}
 I(X=x: Y) = H(Y) - H(Y| X=x) .
\label{I6}
\end{equation}
Here the quantities $H(Y)$ and $H(Y| X=x)$ on the right-hand side
of the equations are always equal to or less than the
corresponding quantities under the uniform distribution
we analyzed first. The values of the quantities
$I(X=x: Y)$ under the assumption of uniform distribution
of $Y$ and $Y|X=x$ versus any other distribution are not
related by inequality in a particular direction.
The equalities $H(X|X=x) = 0$ and $I(X=x: X) = H(X)$
hold under any distribution of the variables. Since
$I(X=x: Y)$ is a function of outcomes of $X$, while $I(Y=y: X)$
is a function of outcomes of $Y$, we do not compare them directly.
However, forming the expectation defined as
\begin{eqnarray*}
{\bf E} (I(X=x: Y)) & = & \sum_x f_1 (x)I(X=x: Y), \\
{\bf E} (I(Y=y: X)) & = & \sum_y f_2(y)I(Y=y: X),
\end{eqnarray*}
and combining Equations~\ref{I5}, \ref{I6},
we see that the resulting quantities are equal. Denoting
this quantity by $I(X;Y)$ and calling
it the %
\it mutual information\index{information!mutual|bold} %
\rm in $X$ and $Y$,
we see that this information is 
{\it symmetric}:\index{information!symmetry of|see{symmetry of information}}\index{symmetry of information!stochastic|bold}
\begin{equation}
I(X; Y) = {\bf E} (I(X=x: Y)) = {\bf E} (I(Y=y: X)).
\end{equation}
Writing this out we find
that the 
 {\em mutual information} $I(X;Y)$
is defined by:
\begin{equation}\label{eq.mutinfprob}
 I(X;Y) = \sum_x \sum_y f(x,y) \log \frac{f(x,y)}{f_1(x)f_2(y)} .
\end{equation}
Another way to express this is as follows: a well-known criterion for
the difference between a given distribution $f(x)$ and a distribution
 $g(x)$ it is compared with is the so-called
{\em Kullback-Leibler divergence}
\begin{equation}\label{eq.kl}
D(f \parallel g ) = \sum_x f(x) \log f(x)/g(x). 
\end{equation}
It has the important property that 
\begin{equation}\label{eq.ii}
D (f \parallel g ) \geq 0
\end{equation}
with equality only  iff $f(x)=g(x)$ for all $x$. This is called the
{\em information inequality} in \cite{CT91}, p. 26.
Thus, the mutual information is the Kullback-Leibler divergence between
the joint distribution and the product 
$f_1(x)f_2(y)$ of the two marginal distributions. If this quantity is 0
then $f(x,y)=f_1(x)f_2(y)$ for every pair $x,y$, which is the same as
saying that $X$ and $Y$ are independent random variables.
\begin{example}
\rm
\label{ex:mutual}
Suppose we want to exchange the information about the outcome $X=x$
and it is known already that outcome $Y=y$ is the case, 
that is, $x$ has property $y$.
Then we require (using the Shannon-Fano code) about
$ \log 1/ P(X=x | Y=y)$ bits to communicate $x$. On average, over the
joint distribution $P(X=x, Y=y)$ we use $H(X|Y)$ bits,
which is optimal by Shannon's noiseless coding theorem.
In fact, exploiting the mutual information paradigm,
the expected information $I(Y ; X)$ 
that outcome $Y=y$ gives about outcome $X=x$
is the same as the expected information that $X=x$ gives about $Y=y$,
and is never negative. Yet
there may certainly exist
{\em individual\/} $y$ such that $I(Y=y : X)$ is negative. For example, we may
have ${\cal X } = \{0,1\}$, ${\cal Y} = \{0,1\}$, $P(X=1 | Y=0) = 1$,
$P(X=1 | Y = 1) = 1/2$, $P(Y=1) = \epsilon$. Then $I(Y; X) =
H(\epsilon,1- \epsilon)$ whereas 
$I(Y=1 : X) = H(\epsilon,1-\epsilon) + \epsilon - 1$. For small
$\epsilon$, this quantity is smaller than $0$. 
\end{example}

\ 
\\
{\bf Problem and Lacuna:}
The quantity $I(Y; X)$ 
symmetrically characterizes to what extent random
variables $X$ and $Y$ are correlated. An inherent problem
with probabilistic definitions 
is that --- as we have just seen --- although $I(Y; X) = {\bf E} (I(Y
= y: X))$
is always positive, for some probability
distributions 
and some $y$, $I(Y=y: X)$
can turn out to be negative---which
definitely contradicts our naive notion of information content.
\commentout{
How is this possible? The
concept of information as used in the theory of communication
is a probabilistic notion, which is natural for
information transmission over communication channels.
Nonetheless, 
we tend to
identify
\it probabilities %
\rm of messages with 
\it frequencies %
\rm of messages in a sufficiently
long sequence, which under some conditions on the stochastic
source can be rigorously justified.
For instance, Morse code\index{code!Morse} transmissions of English
telegrams over a communication channel
can be validly treated by probabilistic
methods even if we (as is usual) use empirical
frequencies for probabilities. The great probabilist,
Kolmogorov, remarks, ``If something goes wrong here,
\index{Kolmogorov, A.N.}
the problem lies in the vagueness of our ideas of
the relation between mathematical probability theory
and real random events in general.'' 
}
The {\em algorithmic\/} mutual information we introduce below can {\em
  never\/} be negative, and in this sense is closer to the intuitive
notion of information content.

\subsection{Algorithmic Mutual Information}
\label{sec:algmi}
For individual objects the information about one another
is possibly even more fundamental than for random sources.
Kolmogorov \cite{Ko65}:
\begin{quote}
Actually, it is most fruitful to discuss the quantity of information
``conveyed by an object'' ($x$) ``about an object'' ($y$). It is not
an accident that in the probabilistic approach this has led
to a generalization to the case of continuous variables, for which
the entropy is finite but, in a large number of cases,
\[
I_W(x,y) = \int \int P_{xy}(dx \; dy)\log_2 
\frac{P_{xy}(dx \; dy)}{P_x(dx) P_y(dy)}
\]
is finite.
The real objects that we study are very (infinitely) complex,
but the relationships between two separate objects diminish as the 
schemes used to describe them become simpler.
While a map yields a considerable amount of information about a region
of the earth's surface, the microstructure of the paper and the ink
on the paper have no relation to the microstructure 
of the area shown on the map.''
\end{quote} 
In the discussions on Shannon mutual information, we first
needed to introduce a conditional version of entropy. Analogously, to
prepare for the definition of algorithmic mutual information, we need
a notion of conditional Kolmogorov complexity. 

Intuitively, the
conditional prefix Kolmogorov complexity $K(x|y)$ of $x$ given $y$ can
be interpreted as the shortest prefix program $p$ such that, when $y$
is given to the program $p$ as input, the program prints $x$ and then
halts. The idea of providing $p$ with an input $y$ is realized by putting
$\langle p,y \rangle$ rather than just $p$ on the input tape of the
universal prefix machine $U$.
\begin{definition}\label{def.KolmKb}
\rm
The {\em conditional prefix Kolmogorov complexity} of $x$ given $y$ (for
free) is
\[K(x|y) = \min_{p}\{l(p): U(\langle p,y \rangle )=x , p \in \{0,1\}^*\}. \]
We define 
\begin{equation}
\label{eq:redefine}
K(x)=K(x|\epsilon).
\end{equation}
\end{definition}
Note that we just redefined $K(x)$ so that the unconditional
Kolmogorov complexity is {\em exactly\/} equal to the conditional
Kolmogorov complexity with empty input. This does not contradict our
earlier definition: we can choose a reference prefix machine $U$ such
that $U(\langle p,\epsilon \rangle) = U(p)$. Then (\ref{eq:redefine})
holds automatically.

We now have the technical apparatus to express the relation between 
entropy inequalities and Kolmogorov complexity inequalities.
Recall that the entropy expresses the expected information
to transmit an outcome of a known random source, 
while the Kolmogorov complexity
of every such outcome expresses the specific information
contained in that outcome. This makes us wonder to what extend the
entropy-(in)equalities hold for the corresponding Kolmogorov
complexity situation. In the latter case the corresponding (in)equality
is a far stronger statement, implying the same (in)equality in the
entropy setting. It is remarkable, therefore, that similar inequalities
hold for both cases, where the entropy ones hold exactly while the 
Kolmogorov complexity ones hold up to a logarithmic, and in some cases
$O(1)$,
additive precision. 
 
\paragraph{Additivity:}
By definition, $K(x,y) = K(\langle x,y \rangle)$.
Trivially, the symmetry property holds: $K(x,y) \eqa K(y,x)$.
Another interesting property is the ``Additivity of Complexity''
property that, as we explain further below, 
is equivalent to the ``Symmetry of 
Algorithmic Mutual Information'' property. Recall that 
$x^*$ denotes the first (in a standard enumeration order) 
shortest prefix program that
generates $x$ and then halts.
\begin{theorem}[Additivity of Complexity/Symmetry of Mutual Information]
\label{thm:additive}
\begin{equation}\label{eq.soi}
  K(x, y) \eqa K(x) + K(y \mid x^*) \eqa K(y) + K(x \mid y^*).
 \end{equation}
\end{theorem}
This is the Kolmogorov complexity equivalent of the entropy equality
(\ref{I5}). That this latter equality holds is true
by simply rewriting both sides of the equation according to the
definitions of averages of joint and marginal probabilities. 
In fact, potential individual differences are averaged out.
But in the Kolmogorov complexity case we do nothing like that:
it is truly remarkable that additivity of algorithmic information
holds for individual objects. It was first proven by Kolmogorov
and Leonid A. Levin for the plain (non-prefix) version
of Kolmogorov complexity, where it holds up to an additive logarithmic
term, and reported in \cite{ZvLe70}.
The prefix-version (\ref{eq.soi}), holding up to an $O(1)$
additive term is due to \cite{Ga74}, can be found
as Theorem 3.9.1 in~\cite{LiVi97}, and has a difficult proof.
\paragraph{Symmetry:}
To define the algorithmic mutual information between
two individual objects $x$ and $y$ with no
probabilities involved, it is instructive to first recall
the probabilistic notion (\ref{eq.mutinfprob}).
Rewriting (\ref{eq.mutinfprob})
as
\[ \sum_x \sum_y f(x,y) [  \log 1/f(x) + \log 1/ f(y) - \log 1/f(x,y) ] , \]
and noting that $ \log 1/ f ( s )$ is
very close to the length of the
prefix-free Shannon-Fano code for $s$, we are led to the following
definition.
The
{\em information in  $y$ about $x$}
 is defined as
 \begin{equation}\label{def.mutinf}
   I(y : x) = K(x) - K(x  \mid  y^*) \eqa K(x) + K(y) - K(x, y),
 \end{equation}
where the second equality is a consequence of~(\ref{eq.soi})
and states that this information is symmetrical,
$I(x:y) \eqa I(y:x)$, and therefore we can talk about
{\em mutual information}.\footnote{The notation of the
algorithmic (individual)  
 notion $I(x:y)$ distinguishes it from the probabilistic
(average) notion  
$I(X; Y)$.  We deviate slightly from~\cite{LiVi97}
where $I(y : x)$ is defined as $K(x) - K(x \mid y)$.}
\paragraph{Precision -- $O(1)$ vs. $O(\log n)$:}
The version of (\ref{eq.soi})  with just $x$ and $y$ in the 
conditionals doesn't
hold with $\eqa$, but holds up to additive logarithmic terms 
that cannot be eliminated. To gain some further insight in this
matter, first consider the following lemma:
\begin{lemma}
$x^*$ has the same information as the
pair $x,K(x)$, that is, $K(x^* \mid x,K(x)),K(x,K(x) \mid x^*)=O(1)$.
\end{lemma}
\begin{proof}
Given $x,K(x)$ we can run all programs simultaneously in
dovetailed fashion and select the first program of length $K(x)$
that halts with output $x$ as $x^*$. (Dovetailed fashion means that
in phase $k$ of the process we run all programs $i$ for $j$ steps
such that $i+j=k$, $k=1,2, \ldots$) 
\end{proof}

\noindent
Thus, $x^*$ provides more information than $x$. Therefore, we have to
be very careful when extending Theorem~\ref{thm:additive}. 
For example, the conditional version of (\ref{eq.soi}) is:
 \begin{equation}\label{eq.soi-cond}
  K(x, y \mid z) \eqa K(x \mid z) + K(y \mid x, K(x \mid z), z).
 \end{equation}
Note that a naive version
 \[
  K(x, y \mid z) \eqa K(x \mid z) + K(y \mid x^{*}, z)
 \]
is incorrect: taking $z = x$, $y = K(x)$,
the left-hand side equals $K(x^{*} \mid x)$ which can be as large as
$\log n - \log \log n + O(1)$, and the right-hand side
equals $K(x \mid x) + K(K(x) \mid x^{*}, x) \eqa 0$.

But up to logarithmic precision we do not need to
be that careful. In fact, it turns out that {\em every}
linear entropy inequality holds for the corresponding Kolmogorov
complexities within a logarithmic additive error, \cite{HRSV00}:
\begin{theorem}
All linear (in)equalities that are valid for Kolmogorov complexity
are also valid for Shannon entropy and vice versa---provided
we require the 
Kolmogorov complexity (in)equalities to hold up to additive
logarithmic precision only.
\end{theorem}
\subsection{Expected Algorithmic Mutual Information Equals Probabilistic Mutual Information}
Theorem~\ref{theo.eq.entropy} 
gave the relationship between entropy and ordinary Kolmogorov
complexity; it showed that the entropy of distribution $P$ is
approximately equal to the expected (under $P$) Kolmogorov
complexity. Theorem~\ref{thm:mutinf} gives the analogous result for
the mutual information (to facilitate comparison to
Theorem~\ref{theo.eq.entropy}, note that $x$ and $y$ in
(\ref{eq.eqamipmi}) below may stand for strings of arbitrary length $n$).
\begin{theorem}
\label{thm:mutinf}
Given a computable probability distribution $f(x,y)$ over $(x,y)$
we have
\begin{align}\label{eq.eqamipmi}
I(X; Y) - K(f) & \lea  \sum_x \sum_y f(x,y) I(x:y)
\\& \lea I(X;Y) + 2 K(f) ,
\nonumber
\end{align}
\end{theorem}
\begin{proof}
Rewrite the expectation
\begin{align*}
\sum_x \sum_y f(x,y) I(x:y) \eqa
\sum_x \sum_y & f(x,y)  [K(x)
\\& + K(y) - K(x, y)].
\end{align*}
 Define
$\sum_y f(x,y) = f_1 (x)$
and $\sum_x f(x,y) = f_2(y)$
to obtain
\begin{align*}
\sum_x \sum_y f(x,y) I(x:y) \eqa
 \sum_x & f_1 (x) K(x)
 + \sum_y f_2 (y) K(y)
\\& - \sum_{x,y} f(x,y) K(x, y).
\end{align*}
Given the program that computes $f$, we can approximate $f_1 (x)$
by  $q_1 (x,y_0) = \sum_{y \leq y_0} f(x,y)$, and
similarly for $f_2$. That is, the
distributions $f_i$ ($i=1,2$) are lower semicomputable.
Because they sum to 1 it can be shown
they must also be computable. 
By Theorem~\ref{theo.eq.entropy},
we have $H(g) \lea \sum_x g(x) K(x) \lea H(g) + K(g)$
for every computable probability mass function $g$.
                                                                                
Hence, $H(f_i) \lea \sum_x f_i (x) K(x) \lea H(f_i) + K(f_i)$
($i=1,2$), and $H(f) \lea \sum_{x,y} f (x,y) K(x,y) \lea H(f) + K(f)$.
On the other hand, the probabilistic mutual information
 (\ref{eq.mutinfprob}) is expressed in the entropies by
$I(X;Y) = H(f_1) + H(f_2) - H(f)$.
By construction of the $f_i$'s above,
we have $K(f_1), K(f_2) \lea K(f)$. Since the complexities
are positive, substitution
establishes the lemma.
\end{proof}
                                                                                
Can we get rid of the $K(f)$ error term? The answer is affirmative;
by putting $f(\cdot)$ in the conditional, and
applying \eqref{eq.condentropy}, we can even get rid of
the computability requirement.
                                                                                
\begin{lemma}
Given a joint probability distribution $f(x,y)$ over $(x,y)$
(not necessarily computable) we have
\[ I(X;Y)  \eqa  \sum_x \sum_y f(x,y) I(x:y \mid f) , \]
where the auxiliary $f$ means that we can directly access the
values $f(x,y)$ on the
auxiliary conditional information tape of the reference
universal prefix machine.
\end{lemma}
                                                                                
\begin{proof}
The lemma follows from the definition of conditional
algorithmic mutual information, 
if we show that $\sum_{x}
f(x) K(x \mid f) \eqa H(f)$,
where the $O(1)$ term implicit in the $\eqa$ sign
is independent of $f$.
                                                                                
Equip the reference universal prefix machine,
with an $O(1)$ length
program to compute a Shannon-Fano code from the auxiliary table
of probabilities.
Then, given an input $r$, it can determine
whether $r$ is the Shannon-Fano code word for some $x$.
Such a code word
has length $\eqa  \log 1/f(x)$.
If this is the case, then the machine
outputs $x$, otherwise it halts without output. Therefore,
$K(x   \mid f) \lea  \log 1/ f(x)$.
This shows
the upper bound on the expected prefix complexity.
The lower bound follows as usual
from the Noiseless Coding Theorem.
\end{proof}

Thus, we see that the expectation of the algorithmic mutual
information $I(x:y)$ is close to the probabilistic mutual information
$I(X; Y)$ --- which is important: if
this were not the case then the algorithmic notion would not
be a sharpening of the probabilistic notion to individual objects,
but something else.

\section{Mutual Information Non-Increase}
\label{sect.mini}
\subsection{Probabilistic Version}
Is it possible to increase the mutual information between
two random variables, by processing the outcomes in some deterministic
manner? The answer is negative:
For every function $T$
we have
\begin{equation}\label{eq.infnonincrprob}
I(X ; Y) \geq I( X ; T(Y)),
\end{equation}
that is, mutual information between two random variables
cannot be increased by processing their outcomes in any deterministic way.
The same holds in an appropriate sense for randomized processing
of the outcomes of the random variables.
This fact is called the {\em data processing inequality} \cite{CT91},
Theorem 2.8.1. The reason why it holds is that \eqref{eq.mutinfprob}
is expressed in terms of probabilities $f(a,b), f_1(a), f_2(b)$,
rather than in terms of the arguments.
Processing the arguments $a,b$ will not increase the value
of the expression in the right-hand side. If the processing of the arguments
just renames them in a one-to-one manner then the expression
keeps the same value. If the processing eliminates or merges arguments
then it is easy to check from the formula
that the expression value doesn't increase.

\subsection{Algorithmic Version}
\label{sect:minialg}
In the algorithmic version of mutual information, the notion
is expressed in terms of the individual arguments instead of
solely in terms of the probabilities as in the probabilistic version.
Therefore, the reason for \eqref{eq.infnonincrprob} to hold is
not valid in the algorithmic case. Yet it turns out that the 
data processing inequality also holds between individual objects,
by far more subtle arguments and not precisely but with a small
tolerance. The first to observe this fact was Leonid A. Levin
who proved his ``information non-growth,'' and ``information
conservation inequalities'' for both finite and infinite sequences
under both deterministic and randomized data processing,
\cite{Le74,Le84}.

\subsubsection{A Triangle Inequality}
We first discuss some useful technical lemmas.
The  additivity of complexity (symmetry of information)
\eqref{eq.soi} can be used to
derive a ``directed triangle inequality'' from \cite{GTV01}, 
that is needed later.
 \begin{theorem}\label{lem.magic}
For all $x,y,z$, 
 \[
  K(x \mid y^*) \lea K(x, z \mid y^{*}) \lea K(z \mid y^*) + K(x \mid z^*).
 \]
 \end{theorem}

\begin{proof}
Using~(\ref{eq.soi}), an evident inequality introducing
an auxiliary object $z$, and twice (~\ref{eq.soi}) again:
 \begin{align*}
  K(x, z \mid y^*) &\eqa 
    K(x,y,z) - K(y) 
\\ & \lea K(z) + K(x \mid z^*) + K(y \mid z^*) - K(y)
\\ &\eqa K(y,z) - K(y) + K(x \mid z^*) 
\\ & \eqa K(x \mid z^*) + K(z \mid y^*).
 \end{align*}

\end{proof}

\begin{remark}
\rm
This theorem has bizarre consequences. These  consequences are not
simple unexpected artifacts of our definitions, but, to the contrary,
they show the power and the genuine contribution to our understanding
represented by the deep and important mathematical relation
(\ref{eq.soi}).
 
Denote $k=K(y)$ and substitute $k=z$ and $K(k)=x$
to find the following counterintuitive corollary: To determine the complexity
of the complexity of an object $y$ it suffices to give both $y$ and
the complexity of $y$. This is counterintuitive since in general
we cannot compute the complexity of an object from the object itself;
if we could this would also solve the 
so-called ``halting problem'', \cite{LiVi97}. This noncomputability
can be quantified in terms of $K(K(y) \mid y )$ which can rise to
almost $K(K(y))$ for some $y$.
But in the
seemingly similar, but subtly different, setting below it is possible.

\begin{corollary}
As above, let $k$ denote $K(y)$. Then,
$K(K(k) \mid  y,k) \eqa K(K(k) \mid y^*)  \lea K(K(k) \mid k^*)+K(k \mid y,k) \eqa 0$.
\end{corollary}
\end{remark}

Now back to whether mutual information in one object
about another one cannot be increased. In the probabilistic
setting this was shown to hold for random variables. But does
it also hold for individual outcomes? 
In \cite{Le74,Le84} it was shown that
the information in one individual string about another
cannot be increased by any deterministic algorithmic method
by more than a constant. With added randomization this holds
with overwhelming probability.
Here, we follow the proof method
of \cite{GTV01} and 
use the triangle inequality of Theorem~\ref{lem.magic} to recall,
and to give proofs of this information non-increase.


\subsubsection{Deterministic Data Processing:}
Recall the definition~\ref{def.mutinf} and Theorem~\ref{eq.eqamipmi}.
We prove a strong version of the information non-increase law
under deterministic processing (later we need the attached corollary):

\begin{theorem}
Given $x$ and $z$, let $q$ be a program 
computing $z$ from $x^*$.
Then
 \begin{equation}\label{eq.nonincrease2}
   I(z : y) \lea I(x : y) + K(q).
 \end{equation}
\end{theorem}

\begin{proof}
By the triangle inequality,
 \begin{align*}
   K(y \mid x^{*}) & \lea K(y \mid z^{*}) + K(z \mid x^{*})
 \\&  \eqa K(y \mid z^{*})+ K(q).
 \end{align*}
Thus,
 \begin{align*}
   I(x : y) & = K(y) - K(y \mid x^{*})
\\ & \gea K(y) - K(y \mid z^{*}) - K(q)
 \\ &  = I(z : y) - K(q).
 \end{align*}
\end{proof}

This also implies the slightly weaker but intuitively
more appealing statement that the mutual information between strings 
$x$ and $y$ cannot be increased by processing $x$ and $y$ separately by
deterministic computations.
 \begin{corollary} Let $f, g$ be recursive functions.
Then
 \begin{equation}\label{eq.nonincrease}
   I(f(x) : g(y)) \lea I(x : y) + K(f)+K(g).
 \end{equation}
 \end{corollary}
 \begin{proof}
It suffices to prove the case $g(y) = y$ and apply it twice.
The proof is by replacing the program $q$ that computes 
a particular string $z$ 
from a particular $x^*$ in (\ref{eq.nonincrease2}). There, $q$
possibly depends on $x^*$ and $z$. Replace it by a program $q_f$ that first
computes $x$ from $x^*$, followed  by computing a
recursive function 
$f$, that is,  $q_f$ is independent of $x$.
Since we only require an $O(1)$-length program to compute
$x$ from $x^*$ we can choose $l(q_f) \eqa K(f)$.

By the triangle inequality,
 \begin{align*}
   K(y \mid x^{*}) & \lea K(y \mid f(x)^{*}) + K(f(x) \mid x^{*})
 \\&  \eqa K(y \mid f(x)^{*})+ K(f).
 \end{align*}
Thus,
 \begin{align*}
   I(x : y) & = K(y) - K(y \mid x^{*}) 
\\ & \gea K(y) - K(y \mid f(x)^{*}) - K(f)
 \\ &  = I(f(x) : y) - K(f).
 \end{align*}
 \end{proof}

\subsubsection{Randomized Data Processing:}
It turns out that furthermore, randomized computation can increase
information only with negligible probability.
Recall from Section~\ref{sec:m} that the 
{\em universal probability} $\m(x) = 2^{-K(x)}$ is 
maximal within a multiplicative constant among lower semicomputable
semimeasures.
So, in particular, for each computable measure $f(x)$ we have
$f(x) \leq c_1 \m(x)$, where the constant factor $c_1$ depends on $f$.
This property also holds when we have an extra parameter, like $y^*$,
in the condition.

Suppose that $z$ is obtained from $x$ by some randomized computation.
We assume that 
the probability $f(z \mid x)$ of obtaining $z$ from $x$ is a semicomputable
distribution over the $z$'s.
Therefore it is upperbounded by
$\m(z \mid x) \leq c_2 \m(z \mid x^{*}) = 2^{-K(z \mid x^{*})}$.
The information increase $I(z : y) - I(x : y)$ satisfies the theorem below.
                                                                                
 \begin{theorem} There is a constant $c_3$ such that
for all $x,y,z$ we have
 \[
  \m(z \mid x^{*}) 2^{I(z : y) - I(x : y)}
  \leq c_3 \m(z \mid x^{*}, y, K(y \mid x^{*})).
 \]
 \end{theorem}
                                                                                
\begin{remark}
\rm
For example, the probability of an increase of mutual information
by the amount $d$ is $O( 2^{-d})$.
The theorem
implies $\sum_{z} \m(z \mid x^{*}) 2^{I(z : y) - I(x : y)} =O(1)$,
the $\m(\cdot \mid x^{*})$-expectation of the exponential of the increase
is bounded by a constant.
\end{remark}
                                                                                
 \begin{proof}
We have
 \begin{align*}
 I(z : y) - I(x : y) & = K(y) - K(y \mid z^{*}) - (K(y) - K(y \mid x^{*}))
 \\&  = K(y \mid x^{*}) - K(y \mid z^{*}).
 \end{align*}
 The negative logarithm of the left-hand side in the theorem is therefore
 \[
  K(z \mid x^{*}) + K(y \mid z^{*}) - K(y \mid x^{*}).
 \]
Using Theorem~\ref{lem.magic}, and the conditional
additivity (\ref{eq.soi-cond}), this is
 \[
   \gea K(y, z \mid x^{*}) - K(y \mid x^{*}) \eqa
        K(z \mid x^{*}, y, K(y \mid x^{*})).
 \]
 \end{proof}

\begin{remark}
  \rm An example of the use of algorithmic mutual information is as
  follows \cite{Le02}.  A celebrated result of K. G\"odel states that
  Peano Arithmetic is incomplete in the sense that it cannot be
  consistently extended to a complete theory using recursively
  enumerable axiom sets.  (Here `complete' means that every sentence of
  Peano Arithmetic is decidable within the theory; for further
  details on the terminology used in this example, we refer to
  \cite{LiVi97}). The essence is the non-existence of total recursive
  extensions of a universal partial recursive predicate. This is
  usually taken to mean that mathematics is undecidable.
  Non-existence of an algorithmic solution need not be a problem when
  the requirements do not imply unique solutions. A perfect example is
  the generation of strings of high Kolmogorov complexity, say of half
  the length of the strings. There is no deterministic effective process that can
  produce such a string; but repeatedly flipping a fair coin we
  generate a desired string with overwhelming probability. Therefore,
  the question arises whether randomized means allow us to bypass
  G\"odel's result.  The notion of mutual information between two
  finite strings can be refined and extended to infinite sequences, so
  that, again, it cannot be increased by either deterministic or
  randomized processing.  In \cite{Le02} the existence of an infinite
  sequence is shown that has infinite mutual information with all
  total extensions of a universal partial recursive predicate. As
  Levin states ``it plays the role of password: no substantial
  information about it can be guessed, no matter what methods are
  allowed.''  This ``forbidden information'' is used to extend the
  G\"odel's incompleteness result to also hold for consistent
  extensions to a complete theory by randomized means with
  non-vanishing probability.
\end{remark}

\ 
\\
\paragraph{Problem and Lacuna:}
Entropy, Kolmogorov complexity and mutual
(algorithmic) information are concepts that do not distinguish 
between different {\em   kinds\/} of information (such as `meaningful' and `meaningless'
information). In the remainder of this paper, we  show how these more
intricate notions 
can be arrived at, typically by {\em constraining\/} the description
methods with which strings are allowed to be encoded
(Section~\ref{sec:algsuf}) and by considering {\em lossy\/} rather
than lossless compression (Section~\ref{sect.rdsf}). Nevertheless, the basic
notions entropy, Kolmogorov complexity and mutual information continue
to play a fundamental r\^ole. 
\section{Sufficient Statistic}
\label{sect.sufstat}
In introducing the notion of sufficiency in classical
statistics,  Fisher~\cite{Fi22} stated:
\begin{quote}
       ``The statistic chosen should summarize the whole of the relevant
information supplied by the sample. This may be called
       the Criterion of Sufficiency $\ldots$
In the case of the normal curve
of distribution it is evident that the second moment is a
       sufficient statistic for estimating the standard deviation.''
\end{quote}
A ``sufficient'' statistic of the data
contains all information in the data about the model class.
Below we first discuss the standard notion of (probabilistic)
sufficient statistic as employed in the statistical literature. 
We show that this notion has a natural interpretation in terms of
Shannon mutual information, so that we may just as well 
think of a probabilistic sufficient
statistic as a concept in Shannon information theory. Just as in the
other sections of this paper, there is a corresponding notion in the
Kolmogorov complexity literature: the algorithmic sufficient statistic 
which we introduce in Section~\ref{sec:algsuf}. Finally, 
in Section~\ref{sec:relpa} we connect the statistical/Shannon
and the algorithmic notions of sufficiency. 
\subsection{Probabilistic Sufficient Statistic}
\label{sec:probstat}
Let $\{ P_\theta \}$ be a family of distributions,
also called a {\em model class}, of a
random variable $X$ that takes values in a finite or countable
{\em set of data} ${\cal X}$. 
Let ${\mathbf \Theta}$
be the set of parameters $\theta$ parameterizing the family
$\{ P_\theta \}$. Any function $S: {\cal X} \rightarrow {\cal S}$
taking values in some set  ${\cal S}$ is said to be a {\em statistic}
of the data in ${\cal X}$. A {\em statistic}
$S$ is said to be {\em sufficient} for the family $\{
P_{\theta} \}$ if,
for every $s \in {\cal S}$,
the conditional distribution
\begin{equation}
\label{eq:cond}
P_{\theta}(X= \cdot \mid S(x) = s)
\end{equation}
is invariant under changes of $\theta$. This is the
standard definition 
in the statistical literature, see  
for example \cite{CoxH74}. 
Intuitively, (\ref{eq:cond}) means that all information about $\theta$
in the observation $x$ is present in the (coarser) observation $S(x)$,
in line with Fisher's quote above.

The notion of `sufficient statistic'
can be equivalently expressed in terms
of probability mass functions. Let $f_{\theta}(x) = P_{\theta} (X=x)$ 
denote the
probability mass of $x$ according to $P_{\theta}$. 
We identify distributions $P_{\theta}$ with their
mass functions $f_{\theta}$ and denote the model class $\{P_{\theta} \}$ by
$\{f_{\theta}\}$. Let
$f_{\theta}(x | s)$ denote the
probability mass function of the conditional distribution (\ref{eq:cond}), defined as
in Section~\ref{sec:preliminaries}. That is, 
$$
f_{\theta}(x|s) 
=
\begin{cases}
f_{\theta}(x) / \sum_{x \in{\cal X}: S(x) = s} f_{\theta}(x) 
& \text{\ if\ } S(x) = s \\
0 & \text{\ if\ } S(x) \neq s .
\end{cases}
$$
The requirement of $S$ to be sufficient is equivalent to the existence
of a function 
$g: {\cal X} \times {\cal S} \rightarrow {\cal R}$ such
that
\begin{equation}
\label{eq:standarddef}
g(x \mid s) = f_{\theta}(x \mid s),
\end{equation}
for every $\theta \in {\mathbf \Theta}$, $s \in {\cal S}$, 
$x \in {\cal X}$. (Here we change the common
notation `$g(x,s)$' to  `$g(x \mid s)$' which is more expressive for
our purpose.)

\begin{example}
\rm
Let ${\cal X} = \{ 0,1\}^{n}$, let $X = (X_1, \ldots, X_n)$. Let
$\{P_{\theta} : \theta \in (0,1 ) \}$ be the set of $n$-fold
Bernoulli
distributions on
${\cal X}$ with parameter $\theta$. That is,
$$
f_{\theta}(x) = f_{\theta}(x_1 \ldots x_n) = \theta^{S(x)}(1- \theta)^{n-S(x)}$$
where $S(x)$ is the number of $1$'s in $x$. Then $S(x)$ is a
sufficient statistic for $\{ P_{\theta} \}$. Namely, fix an arbitrary
$P_{\theta}$ with $\theta \in (0,1)$ and an arbitrary $s$ with $0 < s<
n$.
Then all $x$'s with $s$ ones and $n-s$ zeroes are equally probable.
The number of such $x$'s is $\binom{n}{s}$. Therefore, the
probability $P_\theta(X=x \mid S(x) = s)$ is equal to
$1/ \binom{n}{s}$, and this does not depend on the parameter
$\theta$.
Equivalently, for all $\theta \in (0,1)$,
\begin{equation}
\label{eq:berndef}
f_{\theta}(x \mid s) 
= \begin{cases}  1/ \binom{n}{s} & \text{\ if\ } S(x) = s
  \\
0 & \text{\ otherwise.}
\end{cases}
\end{equation}
Since (\ref{eq:berndef}) satisfies
(\ref{eq:standarddef}) (with $g(x|s)$ the uniform distribution on all
$x$ with exactly $s$ ones), $S(x)$
is a sufficient statistic relative to the model class $\{P_{\theta}\}$. 
In the Bernoulli case, 
$g(x|s)$ can be
obtained by starting from the {\em uniform\/}
distribution on ${\cal X}$ ($\theta = \frac{1}{2}$),
and conditioning on $S(x)=s$. But 
$g$ is not necessarily uniform. For example, for the 
Poisson model class, where $\{f_{\theta}\}$
represents the set of Poisson distributions on $n$ observations, 
the observed mean is a sufficient statistic and the corresponding $g$
is far from uniform.
All information 
about the parameter $\theta$ in
the observation $x$ is already contained in $S(x)$. 
In the Bernoulli case, once we know
the number $S(x)$ of $1$'s in $x$, all further details
of $x$ (such as the order of $0$s and $1$s) are irrelevant 
for determination of the  Bernoulli parameter $\theta$.

To give an example of a 
statistics that is not sufficient for the Bernoulli model class, consider
the statistic $T(x)$ which counts the number of 1s in $x$ that are
followed by a $1$. On the other hand, for every statistic $U$, the
combined statistic $V(x) := (S(x),
U(x))$ with $S(x)$ as before, is sufficient, since it contains all
information in $S(x)$. But in contrast to $S(x)$, a 
statistic such as $V(x)$ is typically not
{\em minimal}, as explained further below.
\end{example}

It will be useful to rewrite
(\ref{eq:standarddef}) as
\begin{equation}
\label{eq:indivdef}
 \log 1/f_{\theta}(x \mid s)  =  \log 1/ g(x| s).
\end{equation}
\begin{definition}\label{def.wss}
\rm
A function  $S: {\cal X} \rightarrow {\cal S}$ is
  a {\em probabilistic sufficient statistic} for $\{f_{\theta}\}$ if
  there exists a function $g : {\cal X} \times {\cal
    S} \rightarrow {\cal R}$ such that (\ref{eq:indivdef}) holds for every
  $\theta \in {\mathbf \Theta}$, every $x \in {\cal X}$, every $s \in
  {\cal S}$ (Here we use the convention $\log 1/0 = \infty$).
\end{definition}
\paragraph{Expectation-version of definition:}
The standard definition of probabilistic
sufficient statistics is ostensibly of the
`individual-sequence'-type: for $S$ to be sufficient,
(\ref{eq:indivdef}) has to hold for
{\em every} $x$, rather than merely in expectation or with high probability.
However, 
the definition turns out to be equivalent to an expectation-oriented
form, as shown in Proposition~\ref{prop:suff}. We first introduce an a priori distribution 
over $\Theta$, the parameter set for our model class $\{ f_\theta\}$. We
denote the probability density of this distribution by $p_1$.
This way we can define a joint distribution
$p(\theta,x) = p_1 (\theta) f_{\theta}(x)$.
\begin{proposition}
\label{prop:suff}
\rm
The following two statements are equivalent to
Definition~\ref{def.wss}: (1)
For every $\theta \in  {\mathbf \Theta}$,
\begin{equation}
\label{eq:expdef}
\sum_x f_\theta(x)  \log 1/f_\theta(x \mid S(x)) =
 \sum_{x} f_\theta(x) \log 1/g(x \mid S(x)) \; .
\end{equation}
(2) For {\em every} prior $p_1(\theta)$ on ${\mathbf \Theta}$,
\begin{equation}
\label{eq:expdef2}
\sum_{\theta,x} p(\theta,x)  \log 1/f_\theta(x \mid S(x))
= \sum_{\theta,x} p(\theta,x) \log 1/g(x \mid S(x)) \; .
\end{equation}
\end{proposition}
\begin{proof}
  {\em Definition~\ref{def.wss} $\Rightarrow$ \eqref{eq:expdef}:} 
Suppose (\ref{eq:indivdef}) holds for every $\theta
  \in {\mathbf \Theta}$, every $x \in {\cal X}$, every $s \in {\cal S}$.
  Then it also holds in expectation for every $\theta \in {\mathbf
    \Theta}$:
\begin{equation}
\label{eq:propsuff1}
\sum_{x} f_\theta(x) \log 1/f_{\theta}(x|S(x)) = \sum_{x}
f_\theta(x) \log 1/ g(x| S(x))].
\end{equation}
                                                                                
  {\em \eqref{eq:expdef}$\Rightarrow$   Definition~\ref{def.wss}:}
Suppose that for every $\theta \in {\mathbf \Theta}$,
(\ref{eq:propsuff1}) holds. 
Denote
\begin{equation}\label{eq:overload}
f_{\theta}(s) = \sum_{y \in {\cal X}: S(y)=s} f_{\theta} (y).
\end{equation}
By adding 
$\sum_x f_{\theta}(x) \log 1/f_\theta(S(x))$ to both
sides of the equation, (\ref{eq:propsuff1}) can be rewritten as
\begin{equation}
\label{eq:suffent}
\sum_x f_{\theta} (x) \log 1/ f_{\theta}(x) =
\sum_x f_{\theta}(x)  \log1/ g_{\theta}(x),
\end{equation}
with
$g_{\theta}(x) = f_{\theta}(S(x)) \cdot g(x| S(x))].$
By the information inequality \eqref{eq.ii}, the equality
(\ref{eq:suffent}) can
only hold if $g_{\theta}(x) =
f_{\theta}(x)$ for every $x \in {\cal X}$.
Hence, we have established \eqref{eq:indivdef}.
                                                                                
{\em  \eqref{eq:expdef} $\Leftrightarrow$ \eqref{eq:expdef2}:} 
follows by linearity of expectation.
\end{proof}

\paragraph{Mutual information-version of definition:}
After some rearranging of terms, the characterization
(\ref{eq:expdef2}) gives rise to the intuitively appealing definition
of probabilistic sufficient statistic in terms of mutual information
\eqref{eq.mutinfprob}. The resulting formulation of sufficiency 
is as follows \cite{CT91}:
 $S$ is sufficient for $\{ f_{\theta} \}$ iff for all
priors $p_1$ on ${\mathbf \Theta}$:
\begin{equation}\label{eq.suffstatprob}
I(\Theta ; X) = I( \Theta ; S(X))
\end{equation}
 for all distributions
of $\theta$.

Thus, a statistic $S(x)$ is 
sufficient if the probabilistic mutual
information is invariant under taking the statistic \eqref{eq.suffstatprob}.
\paragraph{Minimal Probabilistic Sufficient Statistic:}
A sufficient statistic may contain information 
that is not relevant: for a normal distribution the sample mean
is a sufficient statistic, but the pair of functions
which give the
mean of the even-numbered samples and the odd-numbered samples
respectively, is also a sufficient statistic.
A statistic $S(x)$ is a {\em minimal} sufficient statistic
with respect to an indexed
model class $\{f_{\theta}\}$, if it is a
function of all other sufficient statistics: it contains no
irrelevant information and maximally compresses the information 
in the data about
the model class.
For the family of normal distributions
the sample mean is a minimal sufficient statistic, but the
sufficient statistic consisting of the mean of the even samples
in combination with the mean of the odd samples is not minimal.
Note that one cannot improve on sufficiency:
The data processing inequality \eqref{eq.infnonincrprob} states that
$
I(\Theta ; X) \geq I( \Theta ; S(X)),
$
for every function $S$, and that
for randomized functions $S$ an appropriate related expression holds.
That is, mutual information between data random variable and model random
variable 
cannot be increased by processing the data sample in any way.
\paragraph{Problem and Lacuna:}
We can think of the probabilistic sufficient statistic as extracting
those patterns in the data that are relevant in determining the
parameters of a statistical model class. But what if we do not want to
commit ourselves to a simple finite-dimensional parametric model
class? In the most general context, we may consider the model class of all computable
distributions, or all computable sets of which the observed data is an
element. Does there exist an analogue
of the sufficient statistic that  automatically summarizes {\em all\/}
information in the sample $x$ that is relevant for determining the
``best'' (appropriately defined)
model for $x$ within this enormous class of models? Of course, 
we may consider the
literal data $x$ as a statistic of $x$, but that would not be
satisfactory: we would still like our generalized statistic, at least
in many cases, to be
considerably coarser, and much more concise, than the data $x$ itself. 
It turns
out that, to some extent, this is achieved by
the {\em algorithmic\/} sufficient statistic 
of the data: it 
summarizes {\em all\/} conceivably relevant information in the
data $x$; at the same time, many types of data $x$ admit an algorithmic
sufficient statistic that is concise in the sense that it has very small
Kolmogorov complexity.
\subsection{Algorithmic Sufficient Statistic}
\label{sec:algsuf}
\subsubsection{Meaningful Information}\index{meaningful information}
\label{sect.meaning}
The information contained in an individual
finite object (like a finite binary string) is measured
by its Kolmogorov complexity---the length of the shortest binary program
that computes the object. Such a shortest program contains no redundancy:
every bit is information; but is it meaningful information?
If we flip a fair coin to obtain a finite binary string, then with overwhelming
probability that string constitutes its own shortest program. However,
also with overwhelming probability all the bits in the string are meaningless
information, random noise. On the other hand, let an object
$x$ be a sequence of observations of heavenly bodies. Then $x$
can be described by the binary
string $pd$, where $p$ is the description of
the laws of gravity and the observational
parameter setting, while $d$ accounts for the measurement errors:
we can divide the information in $x$ into
meaningful information $p$ and accidental information $d$.
The main task for statistical inference and learning theory is to
distill the meaningful information present in the data. The question
arises whether it is possible to separate meaningful
information from accidental information, and if so, how.
The essence of the solution to this problem is revealed when we
write Definition~\ref{def.KolmK}
as follows:
             \begin{equation}\label{eq.kcmdl}
K(x)  =
\min_{p,i} \{K(i)+l(p):T_i(p) =x\}+O(1),
\end{equation}
where the minimum is taken over
$p \in \{0,1\}^*$ and $i \in \{1,2, \ldots\}$. 
The justification is that for the fixed reference 
universal prefix Turing machine
$U(\langle i,p \rangle)=T_i(p)$ for all $i$ and $p$. Since $i^*$ 
denotes the shortest self-delimiting program for $i$, we have
$|i^*|=K(i)$.
The expression \eqref{eq.kcmdl}
 emphasizes the two-part code nature of Kolmogorov complexity.
In a randomly truncated initial segment of a time series
$$x = 10101010101010101010101010,$$
we can encode $x$ by a small Turing machine printing a specified
number of copies of the pattern ``01.'' 
This way, $K(x)$ is viewed  as the shortest length of
a two-part code for $x$, one part describing a Turing machine $T$,
or {\em model}, for the {\em regular} aspects of $x$,
and the second part describing
the {\em irregular} aspects of $x$ in the form
of a program $p$ to be interpreted by $T$.
The regular, or ``valuable,'' information in $x$ is constituted
by the bits in the ``model'' while the random or ``useless''
information of $x$ constitutes the remainder.
This leaves open the crucial question: How to choose
$T$ and $p$ that together describe $x$? In general, many
combinations of $T$ and $p$ are possible, but we want to find 
a $T$ that describes the meaningful aspects of $x$. 

\subsubsection{Data and Model}
\index{data}\index{model} We consider only finite binary data strings
$x$.  Our model class consists of Turing machines $T$ that enumerate a
finite set, say $S$, such that on input $p \leq |S|$ we have $T(p)=x$
with $x$ the $p$th element of $T$'s enumeration of $S$, and $T(p)$ is
a special {\em undefined} value if $p>|S|$.  The ``best fitting''
model for $x$ is a Turing machine $T$ that reaches the minimum
description length in (\ref{eq.kcmdl}).  There may be many such $T$,
but, as we will see, if chosen properly, such a machine $T$ embodies
the amount of useful information contained in $x$. Thus, we have
divided a shortest program $x^*$ for $x$ into parts $x^*=T^*(p)$ such
that $T^*$ is a shortest self-delimiting program for $T$.  Now suppose
we consider only low complexity finite-set models, and under these
constraints the shortest two-part description happens to be longer
than the shortest one-part description.  For example, this can happen
if the data is generated by a model that is too complex to be in the
contemplated model class.  Does the model minimizing the two-part
description still capture all (or as much as possible) meaningful
information? Such considerations require study of the relation between
the complexity limit on the contemplated model classes, the shortest
two-part code length, and the amount of meaningful information
captured.

In the following we will distinguish between ``models'' that are
finite sets, and the ``shortest programs'' to compute those models
that are finite strings. The latter will be called `algorithmic statistics'. 
In a way the distinction between ``model'' and ``statistic'' is
artificial, but for now we prefer clarity and unambiguousness in the
discussion.  Moreover, the terminology is customary in the literature
on algorithmic statistics.  Note that strictly speaking, neither an
algorithmic statistic nor the set it defines is a statistic in the
probabilistic sense: the latter was defined as a {\em function\/} on
the set of possible data samples of given length. Both notions are
unified in Section~\ref{sec:relpa}.
\subsubsection{Typical Elements}
\index{typical data}\index{random data}
Consider a string $x$
of length $n$ and prefix complexity $K(x)=k$.
For every finite set $S  \subseteq \{0,1\}^*$ containing
$x$ we have $K(x | S)\le\log|S|+O(1)$.
Indeed, consider the prefix code of $x$
consisting of its $\lceil\log|S|\rceil$ bit long index
of $x$ in the lexicographical ordering of $S$.
This code is called
\emph{data-to-model code}. 
We identify the {\em structure} or {\em regularity} in $x$ that are
to be summarized with a set $S$
of which $x$ is a {\em random} or  {\em typical} member:
given $S$ containing $x$, 
the element $x$ cannot
be described significantly shorter than by its maximal length index in $S$,
that is, $ K(x \mid S) \geq \log |S| +O(1) $. 

\begin{definition}
\rm
Let $\beta \ge 0$ be an agreed upon, fixed, constant.
A finite binary string $x$
is a {\em typical} or {\em random} element of a set $S$ of finite binary
strings, if $x \in S$ and
\begin{equation}\label{eq.deftyp}
 K(x \mid S) \ge \log |S| - \beta.
\end{equation}
We will not indicate the dependence on $\beta$ explicitly, but the
constants in all our inequalities ($O(1)$) will be allowed to be functions
of this $\beta$.
\end{definition}

This definition requires a finite $S$.
In fact, since
$K(x \mid S) \leq K(x)+O(1) $, it limits the size of $S$ to $O(2^k)$.
Note that the notion of typicality is not absolute
but depends on fixing the constant implicit in the $O$-notation.

\begin{example}\label{xmp.typical}
\rm
Consider the set $S$ of binary strings of length $n$
whose every odd position is 0.
Let $x$ be an element of this set in which the subsequence of bits in
even positions is an incompressible string.
Then $x$ is a typical element of $S$ (or by with some abuse
of language we can say $S$ is typical for $x$).
But $x$ is also a typical element of the set $\{x\}$.
\end{example}

\subsubsection{Optimal Sets}
\index{optimal model}
Let $x$ be a binary data string of length $n$.
For every finite set $S \ni x$, we have
$K(x) \leq K(S) + \log |S| + O(1)$,
since we can describe $x$ by giving $S$ and the index of $x$
in a standard enumeration of $S$. Clearly this can be implemented
by a Turing machine computing the finite set $S$ and a program
$p$ giving the index of $x$ in $S$.
The size of a set containing $x$ measures intuitively the number of
properties of $x$ that are represented: 
The largest set is $\{0,1\}^{n}$ and represents only one property
of $x$, namely, being of length $n$. It clearly ``underfits''
as explanation or model for $x$. The smallest set containing $x$
is the singleton set $\{x\}$ and represents all conceivable properties
of $x$. It clearly ``overfits'' as explanation or model for $x$.  

There are two natural measures of suitability of such a set as
a model for $x$.
We might prefer either the simplest set, or the smallest set, as
corresponding to the most likely structure `explaining' $x$.
Both the largest set $\{0,1\}^n$ (having low complexity of about $K(n)$)
and the  singleton set $\{x\}$ (having high complexity of about $K(x)$), 
while certainly statistics for $x$,
would indeed be considered poor explanations.
\index{two-stage description}
We would like to balance simplicity of model versus size of model.
Both measures relate to the optimality of a two-stage description of
$x$ using a finite set $S$ that contains it. Elaborating on
the two-part code:
\begin{align}\label{eq.twostage}
 K(x) \leq K(x,S)  & \leq  K(S) + K(x \mid S) +O(1)
\\ & \leq K(S) + \log |S| +O(1),
\nonumber
\end{align}
where only the final substitution of $K(x \mid S)$ by $\log |S|+O(1)$
uses the fact that $x$ is an element of $S$.
The closer the right-hand side of \eqref{eq.twostage} gets
to the left-hand side, the better the description of $x$ is in terms
of the set $S$.
This implies a trade-off between meaningful model information, $K(S)$,
and meaningless ``noise'' $\log |S|$. 
A set $S$ (containing $x$)
 for which \eqref{eq.twostage} holds with equality
\begin{equation}\label{eq.optim}
K(x) = K(S) + \log |S| +O(1),
\end{equation}
is called {\em optimal}. 
A data string $x$ can be typical for a set $S$ without that set $S$
being optimal for $x$. This is the case precisely when $x$ is
typical for $S$ (that is $K(x|S)=\log S +O(1)$)
while $K(x,S)>K(x)$.


\subsubsection{Sufficient Statistic}
\label{sect.ss}
 Intuitively, a model expresses the essence
of the data if the two-part code describing the data consisting of the
model and the data-to-model code is as concise as 
the best one-part description.

Mindful of our distinction between a finite set $S$ and a
program that describes $S$ in a required representation format,
we call a shortest program for an optimal set with respect to $x$
an {\em algorithmic sufficient statistic} for $x$.
Furthermore, among optimal sets,
there is a direct trade-off between complexity and log-size, which together
sum to $ K(x)+O(1)$.

\begin{example}\label{xmp.optimal}
\rm
It can be shown that the set $S$ of Example~\ref{xmp.typical} is also
optimal, and so is $\{x\}$.
Sets for which $x$ is typical form a much wider class than optimal 
sets for $x$: the set
$\{x,y\}$ is still typical for
$x$ but with most $y$, it will be too complex to be optimal for $x$.

For a perhaps less artificial example, consider complexities conditional
on the length $n$ of strings.
Let $y$ be a random string of length $n$, let
$S_{y}$ be the set of strings of length $n$ which have 0's exactly
where $y$ has, and let $x$ be a random element of $S_{y}$.
Then $x$ has about 25\%
1's,
so its complexity is much less than $n$.
The set $S_{y}$ has $x$ as a typical element,
but is too complex to be optimal,
since its  complexity (even conditional on $n$) is still $n$.
\end{example}


An algorithmic sufficient statistic 
\index{sufficient statistic, algorithmic}
\index{sufficient statistic, algorithmic minimal}
\index{sufficient statistic, probabilistic}
\index{sufficient statistic, probabilistic minimal}
is a sharper individual notion than a probabilistic sufficient
statistic. An optimal set $S$ associated with $x$ (the shortest
program computing $S$ is the corresponding
sufficient statistic associated with $x$) is chosen such that
$x$ is maximally random with respect to it. That is, the
information in $x$ is divided in a relevant structure expressed
by the set $S$, and the remaining randomness with respect
to that structure, expressed by $x$'s index in $S$ of $\log |S|$
bits. The shortest program for $S$ is itself alone an algorithmic
definition of structure, without a probabilistic interpretation.

Those optimal sets that admit the shortest possible program  
are called {\em algorithmic minimal sufficient statistics\/} of
$x$. They will play a major role in the next section on the Kolmogorov
structure function. Summarizing:
\begin{definition}[Algorithmic sufficient statistic, algorithmic
  minimal sufficient statistic]
\label{def:algsufstat}
An {\em algorithmic sufficient statistic\/} of $x$ is a shortest program for
a set $S$ containing $x$ that is optimal, i.e. it satisfies (\ref{eq.optim}).
An algorithmic sufficient statistic with optimal set $S$ is {\em
  minimal\/} if there exists no optimal set $S'$ with $K(S') < K(S)$.
\end{definition}
\begin{example}
\rm
Let $k$ be a number in the range $0,1,\dots,n$
of complexity $\log n+ O(1)$ given $n$ and let $x$ be a string of length
$n$ having $k$ ones of complexity $K(x \mid n,k) \geq \log {n \choose k}$
given $n,k$. This $x$ can be viewed as a typical result of
tossing a coin with a bias about $p=k/n$.
A two-part description
of $x$ is given by
the number $k$ of 1's in $x$ first, followed by the index
$j \leq \log |S|$  of $x$
in the set $S$ of strings of length $n$ with $k$ 1's.
This set is optimal, since
$K(x \mid n)=K(x,k \mid n)=K(k \mid n)+K(x \mid k,n)= K(S)+ \log|S|$.

Note that $S$ encodes the number of $1$s in $x$. The shortest program
for $S$  is an
algorithmic minimal sufficient statistic for {\em most\/} $x$ of
length $n$ with $k$ $1$'s, since only a fraction of at most $2^{-m}$
$x$'s of length $n$ with $k$ $1$s can have $K(x) < \log | S| - m$
(Section~\ref{sec:kolmogorov}). But of course there exist $x$'s with
$k$ ones which have much more regularity. An example is the string
starting with $k$ $1$'s followed by $n-k$ $0$'s. For such strings, $S$ is
still optimal and the shortest program for $S$ is still an algorithmic
sufficient statistic, but not a minimal one.
\end{example}

\subsection{Relating Probabilistic and Algorithmic Sufficiency}
\label{sec:relpa}
We want to relate `algorithmic sufficient statistics' (defined
independently of any model class $\{f_\theta\}$) to probabilistic sufficient
statistics (defined relative to some model class
$\{f_\theta\}$ as in Section~\ref{sec:probstat}). We will show that,
essentially, algorithmic sufficient statistics are probabilistic
nearly-sufficient statistics with respect to {\em all\/} model families $\{
f_{\theta} \}$. Since the notion of
algorithmic sufficiency is only defined to within additive constants,
we cannot expect algorithmic sufficient statistics to satisfy
the requirements (\ref{eq:indivdef}) or (\ref{eq:expdef}) for probabilistic sufficiency {\em
  exactly}, but only `nearly\footnote{We use `nearly' rather than
  `almost' since `almost' suggests things like `almost
  everywhere/almost surely/with probability 1'. Instead, `nearly' means, roughly speaking,  `to within $O(1)$'.}'.
\paragraph{Nearly Sufficient Statistics:}
Intuitively, we may consider a probabilistic statistic $S$ to be
nearly sufficient if (\ref{eq:indivdef}) or (\ref{eq:expdef}) holds to
within some constant. For long sequences $x$, this constant will then
be negligible compared to the two terms in
(\ref{eq:indivdef}) or (\ref{eq:expdef}) which, for most practically
interesting statistical model classes, typically grow linearly in the
sequence length. But now we encounter a difficulty: 
\begin{quote}
whereas
(\ref{eq:indivdef}) and (\ref{eq:expdef}) are equivalent if they are
required to hold exactly, they express something substantially
different if they are only required to hold within a constant.
\end{quote}
Because of our observation
above, when relating probabilistic and algorithmic statistics 
we have to be very careful about what
happens if $n$ is allowed to change. Thus, we need to extend probabilistic and algorithmic statistics to strings of arbitrary length. This
leads to 
the following generalized definition of a statistic:
\begin{definition}
\rm  
\label{def:seqstat}
A {\em sequential statistic\/} is a function $S: \{0,1\}^* \rightarrow
2^{\{0,1 \}^*}$, such that for all $n$, all $x \in \{0,1\}^n$,
(1) $S(x) \subseteq \{0,1\}^n$, and (2) $x \in S(x)$, and (3) for all $n$, the set 
$$
\{ s \; | \; \text{There exists $x \in \{0,1\}^n$ with 
$S(x) = s $} \ \} 
$$
is a partition of $\{0,1\}^n$.
\end{definition}  
Algorithmic statistics are defined relative to
individual $x$ of some length $n$. 
Probabilistic statistics are defined as functions, hence for all $x$
of given length, but still relative to given length $n$. Such
algorithmic and probabilistic statistics can be
extended to  each $n$ and each $x \in \{0,1\}^n$ in a
variety of ways; the three conditions in Definition~\ref{def:seqstat}
ensure that the extension is done in a reasonable way.
%
Now let $\{ f_{\theta} \}$ be a model class of sequential
information sources (Section~\ref{sec:preliminaries}), i.e. a
statistical model class defined for sequences of arbitrary length rather
than just fixed $n$.  As before, $f^{(n)}_{\theta}$ denotes the
marginal distribution of $f_\theta$ on $\{0,1\}^n$.
\begin{definition}  
\label{def:nearsuff}
\rm
We call sequential statistic  $S$ 
{\em nearly-sufficient for
  $\{f_\theta\}$ in the probabilistic-individual sense} if
there exist functions $g^{(1)}, g^{(2)}, \ldots$ and a constant $c$   such that
for all $\theta$,
all $n$,  every $x \in \{0,1\}^n$,
\begin{equation}
\label{eq:nindivdef}
\biggl|  \log 1/f^{(n)}_{\theta}(x \mid S(x))
- \log 1/ g^{(n)}(x
| S(x))
\bigr] \biggr|
\leq c.
\end{equation}
We say $S$ is {\em nearly-sufficient  for
 $\{f_\theta\}$ in the probabilistic-expectation sense\/} if
there exists functions $g^{(1)}, g^{(2)}, \ldots$ and a constant $c'$ such that
for all $\theta$, all $n$,
\begin{equation}
\label{eq:nexpdef}
\biggl| \sum_{x \in \{0,1\}^n} 
f^{(n)}_{\theta}(x) \bigl[ \log 1/ f^{(n)}_{\theta}(x \mid
S(x)) - \log 1/ g^{(n)}(x| S(x))
\bigr]\;
\biggr|
\leq c'.
\end{equation}
\end{definition}
Inequality \eqref{eq:nindivdef} may
be read as `(\ref{eq:indivdef}) holds within a constant', whereas
(\ref{eq:nexpdef}) may be read as `(\ref{eq:expdef}) holds within a
constant'.
                                                                                
\begin{remark}
\rm
Whereas the
individual-sequence definition (\ref{eq:indivdef}) and the
expectation-definition (\ref{eq:expdef}) are equivalent if we
require exact equality, they become quite different if we allow
equality to within a constant as in Definition~\ref{def:nearsuff}.  To
see this, let $S$ be some sequential 
statistic such that for all large $n$, for some $\theta_1,
\theta_2$, for some $x \in \{0,1\}^n$,
$$f^{(n)}_{\theta_1}(x \mid S(x)) \gg
f^{(n)}_{\theta_2}(x \mid S(x)),
$$
while for all $x' \neq x$ of length $n$,
$f^{(n)}_{\theta_1}(x|S(x)) \approx f^{(n)}_{\theta_2}(x| S(x))$. If
$x$ has very small but nonzero probability according to some $\theta
\in \Theta$, then with very small $f_{\theta}$-probability, the
difference between the left-hand and right-hand side of
(\ref{eq:indivdef}) is very large, and with large
$f_{\theta}$-probability, the difference between the left-hand and
right-hand side of (\ref{eq:indivdef}) is about $0$.  Then $S$ will be
nearly sufficient in expectation, but not in the individual sense.
\end{remark}
In the theorem below we focus on probabilistic statistics that are
`nearly sufficient in an expected sense'. We connect these to
algorithmic sequential statistics, defined as follows:
\begin{definition}
\rm 
%
  A sequential statistic $S$ is {\em
    sufficient in the algorithmic sense\/} 
if there is a constant $c$
  such that for all $n$, all $x \in \{0,1\}^n$, the program generating
  $S(x)$ is an algorithmic sufficient statistic for $x$ (relative
  to constant $c$), i.e. 
\begin{equation}
\label{eq:seqsuf}
K(S(x)) + \log |S(x)|  \leq K(x) + c.
\end{equation}
\end{definition}
In Theorem~\ref{thm:wiske} we relate
algorithmic to probabilistic sufficiency. 
In the theorem, $S$
represents a sequential statistic, $\{f_{\theta}\}$ is a model class of
sequential information sources and $g^{(n)}$ is the conditional
probability mass function arising from the uniform distribution:
$$
g^{(n)}(x |s) = \begin{cases}
1/|\{ x \in { \{0,1\}^{n}} :
S(x) = s \} |  &
\text{\ if $S(x) = s$\ } \\
0 & \text{\ otherwise.}
\end{cases}
$$
\begin{theorem}[algorithmic sufficient statistic is probabilistic
  sufficient statistic]
\label{thm:wiske}
\rm 
Let $S$ be a sequential statistic that is sufficient in the
algorithmic sense. Then for every $\theta$ with $K(f_\theta) < \infty$,
there exists a constant $c$, such that for all $n$, inequality
\eqref{eq:nexpdef} holds with $g^{(n)}$ the uniform distribution. 
Thus, if $\sup_{\theta \in
  {\mathbf \Theta}} K(f_\theta) < \infty$, then $S$ is a nearly-sufficient
statistic for $\{ f_\theta \}$ in the probabilistic-expectation sense,
with $g$ equal to the uniform distribution.
\end{theorem}
                                                                           \noindent
\begin{proof}
The definition of algorithmic sufficiency, (\ref{eq:seqsuf}) directly
implies that 
there exists a
constant $c$ such that for all $\theta$, all $n$,
\begin{equation}
\label{eq:thmalg}
 \sum_{x \in \{0,1\}^n }
f^{(n)}_{\theta}(x) \bigl[ K(S(x)) + \log
|S(x)| \bigr] \leq  \sum_{x \in \{0,1\}^n } f^{(n)}_{\theta}(x)
K(x)
+ c.
\end{equation}
Now fix any $\theta$ with $K(f_\theta) < \infty$. It follows (by the
same reasoning as in Theorem~\ref{theo.eq.entropy}) that
for some $c_{\theta} \approx K(f_\theta)$,
for all $n$,
\begin{equation}
\label{eq:dochter}
0 \leq \sum_{x \in \{0,1\}^n} f^{(n)}_{\theta} (x) K(x) -
\sum_{x \in \{0,1\}^n} f^{(n)}_{\theta} (x) 
 \log 1/ f_{\theta}(x) \leq c_{\theta}.
\end{equation}
Essentially, the left inequality follows by the information inequality
(\ref{eq.ii}): no code can be more
efficient in expectation under $f_\theta$ than the Shannon-Fano code
with lengths $\log 1 /f_\theta(x)$; the right inequality follows
because, since $K(f_\theta) < \infty$,
the Shannon-Fano code can be implemented by a
computer program with a fixed-size independent of $n$.
By (\ref{eq:dochter}), 
(\ref{eq:thmalg}) becomes: for all $n$,
\begin{equation}
\label{eq:thmalgb}
\sum_{x \in \{0,1\}^n} f^{(n)}_{\theta}(x) \log 1/ f_{\theta}(x) \leq 
\sum_{x \in \{0,1\}^n }
f^{(n)}_{\theta}(x) \bigl[ K(S(x)) + \log
|S(x)| \bigr] \leq  \sum_x f^{(n)}_{\theta}(x) \log 1/ f_{\theta}(x)
+ c_\theta.
\end{equation}
For $s \subseteq \{0,1\}^n$, we use the notation
$f^{(n)}_\theta(s)$ according to \eqref{eq:overload}.
Note that, 
by requirement (3) in the definition of sequential statistic, 
$$\sum_{s: \exists x \in \{0,1\}^n : S(x) = s} f^{(n)}_\theta(s) =
1,
$$
whence $f^{(n)}_\theta(s)$ is a probability mass function on
${\cal S}$, the set of values the statistic $S$ can take on sequences
of length $n$. Thus, we get, once again by the information inequality (\ref{eq.ii}), 
\begin{equation}
\label{eq:zoon}
\sum_{x \in \{0,1\}^n} f^{(n)}_{\theta} (x) K(S(x)) \geq
\sum_{x \in \{0,1\}^n} f^{(n)}_{\theta} (x) 
\log 1/ f^{(n)}_{\theta}(S(x)).
\end{equation}
Now note that for all $n$,
\begin{equation}
\label{eq:extra}
\sum_{x \in \{0,1\}^n} f^{(n)}_{\theta} (x) 
\bigl[ \log 1/ f^{(n)}_{\theta}(S(x)) + 
\log 1/ f^{(n)}_{\theta}(x \mid S(x)) 
\bigr]
= \sum_{x \in \{0,1\}^n} f^{(n)}_{\theta}(x) \log 1/ f_{\theta}(x).
\end{equation}
Consider the two-part code which encodes $x$ by first encoding $S(x)$
using $\log 1 /f^{(n)}_\theta(S(x))$ bits, and then encoding $x$ using
$\log |S(x) |$ bits. By the
information inequality, (\ref{eq.ii}), this code must be less efficient
than the Shannon-Fano code with lengths $\log 1/ f_{\theta}(x)$, so
that if follows from (\ref{eq:extra}) that, for all $n$,
\begin{equation}
\label{eq:kind}
\sum_{x \in \{0,1\}^n} f^{(n)}_{\theta} (x) \log | S(x) | \geq
\sum_{x \in \{0,1\}^n} f^{(n)}_{\theta} (x) 
\log 1/ f^{(n)}_{\theta}(x \mid S(x)).
\end{equation}
Now defining
\begin{eqnarray}
u & = & \sum_{x \in \{0,1\}^n }
f^{(n)}_{\theta}(x) K(S(x))   \nonumber \\
v & = &  \sum_{x \in \{0,1\}^n }
f^{(n)}_{\theta}(x) \log |S(x)| \nonumber \\
u' & = & \sum_{x \in \{0,1\}^n} f^{(n)}_{\theta} (x) 
\log 1/ f^{(n)}_{\theta}(S(x))  
 \nonumber \\
v' & = & \sum_{x \in \{0,1\}^n} f^{(n)}_{\theta} (x)
\log 1/ f^{(n)}_{\theta}(x \mid S(x)) \nonumber \\
w & = & \sum_{x \in \{0,1\}^n} f^{(n)}_{\theta}(x) \log 1/
f_{\theta}(x),
\nonumber
\end{eqnarray}
we
find that (\ref{eq:thmalgb}), (\ref{eq:extra}),
(\ref{eq:zoon}) and
(\ref{eq:kind}) express, respectively, 
that $u + v \eqa w$, $u' + v' = w$, $u \geq u'$, 
$v \geq v'$. It follows that $v \eqa v'$, so that
(\ref{eq:kind}) must actually hold with equality up to a
constant. That is, there exist a $c'$ such that for all $n$,
\begin{equation}
\label{eq:kindb}
\bigl| 
\sum_{x \in \{0,1\}^n} f^{(n)}_{\theta} (x) \log | S(x) | -
\sum_{x \in \{0,1\}^n} f^{(n)}_{\theta} (x) 
\log 1/ f^{(n)}_{\theta}(x \mid S(x)) \bigr| \leq c'.
\end{equation}
The result now follows upon noting that (\ref{eq:kindb}) is just 
(\ref{eq:nexpdef}) with $g^{(n)}$  the uniform distribution.
\end{proof}
\section{Rate Distortion and Structure Function}
\label{sect.rdsf}
We continue the discussion about meaningful information of
Section~\ref{sect.meaning}. This time we a priori restrict the number
of bits allowed for conveying the essence of the information. In the
probabilistic situation this takes the form of allowing only a
``rate'' of $R$ bits to communicate as well as possible, on average,
the outcome of a random variable $X$, while the set ${\cal X}$ of
outcomes has cardinality possibly exceeding $2^R$.  Clearly, not all
outcomes can be communicated without information loss, the average of
which is expressed by the ``distortion''.  This leads to the so-called
``rate--distortion'' theory.  In the algorithmic setting the
corresponding idea is to consider a set of models from which to choose
a single model that expresses the ``meaning'' of the given individual
data $x$ as well as possible. If we allow only $R$ bits to express the
model, while possibly the Kolmogorov complexity $K(x) > R$, we suffer
information loss---a situation that arises for example with ``lossy''
compression.  In the latter situation, the data cannot be perfectly
reconstructed from the model, and the question arises in how far the
model can capture the meaning present in the specific data $x$.  This
leads to the so-called ``structure function'' theory.

The limit of $R$ bits to express
a model to capture the most meaningful information
in the data is an individual version of the average notion
of ``rate''. The remaining less meaningful information in the data 
is the individual version of the average-case notion of ``distortion''.
If the $R$ bits are sufficient to express all meaning in the data
then the resulting model is called a ``sufficient statistic'',
in the sense introduced above. The remaining information in the data
is then purely accidental, random, noise. 
For example, a sequence of
outcomes of $n$ tosses of a coin with computable bias $p$,  typically
has a sufficient statistic
of $K(p)$ bits, while the remaining random information is typically
at least about $pn-K(p)$ bits (up to an $O(\sqrt{n})$ additive term).
\subsection{Rate Distortion}
\label{sec:ratedistortion}
Initially, Shannon \cite{Sh48} introduced rate-distortion 
as follows: ``Practically, we are not interested in exact transmission
when we have a continuous source, but only in transmission to
within a given tolerance. The question is, can we assign a definite
rate to a continuous source when we require only a certain fidelity 
of recovery, measured in a suitable way.'' Later, in \cite{Sh59}
he applied this idea to lossy data compression 
of discrete memoryless sources---our topic below.
As before, we consider a situation in which sender $A$ wants to
communicate the outcome of random variable $X$ to receiver $B$. 
Let $X$ take values in some set ${\cal X}$, and the 
distribution $P$ of $X$ be known to both $A$ and $B$. The change
is that now $A$ is only
allowed to use a finite number, say $R$ bits, to communicate, so that
$A$ can only send $2^R$ different messages. Let us denote by ${Y}$ the
encoding function used by $A$. This ${Y}$ maps ${\cal X}$
onto some set ${\cal Y}$. 
We require that  $|{\cal Y}| \leq 2^R$.
If $|{\cal X}| > 2^R$ or if ${\cal X}$ is continuous-valued, 
then necessarily some information
is lost during the communication. There is no decoding function
$D: {\cal Y} \rightarrow {\cal X}$ such that 
$D({Y}(x)) = x$ for all $x$. Thus, $A$ and $B$ cannot ensure that
$x$ can always be reconstructed. As the next best thing, they may
agree on a code such that for all $x$, the value ${Y}(x)$ contains as much
useful information about $x$ as is possible---what exactly `useful'
means depends on the situation at hand; examples are provided below. 
An easy example would be that ${Y}(x)$ is a finite list of elements,
one of which is $x$.
We assume that the `goodness'
of ${Y}(x)$ is gaged by a {\em distortion function\/} $d:
{\cal X} \times {\cal Y} \rightarrow [0, \infty]$. This distortion
function may be any nonnegative function 
that is appropriate to the situation at hand.
In the example above it could be the logarithm of the number
of elements in the list ${Y}(x)$.
Examples of some common distortion functions are the 
Hamming distance and the squared Euclidean distance. 
We can view $Y$ as a 
a random variable on the space ${\cal Y}$, 
a coarse version of the random variable $X$, defined as taking
value $Y=y$ if $X=x$ with $Y(x)=y$.
Write $f(x) = P(X=x)$ and $g(y)=\sum_{x: Y(x)=y} P(X= x)$. 
Once the distortion function
$d$ is fixed, we define the {\em expected \/} distortion by
\begin{align}
\label{eq:distortion}
{\bf E} [ d(X,{Y}) ] & = \sum_{x \in {\cal X}} f(x) d(x,{Y}(x)) \\
\nonumber
& = \sum_{y \in {\cal Y}} g(y) \sum_{x: Y(x)=y} f(x)/g(y) d(x,y) .
\end{align}
If $X$ is a continuous random variable, the sum should be
replaced by an integral. 
\begin{example}
\label{ex:classicrd}
\rm 
In most standard applications of rate distortion theory,
the goal is to compress $x$ in a `lossy' way, 
such that $x$ can be reconstructed `as well as possible' from 
${Y}(x)$. In that case, ${\cal Y} \subseteq {\cal X}$ and 
writing $\hat{x} = Y(x)$,
the value $d(x,\hat{x})$ measures the similarity
between $x$ and $\hat{x}$.  For example, with ${\cal X}$
is the set of real numbers and ${\cal Y}$ is the set
of integers, the squared difference 
$d(x,\hat{x}) = (x- \hat{x})^2$ is a 
viable distortion function.
We may interpret $\hat{x}$ as an
estimate of $x$, and ${\cal Y}$ as the set of values it can take. The
reason we use the notation ${Y}$ rather than $\hat{X}$ (as in,
for example, \cite{CT91}) is that further below, 
we mostly concentrate on slightly non-standard 
applications where ${\cal Y}$ should {\em not\/} 
be interpreted as a subset of ${\cal X}$.
\end{example}
We want to determine the optimal code $Y$ for communication between A and
B under the constraint that there are no more than $2^R$ messages.
That is, we look for the encoding function ${Y}$ that
minimizes the expected distortion, under the constraint that
$|{\cal Y}| \leq 2^R$. 
Usually, the minimum 
achievable expected distortion 
is nonincreasing as a function of increasing  $R$.
\begin{example}
\label{ex:gauss}
  \rm Suppose $X$ is a real-valued, normally (Gaussian) distributed
  random variable with mean ${\bf E}[X] = 0$ and variance ${\bf E} [ X
  - {\bf E} [X]]^2 = \sigma^2$. Let us use the squared Euclidean
  distance $d(x,y) = (x - y)^2$ as a distortion measure.
  If $A$ is allowed to use $R$ bits, then ${\cal Y}$ can have no
  more than $2^R$ elements, in contrast to ${\cal X}$ that is
  uncountably infinite. We should choose ${\cal Y}$ and the
  function ${Y}$ such that (\ref{eq:distortion}) is minimized.
  Suppose first $R=1$. Then the optimal ${Y}$ turns out to be
$$
{Y}(x) = 
\begin{cases} \sqrt{\frac{2}{\pi}} \sigma^2 & \mbox{\ if $x \geq 0 $} \\ 
  - \sqrt{\frac{2}{\pi}} \sigma^2 & \mbox{\ if $x < 0 $}.
\end{cases}
$$
Thus, the domain ${\cal X}$ is partitioned into two regions, one
corresponding to $x \geq 0$, and one to $x < 0$. By the symmetry of
the Gaussian distribution around $0$, it should be clear that this is
the best one can do. Within each of the two region, one picks a
`representative point' so as to minimize (\ref{eq:distortion}).
This mapping allows $B$ to estimate $x$ as well as possible.

Similarly, if $R=2$, then ${\cal X}$ should be partitioned into 4
regions, each of which are to be represented by a single point such
that (\ref{eq:distortion}) is minimized. An extreme case is $R= 0$:
how can $B$ estimate $X$ if it is always given the same information?
This means that ${Y}(x)$ must take the same value for
all $x$.  The expected distortion (\ref{eq:distortion}) is then
minimized if $Y(x) \equiv 0$, the mean of $X$, giving
distortion equal to $\sigma^2$.
\end{example}
In general, there is no need for the space of estimates ${\cal Y}$
to be a subset of ${\cal X}$. We may, for example, also lossily encode
or `estimate' the actual value of $x$ by specifying a set in which $x$
must lie (Section~\ref{sec:meaningful}) 
or a probability distribution (see below) on ${\cal X}$.
\begin{example}
\label{ex:reconcile}
\rm Suppose receiver $B$ wants to estimate the actual $x$ by a probability
distribution $P$ on ${\cal X}$. Thus, if $R$ bits are allowed to be
used, one of $2^{R}$ different distributions on ${\cal X}$ can be sent to
receiver. The most accurate that can be done is to partition ${\cal
  X}$ into $2^R$ subsets ${\cal A}_1, \ldots, {\cal A}_{2^R}$. 
Relative to any such partition, we introduce a new random variable ${Y}$ 
and  abbreviate the event $x \in {\cal A}_y$ to ${Y}=y$.
Sender observes that ${Y} = y$ for some $y \in 
{\cal Y} = \{1, \ldots, 2^R\}$
and passes this
information on to receiver. The information $y$ actually means that $X$
is now distributed according to the conditional distribution $P(X=x
\mid x \in {\cal A}_y) = P(X= x \mid {Y} = y)$. 

It is now natural to measure the quality of the transmitted  distribution
$P(X=x \mid Y = y)$ by its conditional 
entropy,  i.e. the expected
additional number of bits that sender has to transmit before 
receiver knows the value of $x$ with certainty. This can be achieved
by taking 
\begin{equation}
d(x,y) = 
 \log 1/ P(X=x \mid {Y}= y),
\end{equation} 
which we abbreviate to $d(x,y) =  \log 1/ f(x|y)$.
In words, 
the distortion function is the Shannon-Fano code length for the
communicated distribution. The expected distortion then 
becomes equal to the conditional entropy $H(X \mid Y)$ as defined in
Section~\ref{sec:probmutual} (rewrite according to
\eqref{eq:distortion}, $f(x|y)=f(x)/g(y)$ for 
$P(X=x| Y(x)=y)$ and $g(y)$ defined earlier,
and the definition of conditional probability):
\begin{align}
\label{eq:entdist}
{\bf E}
[d(X,{Y})] 
& = \sum_{y \in {\cal Y}} g(y) \sum_{x: Y(x)=y} (f(x)/g(y)) d(x,y)\\
\nonumber
&=  \sum_{y \in {\cal Y}} g(y) \sum_{x: Y(x)=y} f(x|y) \log 1/f(x|y)\\ 
\nonumber
&=  H(X|{Y}).
\end{align}
How is this related to lossless compression? Suppose
for example that $R= 1$. Then 
the optimal distortion is achieved by
partitioning ${\cal X}$ into two sets ${\cal A}_1, {\cal A}_2$ in the
most `informative' possible way, so that the conditional entropy
$$H(X|Y) = \sum_{y=1,2} P(Y=y) H(X| Y=y)$$
is minimized.  If
$Y$ itself is encoded with the Shannon-Fano code, then $H(Y)$ bits are
needed to communicate $Y$. Rewriting 
$H(X|Y)= \sum_{y \in {\cal Y}} P(Y=y) H(X|Y=y)$ and
$H(X|Y=y)= \sum_{x:Y(x)=y} f(x|y) \log 1/f(x|y)$ with $f(x|y)=P(X=x)/P(Y=y)$
and rearranging, shows  that for all such partitions of ${\cal
  X}$ into $|{\cal Y}|$  subsets defined by $Y:{\cal X} \rightarrow {\cal Y}$
 we have
\begin{equation}
\label{eq:snavel}
H(X|Y) + H(Y) = H(X).
\end{equation} 
The minimum rate 
distortion is obtained by choosing the function
$Y$ that minimizes $H(X|Y)$.
By (\ref{eq:snavel}) this is also the $Y$ maximizing $H(Y)$. Thus, the
average total number of bits we need to send our message in this way
is still equal to $H(X)$---the more we save in the second part, the
more we pay in the first part. 
\end{example}
\paragraph{Rate Distortion and
  Mutual Information:} 
Already in his 1948 paper, Shannon established a
deep relation between mutual information and minimum achievable
distortion for (essentially) {\em arbitrary\/} distortion functions.
The relation is summarized in Theorem~\ref{thm:rd} below. To prepare
for the theorem, we need to slightly extend our setting by considering
{\em independent repetitions of the same scenario}. This can be
motivated in various ways such as (a) it often corresponds to the
situation we are trying to model; (b) it allows us to consider non-integer
rates $R$, and (c) it greatly simplifies the mathematical analysis.

\begin{definition}
\rm
Let ${\cal X}, {\cal Y}$ be two sample spaces.
The  distortion of $y \in {\cal Y}$ with respect 
to $x \in {\cal X}$ is defined
by a nonnegative real-valued function $d(x,y)$ as above. 
We extend the definition to sequences: 
the distortion of $(y_1, \ldots , y_n)$
with respect to $(x_1, \ldots , x_n)$ is
\begin{equation}
\label{eq:avdist}
d((x_1,\ldots,x_n),(y_1, \ldots, y_n)) := \frac{1}{n}
\sum_{i=1}^n d(x_i,y_i).
\end{equation}
\end{definition}
Let $X_1, \ldots, X_n$ be $n$ independent identically
distributed random variables on outcome space ${\cal X}$.
Let ${\cal Y}$ be a set of code words.
We want to find a sequence of functions $Y_1, \ldots , Y_n:{\cal X}
\rightarrow {\cal Y}$ so that the message $(Y_1(x_1), \ldots,
Y_n (x_n)) \in {\cal Y}^n$ gives as much expected 
information about the sequence of outcomes $(X_1=x_1,
\ldots, X_n=x_n)$ as is possible, under the constraint that the message
takes at most $R \cdot n$ bits (so that $R$ bits are allowed on
average per outcome of $X_i$). 
Instead of $Y_1, \ldots , Y_n$ above write
$Z_n: {\cal X}^n \rightarrow  {\cal Y}^n$. 
The {\em expected distortion} ${\bf E}[d(X^n,Z_n)]$ for $Z_n$ is
\begin{equation}
 {\bf E}[d(X^n,Z_n)] = \sum_{(x_1, \ldots , x_n) \in {\cal X}^n}
P(X^n = (x_1, \ldots , x_n)) \cdot \frac{1}{n}
\sum_{i=1}^n d(x_i,Y_i(x_i)).
\end{equation}
Consider functions $Z_n$ 
with range ${\cal Z}_n \subseteq {\cal Y}^n$
satisfying $|{\cal Z}_n| \leq 2^{nR}$.
Let for $n \geq 1$ random variables 
a choice $Y_1, \ldots , Y_n$
minimize the expected distortion
under these constraints, and let the corresponding value $D^*_n (R)$ of
the expected distortion be defined by
\begin{equation}
D^*_n (R) = \min_{Z_n: |{\cal Z}_n| \leq 2^{nR}} {\bf E}(d(X^n,Z_n)) .
\end{equation}

\begin{lemma}
For every distortion measure, and all  $R,n,m \geq 1$,
$(n+m)D^*_{n+m} (R) \leq n D^*_{n}(R)+mD^*_m(R)$.
\end{lemma}
\begin{proof}
Let $Y_1, \ldots , Y_{n}$ achieve $D^*_{n}(R)$ 
and $Y'_1, \ldots , Y'_{m}$ achieve $D^*_{m}(R)$. 
Then, $Y_1, \ldots , Y_{n},Y'_1, \ldots , Y'_m$ 
achieves $(nD^*_n(R) + mD^*_m(R))/(n+m)$. This is an upper
bound on the minimal possible value $D^*_{n+m} (R)$ for $n+m$ random variables.
\end{proof}

It follows that for all $R,n \geq 1$ we have $D^*_{2n}(R) \leq
D^*_n(R)$. 
The inequality is
typically strict; \cite{CT91} gives an 
intuitive explanation of this phenomenon. 
For fixed $R$ the value of $D^*_1(R)$ is fixed and it is finite.
Since also $D^*_n(R)$ is necessarily
positive for all $n$, we have established the existence
of the limit
\begin{equation}
\label{eq:dr}
D^*(R) = \lim\inf_{n \rightarrow \infty} D^*_n (R).
\end{equation}
The value of $D^*(R)$ is the minimum achievable distortion
at rate (number of bits/outcome) $R$. Therefore, $D^*(\cdot)$
It is called the 
{\em distortion-rate function}. 
In our Gaussian
Example~\ref{ex:gauss}, $D^*(R)$ quickly converges to $0$ with
increasing $R$. It turns out that for general $d$, when
we view $D^*(R)$ as a
function of $R\in [0,\infty)$, it is {\em convex and nonincreasing}.
\begin{example}
\label{ex:ber}
\rm
  Let ${\cal X} = \{0,1\}$, and let $P(X= 1) = p$.  Let ${\cal Y} =
  \{0,1\}$  and take the Shannon-Fano distortion
function $d(x,y) =  \log 1/ f(x \mid y)$ with notation as in Example~\ref{ex:reconcile}. 
Let $Y$ be a function that
achieves the minimum expected Shannon-Fano
distortion $D^*_1(R)$. As usual we write $Y$ for the random variable $Y(x)$
induced by $X$. Then, $D^*_1(R)={\bf E}[d(X,Y)] =
{\bf E} [ \log 1/ f(X|Y)] = H(X|Y)$. 
At rate $R = 1$, we
  can set $Y= X$ and the minimum achievable distortion is 
given by $D^*_1(1)=H(X|X) =
  0$. Now consider some rate $R$ with $0 < R < 1$, say $R= \frac{1}{2}$.
 Since we are now 
forced to use less than $2^R < 2$
  messages in communicating, only a fixed message can be sent, no
  matter what outcome of the random variable $X$ is realized. 
This means that no communication is
  possible at all and the minimum achievable distortion is 
$D^*_1(\frac{1}{2}) =H(X) =
  H(p,1-p)$. But clearly, if we consider $n$ repetitions of the same
  scenario and are allowed to send a message out of 
$\lfloor 2^{nR} \rfloor$
  candidates, then some useful information can be communicated after
  all, even if $R < 1$. 
In Example~\ref{ex:berb} we will show that 
if $R > H(p,1-p)$, then $D^*(R) = 0$; if $R \leq
  H(p,1-p)$, then $D^*(R) = H(p,1-p) - R$.
\end{example}
Up to now we studied the minimum achievable distortion $D$ as a function of
the rate $R$.
For technical reasons, it is often more convenient to consider the
minimum achievable rate $R$ as a function of the distortion $D$. 
This is the more  celebrated
version, the {\em rate-distortion function} $R^*(D)$.  
Because $D^*(R)$ is convex
and nonincreasing, $R^*(D): [ 0, \infty) \rightarrow [0,\infty]$ is
just the {\em inverse\/} of the function $D^*(R)$.

It turns out to be possible to relate distortion to the Shannon mutual
information.
This remarkable fact, which Shannon proved already in 
\cite{Sh48,Sh59},
illustrates the fundamental nature of Shannon's concepts.
Up till now, we only considered
{\em deterministic\/} encodings $Y: {\cal X} \rightarrow {\cal Y}$.
But it is hard to analyze the rate-distortion,  and distortion-rate,
functions in this setting. It turns out to be advantageous to follow
an indirect route by bringing
information-theoretic techniques into play.
To this end, we generalize the setting to {\em
randomized\/} encodings.  That is, upon observing $X=x$ with probability
$f(x)$, the
sender may use a randomizing device (e.g. a coin) to decide which 
code word in $y \in {\cal Y}$ he is going to send to the receiver. A
randomized encoding $Y$ thus maps each $x \in {\cal X}$ 
to $y \in {\cal Y}$
with probability
$g_x(y)$, denoted in conditional probability format
as $g(y|x)$. Altogether we deal
with a joint distribution $g(x,y)=f(x)g(y|x)$ on
the joint sample
space ${\cal X} \times {\cal Y}$. (In the deterministic case we have
$g(Y(x) \mid x)=1$ for the given function $Y: {\cal X} \rightarrow {\cal Y}$.)

\begin{definition}
\rm
Let $X$ and $Y$ be joint random variables as above, and let $d(x,y)$
be a distortion measure.
The {\em  expected distortion} $D(X,Y)$ of $Y$ with respect to $X$ is 
defined by
\begin{equation}\label{eq.DXY}
D(X,Y)= 
\sum_{x \in {\cal X}, y \in {\cal Y}} g(x,y) d(x,y).
\end{equation}
\end{definition}
Note that for a given problem
the source probability $f(x)$ of outcome $X=x$ is fixed,
but the randomized encoding $Y$, that is the conditional probability
$g(y|x)$ of encoding source word $x$ by code word $y$,
can be chosen to advantage.
We define the auxiliary notion
of {\em information rate distortion function} $R^{(I)}(D)$ by
\begin{equation}
\label{eq:ird}
R^{(I)}(D) = \inf_{Y : D(X,Y) \leq D}
I(X; Y).
\end{equation}
 That is, for random variable $X$,
among {\em all\/} joint random variables $Y$ with expected distortion
to $X$ less
  than or equal to $D$, the information rate $R^{(I)}(D)$
equals the minimal mutual information with $X$.
\begin{theorem}[Shannon]
\label{thm:rd}
For every random source $X$ and distortion measure $d$:
\begin{equation}
\label{eq:rd}
R^*(D) = R^{(I)}(D)
\end{equation}
\end{theorem} 
This remarkable theorem 
states that the best deterministic code achieves a rate-distortion
that equals the minimal information rate possible for a randomized code, 
that is, the minimal
mutual information between the random source and a
randomized code.
Note that this does not mean that $R^*(D)$
is independent of the distortion measure.
In fact, the source random variable $X$,
together with the distortion measure $d$, determines a random
code $Y$ for which the joint random variables $X$ and $Y$
reach the infimum in \eqref{eq:ird}.
The proof of this theorem is given in 
\cite{CT91}. It is illuminating to see how it goes:
It is shown first that, for a random source $X$ and distortion measure $d$,
every deterministic code $Y$ with distortion $\leq D$ has rate
$R \geq R^{(I)} (D)$. Subsequently, it is shown that there
exists
a deterministic code that, with distortion $ \leq D$,
achieves rate $R^*(D)=R^{(I)} (D)$.
To analyze deterministic $R^*(D)$ therefore, 
we can determine the best randomized
code $Y$ for random source $X$ under distortion constraint $D$,
and then we know that simply $R^*(D)=I(X;Y)$. 
\begin{example} (Example~\ref{ex:ber}, continued)
\label{ex:berb}
\rm Suppose we want to compute $R^*(D)$ for some $D$ between $0$ and
$1$.  If we only allow encodings $Y$ that are deterministic functions
of $X$, then either $Y(x) \equiv x$ or $Y(x) \equiv |1- x|$. 
In both cases ${\bf E }
[d(X,Y)] = H(X| Y) = 0$, so $Y$ satisfies the constraint in
(\ref{eq:ird}). In both cases, $I(X, Y) = H(Y) = H(X)$. With
(\ref{eq:rd}) this
shows that $R^*(D) \leq H(X)$. However, $R^*(D)$ is actually smaller:
by allowing randomized codes, we can define $Y_{\alpha}$ as
$Y_{\alpha} (x) = x$ with probability $\alpha$ and $Y_{\alpha} (x) = |1- x|$
with probability $1- \alpha$. For $0 \leq \alpha \leq \frac{1}{2}$, ${\bf E }
[d(X,Y_{\alpha})] = H(X| Y_{\alpha})$ increases with $\alpha$, while
$I(X;Y_{\alpha})$ decreases with $\alpha$.  Thus, by choosing the
$\alpha^*$ for which the constraint ${\bf E } [d(X,Y_{\alpha})] \leq
D$ holds with equality, we find $R^*(D) = I(X; Y_{\alpha^*})$. Let us
now calculate $R^*(D)$ and $D^*(R)$ explicitly.

Since $I(X,Y) = H(X) - H(X|Y)$, we can rewrite $R^*(D)$ as
$$
R^*(D) = H(X)  - \sup_{Y: D(X,Y) \leq D}
H(X|Y).
$$
In the special case where $D$ is itself the 
Shannon-Fano distortion, this can in turn be rewritten as
$$
R^*(D) = H(X)  - \sup_{Y: H(X|Y) \leq D} H(X \mid Y) 
= H(X) - D.
$$
Since $D^*(R)$ is the inverse of $R^*(D)$, 
we find $D^*(R) = H(X) - R$, as announced in Example~\ref{ex:ber}.
\end{example}

\paragraph{Problem and Lacuna:}
In the Rate-Distortion setting we allow (on average) a rate of
$R$ bits to express the data as well as possible in some way, 
and measure the average of loss by some distortion function.
But in many cases, like lossy compression of images,
one is interested in the individual cases. The average over all 
possible images may be irrelevant for the individual cases one meets.
Moreover, one is not particularly interested in bit-loss,
but rather in preserving the essence of the image as well as possible.
As another example, suppose
the distortion function is simply to supply the remaining
bits of the data. But this can be unsatisfactory: we are given
an outcome of a measurement as a real number of $n$ significant bits. Then 
the $R$ most significant bits carry most of the meaning
of the data, while the remaining $n-R$ bits may be irrelevant.
Thus, we are lead to the elusive notion
of a distortion function that captures
the amount of ``meaning'' that is not included in the $R$ rate bits.
These issues are taken up by Kolmogorov's proposal of the structure function.
This cluster  of ideas puts the notion
of Rate--Distortion in an individual algorithmic (Kolmogorov
complexity) setting, and focuses on the meaningful information
in the data. In the end we can recycle the new insights and
connect them to Rate-Distortion notions to provide new foundations
for statistical inference notions as maximum likelihood (ML)
\cite{Fi22}, 
minimum
message length (MML) \cite{WallaceF87}, 
and minimum description length (MDL) \cite{Ri89}.
\subsection{Structure Function}
\label{sec:structure}
\begin{figure}
\begin{center}
\epsfxsize=8cm
\epsfxsize=8cm \epsfbox{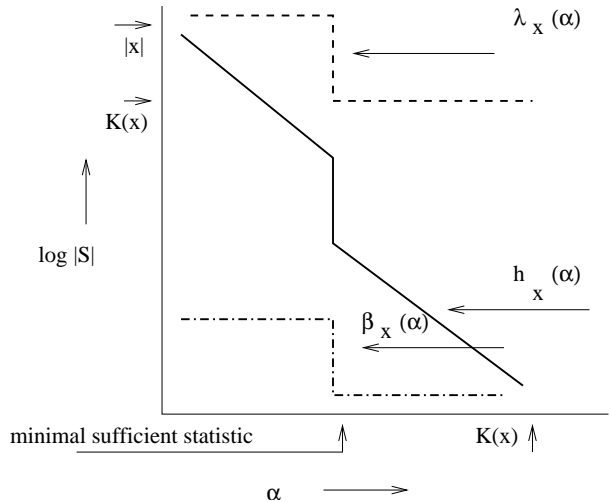}
\end{center}
\caption{Structure functions $h_x(i), \beta_x(\alpha), \lambda_x(\alpha)$,
and minimal sufficient statistic.}
\label{figure.estimator}
\end{figure}

There is a close relation between
functions describing
three, a priori seemingly unrelated, aspects of modeling individual
data, depicted in Figure~\ref{figure.estimator}.
\label{sec:meaningful}
One of these was introduced by
Kolmogorov at a conference in Tallinn 1974 (no written version)
and in a talk at the Moscow Mathematical Society in the same year
of which the abstract \cite{Ko74} 
is as follows (this is the only writing by Kolmogorov about
this circle of ideas):
\begin{quote}
``To each constructive object corresponds a function $\Phi_x(k)$ of a
 natural number $k$---the log of minimal cardinality of $x$-containing
 sets that allow definitions of complexity at most $k$.
 If the element $x$ itself allows a simple definition,
 then the function $\Phi$ drops to $1$ 
 even for small $k$.
 Lacking such definition, the element is ``random'' in a negative sense.
 But it is positively ``probabilistically random'' only when function
 $\Phi$ having taken the value $\Phi_0$ at a relatively small
 $k=k_0$, then changes approximately as $\Phi(k)=\Phi_0-(k-k_0)$.''
\end{quote}
Kolmogorov's $\Phi_x$ is commonly called the ``structure function''
and is here denoted as $h_x$ and defined in \eqref{eq2}.  The
structure function notion entails a proposal for a non-probabilistic
approach to statistics, an individual combinatorial relation between
the data and its model, expressed in terms of Kolmogorov complexity.
It turns out that the structure function determines all stochastic
properties of the data in the sense of determining the best-fitting
model at every model-complexity level, the equivalent notion to
``rate'' in the Shannon theory. A consequence is this: minimizing the
data-to-model code length (finding the ML estimator or MDL estimator),
in a class of contemplated models of prescribed maximal (Kolmogorov)
complexity, {\em always} results in a model of best fit, irrespective
of whether the source producing the data is in the model class
considered.  In this setting, code length minimization {\em always}
separates optimal model information from the remaining accidental
information, and not only with high probability.  The function that
maps the maximal allowed model complexity to the goodness-of-fit
(expressed as minimal ``randomness deficiency'') of the best model
cannot itself be monotonically approximated. However, the shortest
one-part or two-part code above can---implicitly optimizing this
elusive goodness-of-fit.

In probabilistic statistics the goodness of the selection process is
measured in terms of expectations over probabilistic ensembles.  For
current applications, average relations are often irrelevant, since
the part of the support of the probability mass function that will
ever be observed has about zero measure. This may be the case in, for
example, complex video and sound analysis.  There arises the problem
that for individual cases the selection performance may be bad
although the performance is good on average, or vice versa. There is
also the problem of what probability means, whether it is subjective,
objective, or exists at all.  Kolmogorov's proposal strives for the
firmer and less contentious ground of finite combinatorics and
effective computation.

\paragraph{Model Selection:}
It is technically convenient to initially consider the simple model
class of finite sets to obtain our results, just as in
Section~\ref{sec:algsuf}. It then turns out that it is relatively easy
to generalize everything to the model class of computable probability
distributions (Section~\ref{s.prob}). That class is very large
indeed: perhaps it contains every distribution that has ever been
considered in statistics and probability theory, as long as the
parameters are computable numbers---for example rational numbers. Thus
the results are of great generality; indeed, they are so general that
further development of the theory must be aimed at restrictions on
this model class.

Below we will consider various model
selection procedures. These are approaches for finding a model $S$
(containing $x$) for arbitrary data $x$. The goal is to find a model
that captures all meaningful information in the data $x$ . All
approaches we consider are at some level based on coding $x$ by giving
its index in the set $S$, taking $ \log |S|$ bits. This
codelength may be thought of as a particular distortion function, and
here lies the first connection to Shannon's rate-distortion:

\begin{example}\label{rem.rd-ksf1}
\rm
A model selection procedure is a function $Z_n$ mapping binary data of
length $n$ to finite sets of strings of length $n$, containing the
mapped data, $Z_n(x)=S$ ($x \in S$). The range of $Z_n$ satisfies
${\cal Z}_n \subseteq 2^{\{0,1\}^n}$, The distortion function $d$ is
defined to be $d(x,Y(x))= \frac{1}{n} \log |S|$. To define the
rate--distortion function we need that $x$ is the outcome of a random
variable $X$. Here we treat the simple case that $X$ represents $n$
flips of a fair coin; this is substantially generalized in
Section~\ref{sec:esf}. Since each outcome of a fair coin can be
described by one bit, we set the rate $R$ at $0 < R < 1$. Then,
$D_n^*(R) = \min_{Z_n: |{\cal Z}_n| \leq 2^{nR}} \sum_{|x|=n}2^{-n}
\frac{1}{n} \log |Z_n(x)|$ For the minimum of the right-hand side we
can assume that if $y \in Z_n(x)$ then $Z_n(y)=Z_n(x)$ (the distinct
$Z_n(x)$'s are disjoint). Denote the distinct $Z_n(x)$'s by $Z_{n,i}$
with $i=1,\ldots , k$ for some $k \leq 2^{nR}$. Then, $D_n^*(R) = \min
_{Z_n: |{\cal Z}_n| \leq 2^{nR}} \sum_{1=1}^k |Z_{n,i}|2^{-n}
\frac{1}{n} \log |Z_{n,i}|$.  The right-hand side reaches its minimum
for all $Z_{n,i}$'s having the same cardinality and $k=2^{nR}$. Then,
$D_n^*(R) = 2^{nR} 2^{(1-R)n} 2^{-n} \frac{1}{n} \log 2^{(1-R)n} =
1-R$.  Therefore, $D^*(R)= 1-R$ and therefore $R^*(D) = 1-D$.

Alternatively, and more in line with the structure-function
approach below, one may consider repetitions of a random variable $X$
with outcomes in $\{0,1\}^n$. Then,
a model selection procedure is a function $Y$ mapping
binary data of length $n$ to finite sets of strings of length $n$,
containing the mapped data,
$Y(x)=S$ ($x \in S$). The range of $Y$ satisfies
${\cal Y} \subseteq 2^{\{0,1\}^n}$,  The distortion function $d$ is defined
by $d(x,Y(x))= \log |S|$. To define the rate--distortion function
we need that $x$ is the outcome of a random variable $X$, say
a toss of a fair $2^n$-sided coin. Since each outcome of a fair
coin can be described by $n$ bits, we set the rate $R$
at $0 < R < n$. Then, for outcomes $\overline{x}=x_1 \ldots x_m$
($|x_i|=n$), resulting from $m$  i.i.d. random variables $X_1, \ldots , X_m$,
we have $d(\overline{x}, Z_m (\overline{x})) =
\frac{1}{m} \sum_{i=1}^m \log |Y_i (x_i)| =
\frac{1}{m} \log | Y_1(x_1) \times \cdots \times Y_m(x_m)|$. Then, 
$D_m^*(R) = \min_{Z_m: |{\cal Z}_m| \leq 2^{mR}}
\sum_{\overline{x}}2^{-mn} d(\overline{x}, Z_m (\overline{x}))$.
Assume that $\overline{y} \in Z_m(\overline{x})$ if
$Z_m(\overline{y}) = Z_m(\overline{x})$: the distinct
$Z_m(\overline{x})$'s are disjoint and partition $\{0,1\}^{mn}$
into disjoint subsets $Z_{m,i}$, with $i=1, \ldots, k$ for
some $k \leq 2^{mR}$.
Then,
$D_m^*(R) = \min_{Z_m: |{\cal Z}_m| \leq 2^{mR}}
\sum_{i=1,\ldots, k} |Z_{m,i}|2^{-mn} \frac{1}{m}
\log |Z_{m,i}|$.
The right-hand side reaches its minimum for all $Z_{m,i}$'s having
the same cardinality and $k=2^{mR}$, so that
$D_m^*(R) = 2^{(n-R)m} 2^{mR} 2^{-mn} \frac{1}{m} \log 2^{(n-R)m} = n-R$.
Therefore, $D^*(R)= n-R$ and $R^*(D) = n-D$.
In Example~\ref{ex.rd=str} we relate these numbers to the structure
function approach described below.
\end{example}

\paragraph{Model Fitness:}
A distinguishing feature of the structure function approach is that
we want to formalize what it means for an element to  be ``typical''
for a set that contains it. For example, if we flip a fair coin $n$ 
times, then the sequence of $n$ outcomes, denoted by $x$,
 will be an element of the set $\{0,1\}^n$. In fact,
most likely it will be a ``typical'' element in the sense that
it has all properties that hold on average for an element of that set.
For example, $x$ will have $\frac{n}{2} \pm O(\sqrt{n})$ frequency
of 1's, it will have a run of about $\log n$ consecutive 0's,
and so on for many properties. Note that the sequence $x=0 \ldots 01\ldots1$,
consisting of one half 0's followed by one half ones, is very untypical,
even though it satisfies the two properties described explicitly.
The question arises how to formally define ``typicality''. We do
this as follows:
The lack of typicality
of $x$ with respect to a finite set $S$ (the model) containing it,
is the amount by which $K(x|S)$
falls short of the length $\log |S|$ of the data-to-model code (Section~\ref{sec:algsuf}).
Thus, the {\em randomness deficiency} of $x$ in $S$ is defined by
      \begin{equation}\label{eq:randomness-deficiency}
\delta (x | S) = \log |S| - K(x | S),
      \end{equation}
for $x \in S$, and $\infty$ otherwise. Clearly, $x$ can be typical for 
vastly different sets. For example, every $x$ is typical for the singleton
set $\{x\}$, since $\log |\{x\}|=0$ and $K(x \mid \{x\})=O(1)$.
Yet the many $x$'s that have $K(x) \geq n$ are also typical for
$\{0,1\}^n$, but in another way. In the first example, the set is about
as complex as $x$ itself. In the second example, the set is vastly
less complex than $x$: the set has complexity about 
$K(n) \leq \log n + 2 \log \log n$ while $K(x)\geq n$.
Thus, very high complexity data may have simple
sets for which they are typical. As we shall see, 
this is certainly not the case for all high complexity data.
The question arises how typical
data $x$ of length $n$ can be in the best case
for a finite set of complexity $R$
when $R$ ranges from 0 to $n$. The function describing this dependency,
expressed in terms of randomness deficiency to measure the optimal
typicality, as a function of the complexity ``rate'' $R$ ($0 \leq R \leq n$)
of the number of bits we can maximally spend to describe a finite
set containing $x$,
is defined as follows:  


The {\em minimal randomness deficiency} function is
           \begin{equation}
\label{eq1}
\beta_x( R) =
\min_{S} \{ \delta(x| S): S \ni x, \;  K(S) \leq R \},
            \end{equation}
            where we set $\min \emptyset = \infty$.  If $\delta(x |
            S)$ is small, then $x$ may be considered as a {\em
              typical} member of $S$. This means that $S$ is a
            ``best'' model for $x$---a most likely explanation.  There
            are no simple special properties that single it out from
            the majority of elements in $S$.  We therefore like to
            call $\beta_x(R)$ the {\em best-fit estimator}.  This
            is not just terminology: If $\delta (x | S)$ is small,
            then $x$ satisfies {\em all} properties of low Kolmogorov
            complexity that hold with high probability (under the
            uniform distribution) for the elements of $S$. To be
            precise \cite{VV02}: Consider strings of length $n$ and
            let $S$ be a subset of such strings. We view a {\em
              property} of elements in $S$ as a function $f_P: S
            \rightarrow \{0,1\}$. If $f_P(x)=1$ then $x$ has the
            property represented by $f_P$ and if $f_P(x)=0$ then $x$
            does not have the property.  Then: (i) If $f_P$ is a
            property satisfied by all $x$ with $\delta(x | S) \le
            \delta (n)$, then $f_P$ holds with probability at least
            $1-1/2^{\delta(n)}$ for the elements of $S$.
                                                                                
(ii) Let
$f_P$ be any
property
that holds with probability at least
$1-1/2^{\delta (n)}$ for the
elements of $S$. Then, every such $f_P$ holds
simultaneously for every $x \in S$
with $\delta (x | S)\le\delta (n)-K(f_P|S)-O(1)$.

\begin{example}
  \rm {\bf Lossy Compression:} \index{compression, lossy} The function
  $\beta_x( R)$ is relevant to lossy compression (used, for instance,
  to compress images) -- see also Remark~\ref{rem:lossy}.  Assume we
  need to compress $x$ to $R$ bits where $R \ll K(x)$.  Of course this
  implies some loss of information present in $x$.  One way to select
  redundant information to discard is as follows: Find a set $S\ni x$
  with $K(S)\le R$ and with small $\delta(x | S)$, and consider a
  compressed version $S'$ of $S$.  To reconstruct an $x'$, a
  decompresser uncompresses $S'$ to $S$ and selects at random an
  element $x'$ of $S$.  Since with high probability the randomness
  deficiency of $x'$ in $S$ is small, $x'$ serves the purpose of the
  message $x$ as well as does $x$ itself.  Let us look at an example.
  To transmit a picture of ``rain'' through a channel with limited
  capacity $R$, one can transmit the indication that this is a picture
  of the rain and the particular drops may be chosen by the receiver
  at random.  In this interpretation, $\beta_x(R)$ indicates how
  ``random'' or ``typical'' $x$ is with respect to the best model at
  complexity level $R$---and hence how ``indistinguishable'' from the
  original $x$ the randomly reconstructed $x'$ can be expected to be.
\end{example}

\begin{remark}
\rm
This randomness deficiency function quantifies 
the goodness of fit of the best model at complexity $R$
for given data $x$. As far as we know no direct counterpart of this
notion exists in Rate--Distortion theory, or, indeed,
can be expressed in classical theories like Information Theory.
But the situation is different for the next function we define,
which, in almost contradiction to the previous statement, can
be tied to the minimum randomness deficiency function, yet, as will be
seen in Example~\ref{ex.rd=str} and Section~\ref{sec:esf}, 
does have a counterpart in Rate--Distortion theory after all.
\end{remark}

\paragraph{Maximum Likelihood estimator:}
The {\em Kolmogorov structure} function $h_x$ of given data $x$ is defined by
 \begin{equation}\label{eq2}
   h_{x}(R) = \min_{S} \{\log | S| : S \ni x,\; K(S) \leq R\},
\end{equation}
where $S \ni x$ is
a contemplated model for $x$, and $R$ is a nonnegative
integer value bounding the complexity of the contemplated $S$'s.
The structure function uses models that are finite sets and
the value of the structure function is the log-cardinality of the
smallest such set containing the data. Equivalently, we can
use uniform probability mass functions over finite supports (the former
finite set models). The smallest set containing the data then becomes
the uniform probability mass assigning the highest probability
to the data---with the value of the structure function 
the corresponding negative
log-probability. This motivates us to call $h_x$ the {\em maximum likelihood
estimator}. The treatment can be extended from uniform probability
mass functions with finite supports, to  probability models that
are arbitrary computable probability mass functions, keeping
all relevant notions and results essentially unchanged, Section~\ref{s.prob},
justifying the maximum likelihood identification even more.

Clearly, the Kolmogorov structure function is
non-increasing and reaches $\log |\{x\}| = 0$
for the ``rate'' $R = K(x)+c_1$ where $c_1$ is the number of bits required
to change $x$ into $\{x\}$.
It is also easy to see that for argument $K(|x|)+c_2$, where $c_2$
is the number of bits required to compute the
set of all strings of length $|x|$ of $x$ from $|x|$,
the value of the structure function is at most $|x|$; see Figure~\ref{figure.estimator}
\begin{example}\label{ex.rd=str}
\rm
Clearly the structure function measures for individual outcome $x$
a distortion that is related to the
one measured by $D_1^*(R)$ in Example~\ref{rem.rd-ksf1}
for the uniform average of outcomes $x$.
Note that all  strings $x$ of length $n$ satisfy $h_x(K(n)+O(1)) \leq n$
(since $x \in S_n=\{0,1\}^n$ and $K(S_n)=K(n)+O(1)$).
For every $R$ ($0 \leq R \leq n$),
we can describe every $x = x_1x_2 \ldots x_n$ as an element
of the set $A_R = \{x_1 \ldots x_R y_{R+1} \ldots y_n:
y_i \in \{0,1\}, R < i \leq n \}$. Then, $|A_R|=2^{n-R}$
and $K(A_R) \leq R+K(n,R)+O(1) \leq R + O(\log n)$.
This shows that $h_x (R) \leq n-R+O(\log n)$ for every $x$
and every $R$ with $0 \leq R \leq n$; see Figure~\ref{figure.estimator}.

For all $x$'s and $R$'s we can describe $x$ in a two-part code by the set
$S$ witnessing $h_x(R)$ and $x$'s index in that set. The first part
describing $S$ in $K(S)=R$ allows us to generate
$S$, and given $S$ we know $\log |S|$. Then,
we can parse the second part of $\log |S|=h_x(R)$ bits that gives $x$'s
index in $S$. We also need a fixed $O(1)$ bit program to produce $x$
from these descriptions. Since $K(x)$ is the lower bound on 
the length of effective descriptions of $x$, we have $h_x(R)+R \geq K(x)-O(1)$. 
There are $2^n - 2^{n-K(n)+O(1)}$ strings $x$ of complexity $K(x)\geq n$,
\cite{LiVi97}. For all these strings $h_x(R) + R \geq n-O(1)$.
Hence, the expected value $h_x(R)$ equals
$2^{-n} \{ (2^n-2^{n-K(n)+O(1)}) [n-R+O(\log n)]
+ 2^{n-K(n)+O(1)} O(n-R+O(\log n)) \} = n-R + O(n-R/2^{-K(n)})
=  n-R + o(n-R)$
(since $K(n) \rightarrow \infty$ for $n \rightarrow \infty$).
That is, the expectation of $h_x(R)$ equals $(1+o(1))D^*_1(R)
=(1+o(1))D^*(R)$, the Distortion-Rate function, where the
$o(1)$ term goes to 0 with the length $n$ of $x$. In
Section~\ref{sec:esf} we extend this idea to non-uniform distributions
on $X$.
\end{example}

For every $S\ni x$ we have
         \begin{equation}\label{eq.descr}
K(x)\leq K(S)+ \log |S| + O(1).
          \end{equation} 
Indeed,
consider the following \emph{two-part code}
for $x$: the first part is
a shortest  self-delimiting program $p$ of $S$ and the second
part is
$\lceil\log|S|\rceil$ bit long index of $x$
in the lexicographical ordering of $S$. 
Since $S$ determines $\log |S|$ this code is self-delimiting
and we obtain \eqref{eq.descr}
where the constant $O(1)$ is
the length of the program to reconstruct 
$x$ from its two-part code.
We thus conclude that $K(x)\leq R+h_x(R)+O(1)$, that is, the
function $h_x(R)$
never decreases
more than a fixed independent constant below
the diagonal \emph{sufficiency line} $L$ defined by
$L(R)+R = K(x)$,
which is a lower bound on $h_x (R)$
and is approached to within a constant distance by
the graph of $h_x$ for certain $R$'s
(for instance, for $R = K(x)+c_1$).
For these $R$'s we
thus have
$R + h_x (R) = K(x)+O(1)$.
In the terminology we have introduced in Section~\ref{sect.ss} and Definition~\ref{def:algsufstat},
a model corresponding to such an $R$ (witness for
$h_x(R)$) is an optimal set for $x$
and a shortest program to compute this model
is a sufficient statistic. It is
{\em minimal} for the least such $R$ for which the above equality holds.

\paragraph{MDL Estimator:}
The length of the minimal two-part code for $x$ consisting
of the model cost $K(S)$ and the
length of the index of $x$ in
$S$,
the complexity of $S$ upper bounded by $R$, is given by
the {\em MDL (minimum description length) function}:
  \begin{equation}\label{eq.3}
   \lambda_{x}(R) =
\min_{S} \{\Lambda(S): S \ni x,\; K(S) \leq R\},
  \end{equation}
where $\Lambda(S)=\log|S|+K(S) \ge K(x)-O(1)$ is
the total length of two-part code of $x$
with help of model $S$. 
 Clearly,
$\lambda_x (R) \leq  h_x(R)+ R +O(1)$,
but a priori it is still possible that $ h_x(R')+ R'
< h_x(R)+R$ for $R' < R$.
In that case $\lambda_x(R) \leq
 h_x(R')+ R'
< h_x(R)+R$. However, in \cite{VV02} it is shown
that $\lambda_x (R) =  h_x(R)+ R + O(\log n)$
for all $x$ of length $n$. Even so, this doesn't mean that a set
$S$ that witnesses $\lambda_x (R)$ in the sense that $x \in S$,
$K(S) \leq R$, and $K(S)+\log |S|= \lambda_x (R)$,
also witnesses $h_x(R)$. It can in fact be the case that $K(S) \leq R-r$,
and $\log |S|= h_x (R)+r$ for arbitrarily large $r \leq n$.

Apart from being convenient for the technical analysis
in this work, $\lambda_x (R)$ is the
celebrated two-part Minimum Description Length code
length \cite{Ri89} with the
model-code length restricted to at most $R$. 
When $R$ is large enough so that $\lambda_{x}(R) = K(x)$,
then there is a set $S$ that is a sufficient statistic, and
the smallest such $R$ has an associated witness set $S$ that 
is a minimal sufficient statistic. 

The most fundamental result in \cite{VV02}
is the equality
         \begin{equation}\label{eq.eq}
\beta_x (R )  = h_x (R) + R - K(x) = \lambda_x (R)
- K(x)
         \end{equation}
which holds within logarithmic additive terms in argument and value.
Additionally, every set $S$ that witnesses the value $h_x (R )$
(or $\lambda_x(R)$),
also witnesses the value $\beta_x (R)$ (but not vice versa).
It is easy to see that $h_x (R)$ and $\lambda_x(R)$
are
upper semi-computable (Definition~\ref{def.semi});
but we have shown \cite{VV02}
that $\beta_x (R)$ is neither upper nor lower semi-computable
(not even within a great tolerance).
A priori
there is no reason to suppose that
a set that witnesses $h_x (R)$
(or $\lambda_x(R)$) also witnesses $\beta_x (R)$,
for {\em every} $R$.
But the fact that they do, vindicates
Kolmogorov's original proposal
and establishes $h_x$'s pre-eminence over $\beta_x$ -- the
pre-eminence of $h_x$ over $\lambda_x$ is discussed below.
                                                                                
\begin{remark}\label{rem.MLvsMDL}
\rm
 What we call `maximum likelihood' in the form of $h_x$ is really `maximum
likelihood' under a complexity constraint $R$ on the models' as in
$h_x (R)$. In
statistics, it is a well-known fact that maximum likelihood often
fails (dramatically overfits) when the models under consideration are
of unrestricted complexity (for example, with polynomial regression with
Gaussian noise, or with Markov chain model learning, maximum
likelihood will always select a model with $n$ parameters, where $n$ is
the size of the sample---and thus typically, maximum likelihood will
dramatically overfit, whereas for example MDL typically performs
well). The equivalent, in our setting, is that allowing models of unconstrained
complexity  for data $x$, say complexity $K(x)$,
will result in the ML-estimator $h_x (K(x)+O(1))=0$---the witness model
being the trivial, maximally overfitting, set $\{x\}$.
In the MDL case, on the other hand, there may be a long constant
interval with the MDL estimator
 $\lambda_x (R) = K(x)$ ($R \in [R_1 , K(x)]$)
where the length of the two-part code doesn't decrease anymore.
Selecting the least complexity model witnessing this function value
we obtain the, very significant, algorithmic {\em minimal} sufficient
statistic, Definition~\ref{def:algsufstat}.
In this sense, MDL augmented with a bias for the least complex explanation,
which we may call the `Occam's Razor MDL',
is superior to maximum likelihood and resilient to overfitting.
If we don't apply bias in the direction of simple explanations,
then -- at least in our setting -- 
MDL may be just as prone to overfitting as is ML. For example,
if $x$ is a typical random element of $\{0,1\}^n$, then
 $\lambda_x (R) = K(x)+O(1)$ for the entire interval
$K(n)+O(1) \leq R \leq K(x)+O(1) \approx n$.
Choosing the model on the left side, of simplest complexity,
 of complexity $K(n)$
gives us the best fit with the correct model $\{0,1\}^n$.
But choosing a model on the right side, of high complexity, gives us
a model $\{x\}$ of complexity $K(x)+O(1)$ that completely
overfits the data by modeling all random noise in $x$
(which in fact in this example almost completely consists of random noise).
\index{overfit}
                                                                                
Thus, it should be emphasized that 'ML =
MDL' really only holds if complexities are constrained to a value
$R$ (that remains fixed as the sample size grows---note that in the
Markov chain example above, the complexity grows linearly with
the sample size); it certainly
does not hold in an unrestricted sense (not even in the algorithmic setting).
\end{remark}
\begin{remark}
\rm
In a sense, $h_x$ is more strict than $\lambda_x$:
A set that witnesses $h_x(R)$ also witnesses
$\lambda_x(R)$ but not necessarily vice versa. However,
at those complexities $R$ where $\lambda_x (R)$ drops
(a little bit of added complexity in the model allows a
shorter description), the witness set of $\lambda_x$ is
also a witness set of $h_x$. But if $\lambda_x$ stays
constant in an interval $[R_1, R_2]$, then we
can trade-off complexity of a witness set versus its cardinality,
keeping the description length constant. This is of course not possible
with $h_x$ where the cardinality of the witness set at complexity $R$
is fixed at $h_x(R)$.
\end{remark}
                                                                                
The main result  can be taken as a foundation and justification
of common statistical principles in model
selection such as maximum likelihood
or MDL.
The structure functions $\lambda_x,h_x$ and $\beta_x$ can assume all
possible shapes over their full domain of definition (up to
additive logarithmic precision in both argument and value), see \cite{VV02}.
(This establishes
the significance of \eqref{eq.eq}, since it shows that $\lambda_x (R)
\gg K(x)$ is common for $(x, R)$ pairs---in which case the more
or less
easy fact that $\beta_x(R)=0$ for $\lambda_x(R)=K(x)$ is
not
applicable, and it is a priori unlikely that \eqref{eq.eq} holds:
Why should minimizing a set containing
$x$ also minimize its randomness deficiency? Surprisingly, it does!)
 We have exhibited a---to our knowledge first---natural example,
$\beta_x$, of a function that
is not semi-computable but computable with an oracle for the halting problem.

\begin{example}\label{ex.prnr}
\index{randomness, positive}
\index{randomness, negative}
\rm
{\bf ``Positive'' and ``Negative'' Individual Randomness:}
In \cite{GTV01} we showed the existence
of strings for which essentially
the singleton set consisting of the string itself is a minimal
sufficient statistic. While a sufficient
statistic of an object yields a two-part code that is as short as the shortest
one part code, restricting the complexity of the allowed statistic
may yield two-part codes that are considerably longer than the best one-part
code (so the statistic is insufficient).
In fact,
for every object there is a complexity bound below which this happens---but
if that bound is small (logarithmic) we call the object ``stochastic''
since it has a simple satisfactory explanation (sufficient statistic).
Thus, Kolmogorov in \cite{Ko74} 
 makes the important distinction of
an object being random in the ``negative'' sense by having this bound
high (it has high complexity and is not a typical element of
a low-complexity model),
and an object being random in the ``positive,
probabilistic'' sense by both having this bound small and itself
having complexity considerably exceeding this bound
(like a string $x$ of length $n$ with $K(x) \geq n$,
being typical for the
set $\{0,1\}^n$, or the uniform probability distribution over that
set,
while this set or probability distribution
has complexity $K(n)+O(1) = O(\log n)$).
We depict the distinction in Figure~\ref{figure.pos_negrandom}.
In simple terms: High Kolmogorov complexity of a data string
just means that it is random in a {\em negative sense};
but a data string of high Kolmogorov
complexity is {\em positively random} if the simplest satisfactory explanation
(sufficient statistic) has low complexity, 
and it therefore is the typical outcome
of a simple random process. 
                                                                                
\begin{figure}
\begin{center}
\epsfxsize=8cm
\epsfxsize=8cm \epsfbox{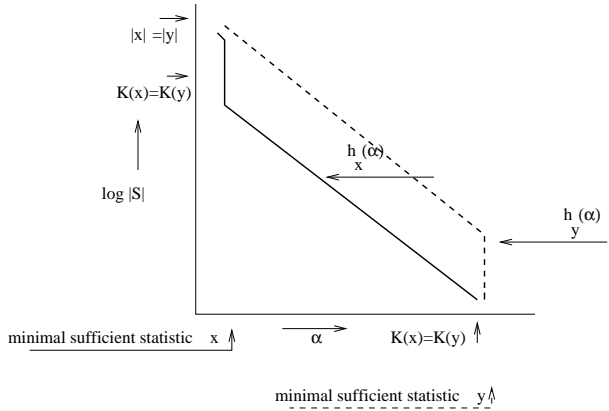}
\end{center}
\caption{Data string $x$ is ``positive random'' or ``stochastic''
 and data string $y$
is just ``negative random'' or ``non-stochastic''.}
\label{figure.pos_negrandom}
\end{figure}

In \cite{VV02} it is shown that for every length $n$ and
every complexity $k \leq n+K(n) + O(1)$ (the maximal complexity
of $x$ of length $n$) and every $R \in [0,k]$,
there are $x$'s of length $n$ and complexity $k$ such that
the minimal randomness deficiency $\beta_x (i) \geq  n-k\pm O(\log
n)$
for every $i \leq R \pm O(\log n)$ and $\beta_x (i) \pm O(\log n)$
for every $i > R \pm O(\log n)$. Therefore, the set of  $n$-length
strings of every complexity $k$ can be partitioned in subsets of strings that
have a Kolmogorov minimal sufficient statistic of complexity
$\Theta (i \log n)$ for $i = 1, \ldots , k/ \Theta (\log n)$.
For instance, there are $n$-length non-stochastic
strings of almost maximal complexity $n -  \sqrt{n}$
having significant $\sqrt{n}\pm O(\log n)$ randomness deficiency with
respect to $\{0,1\}^n$ or, in fact, every other finite set
of complexity less than $n - O(\log n)$!
\end{example}

\subsubsection{Probability Models}
\label{s.prob}
\begin{figure}
\begin{center}
\epsfxsize=8cm
\epsfxsize=8cm \epsfbox{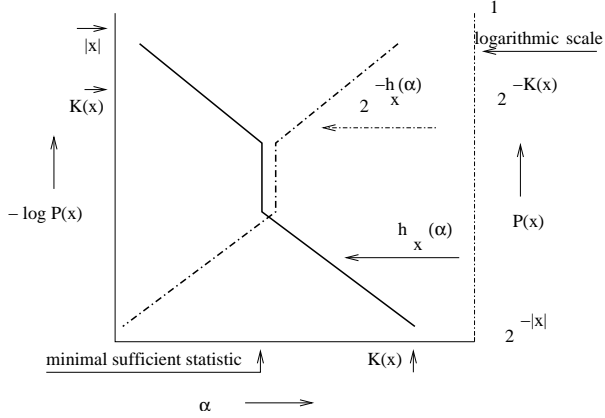}
\end{center}
\caption{Structure function $h_x(i)= \min_f \{ \log 1/ f(x): f(x)>0, \; K(f) \leq
i\}$ with $f$ a computable
probability mass function, with values according to the left
vertical coordinate, and the maximum likelihood estimator $2^{-h_x(i)}=
\max \{f(x): p(x)>0 , \; K(f) \leq i\}$,
with values according to the right-hand side vertical coordinate.}
\label{figure.MLestimator}
\end{figure}

The structure function (and of course the sufficient statistic) use
properties of data strings modeled by finite sets, which amounts to
modeling data by uniform distributions. As already
observed by Kolmogorov himself, it turns out 
that this is no real restriction.
Everything holds also for computable probability mass functions
(probability models), up to additive logarithmic precision.  Another
version of $h_x$ uses probability models $f$ rather than finite set
models. It is defined as $h'_x(R) = \min_{f} \{\log 1/f(x): f(x)>0,
K(f) \leq R\}$.  Since $h'_x(R)$ and $h_x(R)$ are close by
Proposition~\ref{prop.1} below, Theorem~\ref{thm.dresf} and
Corollary~\ref{cor.esf} also apply to $h'_x$ and the distortion-rate
function $D^*(R)$ based on a variation of the 
Shannon-Fano distortion measure
defined by using encodings $Y(x)=f$ with $f$ a computable
probability distribution. In this context, 
the Shannon-Fano distortion measure
is defined by
\begin{equation}\label{eq.sfdf}
d'(x,f)=  \log 1/f(x).
\end{equation}
It remains to show that probability models are essentially the same as
finite set models.  We restrict ourselves to the model class of {\em
  computable probability distributions}. Within the present section,
we assume these are defined on strings of arbitrary length; so they
are represented by mass functions $f: \{0,1\}^* \rightarrow [0,1]$
with $\sum f(x) = 1$ being computable according to
Definition~\ref{def.enum.funct}.  A string $x$ is typical for a
distribution $f$ if the randomness deficiency $ \delta (x \mid f) =
\log 1/ f(x) - K(x \mid f) $ is small. The conditional complexity $K(x
\mid f)$ is defined as follows. Say that a function $A$ approximates
$f$ if $|A(y,\eps)-f(y)|<\eps$ for every $y$ and every positive
rational $\eps$. Then $K(x \mid f)$ is the minimum length of a program
that given every function $A$ approximating $f$ as an oracle prints
$x$.  Similarly, $f$ is $c$-optimal for $x$ if $ K(f) + \log 1/ f(x)
\leq K(x)+c $.  Thus, instead of the data-to-model code length
$\log|S|$ for finite set models, we consider the data-to-model code
length $\log 1/ f(x)$ (the Shannon-Fano code). The value $\log 1/f(x)$
measures also how likely $x$ is under the hypothesis $f$. The
mapping $x\mapsto f_{\min}$ where $f_{\min}$ minimizes $\log 1/f(x)$
over $f$ with $K(f)\le R$ is a \emph{maximum likelihood
  estimator}, see figure~\ref{figure.MLestimator}. Our results thus
imply that that maximum likelihood estimator always returns a
hypothesis with minimum randomness deficiency.
                                                                                
It is easy to show that for every data string $x$
and a contemplated finite set model for it, there
is an almost equivalent computable probability model.
The converse is slightly harder:
for every data string $x$ and a contemplated
computable probability  model for it,
there is a finite set model for $x$ that has no worse complexity,
randomness deficiency, and worst-case data-to-model code for $x$,
up to additive logarithmic precision:

\begin{proposition}\label{prop.1}
(a) For every $x$ and every finite set $S \ni x$ there is
a computable probability
mass function $f$ with $\log 1/f(x) =\log|S|$,
$\delta(x \mid f)=\delta(x \mid S)+O(1)$
and $K(f) = K(S)+ O(1)$.

(b) 
   There are constants $c,C$, such that
    for every string $x$, the following holds:
    For every computable probability
    mass function $f$
    there is a finite set $S \ni x$
    such that $\log |S| <  \log 1/ f(x)+1$, $\delta(x \mid S)
\le \delta(x \mid f)+    2\log K(f)+K(\lfloor \log 1/
f(x)\rfloor)+2\log K(\lfloor \log 1/ f(x)\rfloor)+C$
    and
    $K(S) \leq  K(f) + K(\lfloor \log 1/f(x)\rfloor)+C$.

\end{proposition}
                                                                                
\begin{proof}
(a) Define $f(y)= 1/|S|$ for $y \in S$
and 0 otherwise.

(b) Let $m=\lfloor \log 1/f(x)\rfloor$, that is,
$2^{-m-1}<f(x)\le 2^{-m}$.
Define $S = \{y:  f(y)
> 2^{-m-1}\}$. Then,
$|S|<2^{m+1} \leq 2/f(x)$,
which implies the claimed value for $\log |S|$.
To list $S$ it suffices to compute all consecutive values of $f(y)$ to
sufficient precision
until the combined probabilities exceed $1-2^{-m-1}$.
That is, $K(S) \leq
K(f)+ K(m)+O(1)$.
Finally,
$\delta(x \mid S)=\log|S|-K(x|S^*)< \log 1/f(x)-K(x \mid S^*)+1=
\delta(x \mid f)+K(x \mid f)-K(x \mid S^*)+1\le \delta(x \mid f)+K(S^* \mid f)+O(1)$.
The term $K(S^* \mid f)$ can be upper bounded
as $K(K(S))+K(m)+O(1)\le 2\log K(S)+K(m)+O(1)
\le 2\log (K(f)+K(m))+K(m)+O(1)
\le 2\log K(f)+2\log K(m)+K(m)+O(1)$, which implies the claimed bound for
 $\delta(x \mid S)$.

\end{proof}

How large are the nonconstant additive complexity terms in
Proposition~\ref{prop.1} for strings $x$ of length $n$? In item (b),
we are commonly only interested
in $f$ such that $K(f)\le n+O(\log n)$ and
$\log 1/f(x)\le n+O(1)$.
Indeed, for every $f$ there is $f'$ such that
$K(f')\le \min\{K(f),n\}+O(\log n)$,
$\delta(x \mid f')\le \delta(x \mid f)+O(\log n)$,
$\log 1/f'(x)\le\min\{\log1/ f(x),n\}+1$.
Such $f'$ is defined as follows: If
$K(f)>n$ then $f'(x)=1$ and $f'(y)=0$ for every $y\ne x$;
otherwise $f'=(f+U_n)/2$ where $U_n$ stands for
the uniform distribution
on $\{0,1\}^n$.
Then the additive terms in item (b) are $O(\log n)$.

\subsection{Expected Structure Function Equals Distortion--Rate Function}
\label{sec:esf}
In this section we treat the general relation between the expected
value of $h_x(R)$, the expectation taken on a distribution
$f(x)=P(X=x)$ of the random variable $X$ having outcome $x$, and
$D^*(R)$. This involves the development of a rate-distortion theory
for individual sequences and arbitrary computable distortion measures.
Following \cite{VereshchaginV04}, we outline such a theory in
Sections~\ref{sec:spheres}-~\ref{sec:ssrev}. Based on this theory, we
present in Section~\ref{sec:esfb} a general theorem
(Theorem~\ref{thm.dresf}) relating Shannon's $D^*(R)$ to the expected
value of $h_x(R)$, for arbitrary random sources and computable
distortion measures. This generalizes Example~\ref{ex.rd=str} above,
where we analyzed the case of the distortion function
\begin{equation}\label{eq.lcfs1}
d(x,Y(x))
= \log |Y(x)|,
\end{equation}
where $Y(x)$ is an $x$-containing finite set,
for the uniform distribution. Below we first extend this example to
arbitrary generating distributions, keeping the distortion function
still fixed to (\ref{eq.lcfs1}. This will prepare us for the general
development in Sections~\ref{sec:spheres}--\ref{sec:ssrev}
\begin{example}
In Example~\ref{ex.rd=str}
it transpired that
the distortion-rate function is the expected structure function,
the expectation taken over the distribution on the $x$'s.
If, instead of using the uniform
distribution on $\{0,1\}^n$ we use an arbitrary distribution $f(x)$,
it is not difficult to compute the rate-distortion
function $R^*(D)= H(X)- \sup_{Y:d(X,Y) \leq D} H(X|Y)$ where 
$Y$ is a random vaiable  with outcomes that are finite sets. Since $d$
is a special type of Shannon-Fano distortion, with 
$d(x,y) = P(X=x | Y=y) = \log |y|$ if $x \in y$, and 0 otherwise,   
we have already met
$D^*(R)$ for the distortion measure \eqref{eq.lcfs1} in another guise.
By the conclusion of Example~\ref{ex:berb}, generalized to the random
variable $X$ having outcomes in $\{0,1\}^n$, and $R$ being a rate
in between 0 and $n$, we know that
\begin{equation}\label{eq.DRE}
D^*(R) = H(X)-R.
\end{equation}
\end{example}                                                              In the particular case analyzed above, the code word for a source word
is a finite set containing the source word, and the distortion is the
log-cardinality of the finite set. Considering the set of source words
of length $n$, the distortion-rate function is the diagonal line from
$n$ to $n$.  The structure functions of the individual data $x$ of
length $n$, on the other hand, always start at $n$, decrease at a
slope of at least -1 until they hit the diagonal from $K(x)$ to
$K(x)$, which they must do, and follow the diagonal henceforth. Above
we proved that the average of the structure function is simply the
straight line, the diagonal, between $n$ and $n$. This is the case,
since the strings $x$ with $K(x) \geq n$ are the overwhelming
majority. All of them have a minimal sufficient statistic (the point
where the structure function hits the diagonal from $K(x)$ to $K(x)$.
This point has complexity at most $K(n)$. The structure function for
all these $x$'s follows the diagonal from about $n$ to $n$, giving
overall an expectation of the structure function close to this
diagonal, that is, the probabilistic distortion-rate function for this
code and distortion measure.
\subsubsection{Distortion Spheres}
\label{sec:spheres}
Modeling the data can be viewed as
encoding the data by a model: the data are source words
to be coded, and models are
code words for the data. As before, the set of possible data is
${\cal X} = \{0,1\}^n$. Let ${\cal R}^+$ denote the set
of non-negative real numbers.
For every model class ${\cal Y}$  (particular set of code words)
we choose an appropriate
recursive function
$d: {\cal X} \times {\cal Y} \rightarrow {\cal R}^+$ defining
the {\em distortion} $d(x,y)$ between data $x \in {\cal X}$ and model $y \in {\cal Y}$.
\begin{remark}[Lossy Compression]
\label{rem:lossy}
\rm
The choice of distortion
function is a selection of which aspects of the data are relevant,
or meaningful, and
which aspects are irrelevant (noise).
We can think of the distortion-rate function as measuring how far the model at
each bit-rate
falls short in representing the data. Distortion-rate theory
underpins the practice of lossy compression.
For example, lossy compression of a sound file gives as ``model''
the compressed file where, among others, the very high and
very low inaudible frequencies have been suppressed. Thus,
the rate-distortion function will penalize the deletion of the inaudible
frequencies but lightly because they are not relevant for the auditory
experience.
                                                                                
But in the traditional distortion-rate approach, we average twice:
once because we consider
a sequence of outcomes of $m$ instantiations of the same random variable,
and once because we take the expectation
over the sequences. Essentially, the results deal with typical ``random'' data
of certain simple distributions. This assumes that the data to a certain extent
satisfy the behavior of repeated outcomes of a random source. 
Kolmogorov \cite{Ko65}:
\begin{quote}
The probabilistic approach is natural in the theory of information
transmission over communication channels carrying ``bulk'' information 
consisting of a large number of unrelated or
weakly related messages obeying
definite probabilistic laws. In this type of problem there is
a harmless and (in applied work) deep-rooted tendency to mix up
probabilities and frequencies within sufficiently long time sequence
(which is rigorously satisfied if it is assumed that ``mixing''
is sufficiently rapid). In practice,
for example, it can be assumed that finding the ``entropy''
of a flow of congratulatory telegrams and the channel ``capacity'' required
for timely and undistorted transmission is validly represented by a
probabilistic treatment even with the usual substitution of empirical
frequencies for probabilities. If something goes wrong here,
the problem lies with the vagueness of our ideas of the relationship
between mathematical probabilities and real random events in general.

But what real meaning is there, for example, in asking how much
information is contained in ``War and Peace''?
Is it reasonable to include the novel in the set of ``possible novels'',
or even to postulate some probability distribution for this set? 
Or, on the other hand, must we assume that the individual scenes in 
this book form a random sequence with ``stocahstic relations'' that damp out 
quite rapidly over a distance of several pages?
\end{quote}
Currently, individual data arising in practice are submitted to
analysis, for example sound or video files, where the assumption that
they either consist of a large number of weakly related messages, or
being an element of a set of possible messages that is susceptible to
analysis, is clearly wrong.  It is precisely the global related aspects
of the data which we want to preserve under lossy compression.  The
rich versatility of the structure functions, that is, many different
distortion-rate functions for different individual data, is all but
obliterated in the averaging that goes on in the traditional
distortion-rate function.  In the structure function approach one
focuses entirely on the stochastic properties of one data item.
\end{remark}
Below we follow \cite{VereshchaginV04}, where we developed a
rate-distortion theory for individual data for general computable
distortion measures, with as specific examples the `Kolmogorov'
distortion below, but also Hamming distortion and Euclidean
distortion. This individual rate-distortion theory is summarized in
Sections~\ref{sec:rdrev} and~\ref{sec:ssrev}. In
Section~\ref{sec:esfb}, Theorem~\ref{thm.dresf}.
we connect this indivual rate-distortion theory to Shannon's. We
emphasize that the typical data items of i.i.d. distributed simple
random variables, or simple ergodic stationary sources, which are the
subject of Theorem~\ref{thm.dresf}, are generally unrelated to the
higly globally structured data we want to analyze using our new
rate-distortion theory for individual data. From the prespective of
lossy compression, the typical data have the characteristics of random
noise, and there is no significant ``meaning'' to be preserved under
the lossy compression.  Rather, Theorem~\ref{thm.dresf} serves as a
`sanity check' showing that in the special, simple case of repetitive
probabilistic data, the new theory behaves essentially like Shannon's
probabilistic rate-distortion theory.
\begin{example}\label{ex.11}
\rm
Let us look at various model classes and distortion measures:
                                                                                
(i) The set of models are the finite sets of finite binary strings.
Let $S \subseteq \{0,1\}^*$ and $|S| < \infty$.
We define $d(x,S) = \log |S|$ if $x \in S$, and $\infty$ otherwise.
                                                                                
(ii) The set of models are the computable probability density functions $f$
mapping $\{0,1\}^*$ to $[0,1]$.
We define $d(x,S) =  \log 1/f(x)$ if $f(x) > 0$, and $\infty$ otherwise.
                                                                                
(iii) The set of models are the total recursive functions  $f$
mapping $\{0,1\}^*$ to ${\cal N}$.
We define $d(x,f) = \min \{ l(d): f(d)=x\}$, and $\infty$ if
no such $d$ exists.

All of these model classes and accompanying
distortions \cite{VV02}, together with the ``communication exchange'' models
in \cite{BKVV03}, are loosely called {\em Kolmogorov} models
and distortion, since the graphs of their structure functions (individual
distortion-rate functions) are all within a strip---of width
logarithmic in the binary length of the data---of one another.
\end{example}
If ${\cal Y}$ is a model class, then
we consider {\em distortion spheres} of given
radius $r$ centered on $y \in {\cal Y}$:
\[
B_y(r)= \{x: d(x,y) = r\}.
\]
This way, every model class and distortion measure can be treated
similarly to the canonical finite set case, which, however, is
especially simple in that the radius not variable.
That is, there is only one distortion sphere  centered on a given finite set,
namely the one with radius equal to the log-cardinality of that finite set.
In fact, that distortion sphere equals the finite set on which it is
centered.
                                                                                
\subsubsection{Randomness Deficiency---Revisited}
\label{sec:rdrev}
Let ${\cal Y}$ be a model class and $d$ a distortion measure.
Since in our definition the distortion is recursive,
given a model $y \in {\cal Y}$ and diameter $r$,
the elements in the distortion sphere
of diameter $r$ can be recursively enumerated from the distortion function.
Giving the index of any element $x$ in that enumeration we can find the
element. Hence, $K(x|y,r) \lea \log |B_y(r)|$. On the other hand,
the vast majority of elements $x$ in the distortion sphere have
complexity $K(x|y,r) \gea  \log |B_y(r)|$ since, for every constant $c$,
 there are only
$2^{\log |B_y(r)|-c} - 1$ binary programs of length $ < \log |B_y(r)|-c$
available, and there are $|B_y(r)|$ elements to be described.
We can now reason as in the similar case of finite set models.
With data $x$ and $r=d(x,y)$,
if $K(x|y,d(x,y))
\gea \log |B_y(d(x,y))|$, then $x$ belongs to every large majority of elements
(has the property represented by that majority)
of the distortion sphere $B_y(d(x,y))$, 
provided that property is simple in the
sense of having a description of low Kolmogorov complexity.
\begin{definition}
\rm
The {\em randomness
deficiency} of $x$ with respect to model $y$ under distortion $d$
is defined as
\[
\delta (x \mid y) = \log |B_y (d(x,y))| - K(x|y,d(x,y)).
\]
Data $x$ is {\em typical} for model $y \in {\cal Y}$ (and that model
``typical'' or ``best fitting'' for $x$) if
\begin{equation}\label{eq.typical}
\delta (x \mid y)  \eqa 0.
\end{equation}
\end{definition}
If $x$ is typical for a model $y$, then the shortest way to effectively
describe $x$, given $y$, takes about as many bits as the
descriptions of the great
majority of elements in
a recursive enumeration of the distortion sphere.
So there are no special simple properties that distinguish $x$
from the great majority of elements
in the distortion sphere: they are all typical or random elements
in the distortion sphere (that is, with respect to the contemplated model).
\begin{example}
\rm
Continuing Example~\ref{ex.11} by applying \eqref{eq.typical}
to different model classes:
                                                                                
(i) {\em Finite sets:}
 For finite set models $S$, clearly $K(x|S) \lea \log |S|$.
Together with \eqref{eq.typical} we have that $x$ is typical for $S$,
and $S$ best fits $x$, if the randomness deficiency
according to \eqref{eq:randomness-deficiency} satisfies
$\delta(x|S) \eqa 0$.
                                                                                
(ii) {\em Computable probability density functions:}
Instead of the data-to-model code length $\log|S|$ for
finite set models, we consider the data-to-model code length
$\log 1/f(x)$ (the Shannon-Fano code). The value $\log 1/f(x)$
measures how likely $x$ is under the hypothesis $f$.
 For probability models $f$,
define the conditional complexity
$K(x \mid f, \lceil  \log 1/f(x) \rceil )$ as follows.
Say that a function
$A$ approximates $f$ if $|A(x,\eps)-f(x)|<\eps$
for every $x$ and every positive rational
$\eps$. Then $K(x \mid f , \lceil \log 1/f(x) \rceil)$ is defined as
the minimum length
of a program that, given $\lceil \log 1/f(x) \rceil$
and any function $A$ approximating $f$
as an oracle, prints $x$.
                                                                                
Clearly
$K(x|f, \lceil \log 1/f(x) \rceil ) \lea \log 1/f(x)$.
Together with \eqref{eq.typical}, we have that $x$ is typical for $f$,
and $f$ best fits $x$, if
$K(x|f, \lceil \log 1/f(x) \rceil) \gea \log |\{z:  \log 1/f(z) \leq
 \log 1/f(x)\}|$. The right-hand side set condition is the same
as $f(z) \geq f(x)$, and there can be only $\leq 1/f(x)$ such $z$,
since otherwise the total probability exceeds 1. Therefore,
the requirement, and hence typicality,
is implied by $K(x|f, \lceil \log 1/f(x) \rceil ) \gea \log 1/f(x)$.
Define  the randomness
deficiency by
$
\delta (x \mid f) =   \log 1/f(x) - K(x \mid f, \lceil \log 1/f(x) \rceil).
$
Altogether, a string $x$ is {\em typical for a distribution} $f$,
or $f$ is the {\em best fitting model} for $x$,
if $\delta (x \mid f) \eqa 0$.
if $\delta (x \mid f) \eqa 0$.
                                                                                
(iii) {\em Total Recursive Functions:}
In place of $\log|S|$ for finite set models
we consider the data-to-model code length (actually, the distortion
$d(x,f)$ above)
$$\len xf=\min\{l(d):f(d)=x\}.$$
Define the conditional complexity
$K(x \mid f, \len xf )$ as
the minimum length
of a program that, given $\len xf$ and an oracle for  $f$,
prints $x$.
                                                                                
Clearly, $K(x|f, \len xf ) \lea \len xf$.
Together with \eqref{eq.typical}, we have that $x$ is typical for $f$,
and $f$ best fits $x$, if $K(x|f, \len xf ) \gea \log \{z: \len zf
 \leq  \len xf \}$. There are at most $(2^{\len xf +1} - 1)$-
many $z$ satisfying the set condition since
$\len zf \in \{0,1\}^*$.  Therefore,
the requirement, and hence typicality,
is implied by $K(x|f, \len xf ) \gea \len xf$.
Define  the  randomness
deficiency by
$
\delta (x \mid f) =   \len xf - K(x \mid f, \len xf ).
$
Altogether, a string $x$ is {\em typical for a total recursive
function} $f$, and $f$ is the {\em best fitting recursive function model}
for $x$
if $\delta (x \mid f) \eqa 0$, or written differently,
\begin{equation}\label{eq.typp}
K(x|f, \len xf ) \eqa \len xf.
\end{equation}
Note that since $\len xf$ is given as conditional information,
with $\len xf = l(d)$ and $f(d)=x$, the quantity $K(x|f, \len xf )$
represents the number of bits in a shortest
{\em self-delimiting} description of $d$.
\end{example}

\begin{remark}
\rm
We required $\len xf$ in the conditional in \eqref{eq.typp}.
This is the information about
the radius of the distortion sphere centered on the model concerned.
Note that in the canonical finite set model case, as treated
in \cite{Ko74,GTV01,VV02}, every model has a fixed radius which
is explicitly provided by the model itself. But in the
more general model
classes of computable probability density functions, or
total recursive functions, models can have a variable radius.
There are subclasses of the more general models that
have fixed radiuses (like the finite set models).
                                                                                
(i) In the computable probability density functions one can think of the
probabilities with a finite support, for example $f_n (x) = 1/2^n$
for $l(x)=n$, and $f(x)=0$ otherwise.
                                                                                
(ii) In the total recursive function case one can similarly think
of functions with finite support, for example $f_n (x) = \sum_{i=1}^n x_i$
for $x=x_1 \ldots x_n$, and $f_n(x)=0$ for $l(x) \neq n$.
                                                                                
The incorporation of the radius in the model will increase the
complexity of the model, and hence of the minimal sufficient statistic
below.
\end{remark}
                                                                                
\subsubsection{Sufficient Statistic---Revisited}
\label{sec:ssrev}
As with the probabilistic sufficient statistic
(Section~\ref{sec:probstat}), a statistic is a function mapping the
data to an element (model) in the contemplated model class. With some
sloppiness of terminology we often call the function value (the model)
also a statistic of the data.  A statistic is called sufficient if the
two-part description of the data by way of the model and the
data-to-model code is as concise as the shortest one-part description
of $x$.  Consider a model class ${\cal Y}$.
\begin{definition}
A model $y \in {\cal Y}$ is a {\em sufficient statistic} for $x$ if
\begin{equation}\label{eq.ssm}
K(y, d(x,y))+ \log |B_y(d(x,y))| \eqa K(x).
\end{equation}
\end{definition}
                                                                                
\begin{lemma}\label{lem.V2}
If $y$ is a sufficient statistic for $x$, then
$K(x \mid y, d(x,y) \eqa  \log |B_y(d(x,y))|$, that is,
$x$ is typical for $y$.
\end{lemma}
\begin{proof}
We can rewrite
$K(x) \lea K(x,y,d(x,y)) \lea K(y,d(x,y))+K(x|y,d(x,y))
\lea K(y, d(x,y))+ \log |B_y(d(x,y))| \eqa K(x)$.
The first three inequalities are straightforward and
the last equality is by the assumption of sufficiency.
Altogether, the first sum equals the second sum, which implies the lemma.
\end{proof}

Thus, if $y$ is a sufficient statistic for $x$, then $x$ is a typical element
for $y$, and $y$ is the best fitting model for $x$.
Note that the converse implication,  ``typicality'' implies
``sufficiency,'' is not valid. Sufficiency is a special type
of typicality, where the model does not add significant
information to the data, since the preceding proof shows
$K(x) \eqa K(x,y,d(x,y))$. Using the symmetry of information \eqref{eq.soi}
this shows that
\begin{equation}\label{eq.pcondx}
K(y,d(x,y) \mid x ) \eqa K(y \mid x) \eqa 0.
\end{equation}
This means that:
                                                                                
(i) A sufficient statistic  $y$ is determined by the data in the sense
that we need only an $O(1)$-bit program, possibly depending on
the data itself, to compute the model
from the data.
                                                                                
(ii) For each model class and distortion there is a universal constant $c$
such that for every data item $x$ there are at most $c$ sufficient
statistics.
                                                                                
\begin{example}
\rm
{\em Finite sets:}
For the model class of finite sets, a set $S$ is a sufficient statistic
for data $x$ if
\[
K(S)+ \log |S| \eqa K(x).
\]
                                                                                
{\em Computable probability density functions:}
For the model class of computable probability density functions,
a function $f$ is a sufficient statistic
for data $x$ if
\[
K(f)  + \log 1/f(x) \eqa K(x).
\]
For the model class of
{\em total recursive functions}, a function $f$ is a
{\em sufficient statistic} for data $x$
if
\begin{equation}\label{eq.ss}
K(x) \eqa K(f)  + \len xf .
\end{equation}
Following the above discussion, the meaningful information in $x$
is represented by $f$ (the model) in $K(f)$ bits, and the
meaningless information in $x$ is represented by $d$ (the noise in
the data) with $f(d)=x$ in $l(d) = \len xf$ bits. Note that
$l(d) \eqa K(d) \eqa K(d|f^*)$,
since the two-part
code $(f^*,d)$ for $x$
cannot be shorter than the shortest one-part code of $K(x)$ bits,
and therefore the $d$-part must already be maximally compressed.
By Lemma~\ref{lem.V2},  $\len xf \eqa   K(x \mid f^* , \len xf)$,
$x$ is typical for $f$,
and hence $K(x) \eqa K(f)  + K(x \mid f^* , \len xf)$.
\end{example}

\subsubsection{Expected Structure Function}
\label{sec:esfb}
We treat the relation between the expected value
of $h_x(R)$, the expectation taken on a
distribution $f(x)=P(X=x)$ of the random variable $X$ having outcome $x$,
and $D^*(R)$, for arbitrary random sources provided the probability mass
function $f(x)$ is recursive.

\begin{theorem}\label{thm.dresf}
Let $d$ be a recursive distortion
measure.
Given $m$ repetitions of a random variable $X$ with outcomes
$x \in {\cal X}$ (typically, ${\cal X}= \{0,1\}^n$) 
with probability $f(x)$, where $f$ is a total
recursive function, we have
$$
{\bf E} \frac{1}{m} h_{\overline{x}} (mR+K(f,d,m,R)+O(\log n))
\leq D^*_m(R)
\leq {\bf E} \frac{1}{m} h_{\overline{x}} (mR),
$$
the expectations are taken over $\overline{x}
= x_1 \ldots x_m$ where $x_i$ is the outcome of the $i$th repetition
of $X$.
\end{theorem}
\begin{proof}
As before, let $X_1, \ldots, X_m$ be $m$ independent identically
distributed random variables on outcome space ${\cal X}$.
Let ${\cal Y}$ be a set of code words.
We want to find a sequence of functions $Y_1, \ldots , Y_m:{\cal X}
\rightarrow {\cal Y}$ so that the message $(Y_1(x_1), \ldots,
Y_m (x_m)) \in {\cal Y}^m$ gives as much expected
information about the sequence of outcomes $(X_1=x_1,
\ldots, X_m=x_m)$ as is possible, under the constraint that the message
takes at most $R \cdot m$ bits (so that $R$ bits are allowed on
average per outcome of $X_i$).
Instead of $Y_1, \ldots , Y_m$ above write
$\overline{Y}: {\cal X}^m \rightarrow  {\cal Y}^m$.
Denote the cardinality of the range of $\overline{Y}$
by $\rho (\overline{Y})= | \{\overline{Y}(\overline{x}): 
\overline{x} \in {\cal X}^m\}|$.
Consider distortion spheres
\begin{equation}\label{eq.lcfs}
B_{\overline{y}}(d) = \{\overline{x}: d(\overline{x},\overline{y}) = d \},
\end{equation}
with $\overline{x} = x_1 \ldots x_m \in {\cal X}^m$
and $\overline{y} \in {\cal Y}^m$.

{\em Left Inequality:}
Keeping the earlier notation, for $m$ i.i.d.
random variables $X_1, \ldots ,X_m$, and extending $f$ to
the $m$-fold Cartesian product of $\{0,1\}^n$, we obtain
$D_m^*(R) = \frac{1}{m} \min_{ \overline{Y}: \rho (\overline{Y}) \leq 2^{mR}}
\sum_{\overline{x}}f(\overline{x})
d(\overline{x}, \overline{Y} (\overline{x}))$.
By definition of $D_m^*(R)$ it equals the following expression in terms
of a minimal canonical covering of $\{0,1\}^{nm}$ by
disjoint nonempty spheres $B'_{\overline{y}_i}(d_i)$
($1 \leq i \leq k$) obtained from the possibly overlapping
distortion spheres $B_{\overline{y}_i}(d_i)$ as follows.
Every element $\overline{x}$ in the overlap between two or more spheres
is assigned to the sphere with the smallest radius and removed
from the other spheres. If there is more than
one sphere of smallest radius, then
we take the sphere of least index in the canonical covering.
Empty $B'$-spheres are removed from the $B'$-covering.
If $S \subseteq \{0,1\}^{nm}$, then $f(S)$ denotes  $\sum_{x \in S} f(x)$. Now,
we can rewrite
\begin{equation}\label{eq.distpart}
D^*_m(R) =
\min_{\overline{y}_1, \ldots , \overline{y}_k; d_1, \ldots , d_k; k \leq 2^{mR}}
 \frac{1}{m}
\sum_{i=1}^k f(B'_{\overline{y}_i}(d_i)) d_i.
\end{equation}
In the structure function setting we consider some individual
 data $\overline{x}$ residing
in one of the covering spheres.
Given $m,n,R$ and a program to compute $f$ and $d$, we can compute the
covering spheres centers $\overline{y}_1, \ldots, \overline{y}_k$,
and radiuses $d_1, \ldots , d_k$,  
and hence the $B'$-sphere canonical covering. In this
covering we can identify every pair $(\overline{y}_i, d_i)$ by
its index $i \leq 2^{mR}$. Therefore,
$K(\overline{y}_i, d_i) \leq mR + K(f,d,m,R)+O(\log n)$ ($1 \leq i \leq k)$.
For $\overline{x} \in B'_{\overline{y}_{i}}(d_i)$
we have $h_{\overline{x}}(mR + K(f,d,m,R)+O(\log n)) \leq d_i$.
Therefore,
${\bf E} \frac{1}{m}h_{\overline{x}}(mR + K(f,d,m,R)+O(\log n)) \leq  D^*_m(R)$,
the expectation taken over
$f(\overline{x})$ for $\overline{x} \in \{0,1\}^{mn}$.

{\em Right Inequality:}
Consider a covering of $\{0,1\}^{nm}$
by the  (possibly overlapping)
distortion spheres $B_{\overline{y}_i}(d_i)$
satisfying $K(B_{\overline{y}_i}(d_i) | mR) < mR-c$, with $c$ an
appropriate constant choosen so that the remainder of the argument
goes through.   
If there are more than one spheres with different (center, radius)-pairs
representing the same subset of  $\{0,1\}^{nm}$, then
we eliminate all of them except the one with the smallest radius.
If there are more than one such spheres, then we only keep the one
with the lexicographically least center. From this covering we obtain 
a canonical covering 
by nonempty disjoint spheres $B'_{\overline{y}_i} (d_i)$ 
similar to that in the previous paragraph, 
($1 \leq i \leq k$). 

For every $\overline{x} \in \{0,1\}^{nm}$
there is a unique 
sphere $B'_{\overline{y}_i}(d_i) \ni \overline{x}$ ($1 \leq i \leq k$). 
Choose the constant $c$ above so that  
$K(B'_{\overline{y}_i}(d_i) |mR )  < mR$. Then,
$k \leq 2^{mR}$. 
Moreover, by construction, if $B'_{\overline{y}_i} (d_i)$
is the sphere containing $\overline{x}$, then 
$h_{\overline{x}} (mR)= d_i$.
Define functions $\gamma: \{0,1\}^{nm} \rightarrow {\cal Y}^m$, 
$\delta: \{0,1\}^{nm} \rightarrow {\cal R}^+$ defined by
$\gamma(\overline{x}) = \overline{y}_i$ and $\delta (\overline{x}) = d_i$
for $\overline{x}$ in the sphere $B'_{\overline{y}_i}(d_i)$.
Then,
\begin{equation}\label{eq.lb2}
 {\bf E} \frac{1}{m}  h_{\overline{x}} (mR) =
\frac{1}{m} \sum_{\overline{x} \in \{0,1\}^{mn}}
f(\overline{x}) d(\overline{x}, \gamma(\overline{x}))
= \frac{1}{m} \sum_{\overline{y}_1, \ldots , \overline{y}_k; d_1, \ldots , d_k} 
f(B'_{\overline{y}_i}(d_i)) d_i .
\end{equation}
The distortion  $D^*_m (R)$ achieves the minimum of the expression in
right-hand side of \eqref{eq.distpart}.
Since $K(B'_{\gamma( \overline{x})} (\delta(\overline{x}))|mR) < mR$, 
the cover in the right-hand side of \eqref{eq.lb2}
is a possible partition satisfying the expression being
minimized in the right-hand side of
\eqref{eq.distpart}, and hence majorizes the minumum $D^*_m(R)$. Therefore,
${\bf E} \frac{1}{m} h_{\overline{x}} (mR) \geq D^*_m(R)$.
\end{proof}

\begin{remark}
\rm
A sphere 
is a subset of $\{0,1\}^{nm}$. The same subset may correspond
to more than one spheres with different centers and radiuses:
 $B_{\overline{y_0}}(d_0) = B_{\overline{y_1}}(d_1)$ with 
$(y_0,d_0) \neq (y_1,d_1)$. 
Hence, $K(B_{\overline{y}} (d))
\leq K(\overline{y},d)) + O(1)$, but possibly
$K(\overline{y},d)) > K(B_{\overline{y}} (d))+O(1)$.
However, in the proof we constructed the ordered sequence of $B'$
spheres such that every sphere uniquely corresponds to a 
(center, radius)-pair. Therefore, $K(B'_{\overline{y}_i}(d_i)|mR)
\eqa K(\overline{y}_i, d_i | mR)$.
\end{remark}

\begin{corollary}\label{cor.esf}
It follows from the above theorem that, for 
a recursive distortion function $d$:
                                                                                                                                                      (i) $
{\bf E} h_{x} (R+K(f,d,R)+O(\log n))
\leq D^*_1 (R)
\leq {\bf E} h_x (R)
$,
for outcomes of a single repetition of random variable $X =x$
with $x \in \{0,1\}^n$,
the expectation taken over $f(x)=P(X =x)$; and
                                                                                
(ii) $\lim_{m \rightarrow \infty} {\bf E} \frac{1}{m} h_{\overline{x}} (mR)
= D^*(R)$
for outcomes $\overline{x} = x_1 \ldots x_m$
of i.i.d. random variables $X_i =x_i$ with $x_i \in \{0,1\}^n$ for
$1 \leq i \leq m$,
the expectation taken over $f(\overline{x})=P(X_i=x_i, i=1, \ldots, m)$
(the extension of
$f$ to $m$ repetitions of $X$).
\end{corollary}
                                                                                
This is the sense in which the expected value of the structure function
is asymptotically equal to the value of the distortion-rate function,
for arbitrary computable distortion measures.
In the structure function approach we dealt with only two
model classes, finite sets and computable probability density functions,
and the associated quantities to be minimized, the log-cardinality
and the negative log-probability, respectively. Translated into
the distortion-rate setting, the models are code words
and the minimalizable quantities are distortion measures.
 In \cite{VV02}
we also investigate the model class of total recursive functions,
and in
\cite{BKVV03} the model class of communication protocols. The associated
quantities to be minimized are then function arguments and communicated
bits, respectively. All these models are equivalent up to logarithmic
precision in argument and value of the corresponding structure functions,
and hence their expectations are asymptotic to the distortion-rate 
functions of the related code-word set and distortion measure.

\section{Conclusion}
We have compared Shannon's and Kolmogorov's theories of information,
highlighting the various similarities and differences. Some of this
material can also be found in \cite{CT91}, the standard reference for
Shannon information theory, as well as \cite{LiVi97}, the standard
reference for Kolmogorov complexity theory. These books predate much
of the recent material on the Kolmogorov theory discussed in the
present paper, such as \cite{HRSV00} (Section~\ref{sec:algmi}),
\cite{Le02} (Section~\ref{sect:minialg}), \cite{GTV01}
(Section~\ref{sec:algsuf}), \cite{VV02, VereshchaginV04}
(Section~\ref{sec:structure}). The material in Sections~\ref{sec:relpa}
and \ref{sec:esf}
has not been published before. The present paper summarizes these
recent contributions and systematically compares
them to the corresponding notions in Shannon's theory.
  
  \paragraph{Related Developments:} There are two major practical theories 
  which have their roots in both Shannon's and Kolmogorov's notions of
  information: first, {\em universal coding}, briefly introduced in
  Appendix~\ref{sec:universal} below, is a remarkably successful theory for
  practical lossless data compression.  Second, Rissanen's {\em
    Minimum Description Length (MDL) Principle\/}
  \cite{Ri89,Grunwald04} is a theory of inductive inference that
  is both practical and successful. Note that direct practical
  application of Shannon's theory is hampered by the typically
  untenable assumption of a true and known distribution generating the
  data. Direct application of Kolmogorov's theory is hampered by the
  noncomputability of Kolmogorov complexity and the strictly asymptotic
  nature of the results.  Both universal coding (of the individual
  sequence type, Appendix~\ref{sec:universal}) and MDL seek to
  overcome both problems by restricting the description methods used
  to those corresponding to a set of probabilistic predictors (thus
  making encodings and their lengths computable and nonasymptotic);
  yet when applying these predictors, the assumption that any one of
  them generates the data is never actually made. Interestingly, while
  in its current form MDL bases inference on universal codes, in
  recent work Rissanen and co-workers have sought to found the
  principle on a restricted form of the algorithmic sufficient
  statistic and Kolmogorov's structure function as discussed in
  Section~\ref{sec:structure} \cite{RissanenT04}.
  
  By looking at general types of prediction errors, of which
  codelengths are merely a special case, one achieves a generalization
  of the Kolmogorov theory that goes by the name of {\em predictive
    complexity}, pioneered by Vovk, Vyugin, Kalnishkan and others\footnote{See {\tt www.vovk.net} for an overview.} \cite{Vovk01}.  Finally, the
  notions of `randomness deficiency' and `typical set' that are
  central to the algorithmic sufficient statistic
  (Section~\ref{sec:algsuf}) are intimately related to 
 the celebrated Martin-L\"of-Kolmogorov theory of {\em randomness in
    individual sequences}, an overview of which is given in
  \cite{LiVi97}.
\appendix
\section{Appendix: Universal Codes}
\label{sec:universal}
Shannon's and Kolmogorov's idea are not directly applicable to
most actual data compression problems. Shannon's theory is hampered 
by the typically
  untenable assumption of a true and known distribution generating the
  data. Kolmogorov's theory is hampered by the
  noncomputability of Kolmogorov complexity and the strictly asymptotic
  nature of the results. Yet there is
a middle ground that is feasible:  {\em
universal codes\/} that may be viewed as both an
generalized version of Shannon's, and a feasible
approximation to Kolmogorov's theory. In introducing
the notion of universal coding Kolmogorov says \cite{Ko65}:
\begin{quote}
``A universal coding method that permits the transmission of
any sufficiently long message [of length $n$] in an alphabet of $s$ letters
with no more $nh$ [$h$ is the empirical entropy] binary digits is
not necessarily excessively complex; in particular, it is not
essential to begin by determining the frequencies $p_r$ for the entire
message.'' 
\end{quote}

Below we repeatedly use the coding concepts introduced in
Section~\ref{sec:coding}. 
Suppose we are given a recursive enumeration 
of prefix codes $D_1, D_2, \ldots$. Let $L_1, L_2, \ldots$ be the
length functions associated with these codes. That is, $L_i(x) = \min_y
\{ l(y) : D_i(y) = x \}$; if there exists no $y$ with $D_i(y) = x$,
then $L_i(y) = \infty$. We may encode $x$ by first
encoding a natural number $k$ using the standard prefix code
for the natural numbers. 
We then encode $x$ itself using the code $D_k$. This leads to a
so-called {\em two-part code\/} $\tilde{D}$ 
with lengths $\tilde{L}$. By construction, this code is prefix and its lengths satisfy
\begin{equation}
\tilde{L}(x) := \min_{k \in {\cal N}} \  \Lint(k) + L_k(x),
\end{equation}
Let ${\bf x}$ be an infinite binary sequence and let $x_{[1:n]} \in
\{0,1\}^n$ be the initial $n$-bit segment of this sequence.
Since  $L_{\cal N}(k) = O (\log k)$,
we have for all $k$, all $n$:
$$
\tilde{L}(x_{[1:n]}) \leq  L_k(x_{[1:n]}) + O(\log k).
$$
Recall that for
each fixed $L_k$, the fraction of sequences of length $n$ that can be
compressed by more than $m$ bits is less than $2^{-m}$. Thus, 
typically, the codes $L_k$ and the strings $x_{[1:n]}$ will be such
that $L_k(x_{[1:n]})$ grows {\em linearly\/} with $n$. 
This implies that for every ${\bf x}$, 
the newly constructed $\tilde{L}$ is `almost as good'
as whatever code $D_k$ in the list is best for that particular ${\bf x}$: the
difference in code lengths is bounded by a constant depending on $k$ but not on
$n$. In particular, for each
infinite sequence ${\bf x}$, for each fixed $k$,
\begin{equation}
\label{eq:universal}
\lim_{n \rightarrow \infty}
\frac{\tilde{L}(x_{[1:n]})}{L_k(x_{[1:n]})} \leq 1.
\end{equation}
A code satisfying (\ref{eq:universal}) is called a {\em universal
  code\/} relative to the {\em comparison class\/} of codes 
$\{ D_1, D_2, \ldots \}$. 
It is `universal' in the sense that it compresses every
sequence essentially as well as the $D_k$ that compresses that particular
sequence the most.
In general, there exist many types of codes that 
are universal: the 2-part universal code defined above is just
one means of achieving (\ref{eq:universal}). 

\paragraph{Universal codes and Kolmogorov:}
%
In most practically interesting cases we may assume that  
for all $k$, the decoding function $D_k$ is computable,
i.e. there exists a prefix Turing machine which 
for all $y \in \{0,1\}^*$, when input $y'$ (the prefix-free version of
$y$), outputs $D_k(y)$ and then
halts. Since such a program has finite length, we must have for all $k$,
$$
l(E^*(x_{[1:n]})) = K(x_{[1:n]}) \leq^+ L_k(x_{[1:n]})
$$
where $E^*$ is the encoding function defined in Section~\ref{sec:kolmogorov}, 
with $l(E^*(x))
= K(x)$. Comparing with (\ref{eq:universal}) shows that
the code $D^*$  with encoding function $E^*$ is a universal code relative to $D_1, D_2,
\ldots$. Thus, we see that the Kolmogorov complexity $K$ is just the length function
of the universal code $D^*$. Note that $D^*$ is an example of a universal
code that is not (explicitly) two-part. 
\begin{example}
\label{ex:universal}
\rm Let us create a universal two-part code that allows us to significantly
compress all  binary strings with frequency of 0's deviating significantly
from $\frac{1}{2}$. For $n_0 < n_1$, let $D_{\langle n,n_0 \rangle }$ be the code that assigns
code words of equal (minimum)  length 
to all strings of length $n$ with $n_0$ zeroes, and no code words to
any other strings. Then $D_{\langle n,n_0 \rangle}$ 
is a prefix-code and $L_{\langle n,n_0 \rangle} (x) = 
\lceil \log \binom{n}{n_0} \rceil$. The universal two part code
$\tilde{D}$ relative to the set of codes $\{ 
D_{\langle i,j\rangle} \; : \; i,j \in {\cal N} \}$ then achieves
the following lengths (to within 1 bit): for all $n$, all $n_0 \in \{0,\ldots,n\}$, all 
$x_{[1:n]}$ with $n_0$ zeroes,
$$
\tilde{L}(x_{[1:n]}) = \log n + \log n_0 + 2 \log \log n + 2 \log \log
n_0 + \log \binom{n}{n_0} = \log  \binom{n}{n_0} + O(\log n)
$$
Using Stirling's approximation of the factorial, $n! \sim
n^{n}e^{-n}\sqrt{2\pi n}$,  we find that 
\begin{multline}
\label{eq:stirling}
\log \binom{n}{n_0} =  
\log n! - \log n_0! + \log (n- n_0)! = \\
n \log n - n_0 \log n_0 - (n-n_0) \log (n- n_0) + O(\log n) = n
H(n_0/n) + O(\log n)
\end{multline}
Note that  $H(n_0/n) \leq 1$, with equality iff $n_0 = n$. Therefore, if
the frequency deviates significantly from $\frac{1}{2}$, $\tilde{D}$
compresses $x_{[1:n]}$ by a factor linear in $n$. In all such cases,
$D^*$ compresses the data by at least the same linear factor.   
Note that (a) each individual code $D_{\langle n,n_0 \rangle}$ is
capable of exploiting a particular type of 
regularity in a sequence to compress that
sequence,
(b) the universal code $\tilde{D}$ may exploit 
{\em many\/} different types of
regularities to compress a sequence, and (c) 
the code $D^*$ with lengths given by
the Kolmogorov complexity asymptotically exploits {\em all\/}
computable regularities so as to maximally compress a sequence.
\end{example}
\paragraph{Universal codes and Shannon:}
If a random variable
$X$ is distributed according to some known probability
mass function $f(x)=P(X=x)$,
then the optimal (in the average sense) code to use is the
Shannon-Fano code. But now suppose it is only known that
$f \in \{f \}$, where $\{ f \}$ is some given (possibly very large,
or even uncountable) set of candidate distributions. Now it is not clear
what code is optimal. We may try the Shannon-Fano code for a particular $f
\in \{ f \}$, but such a code will typically lead to very large
expected code lengths if $X$ turns out to be distributed according to
some $g \in \{ f \}, g \neq f$. 
We may ask whether there exists another
code that is `almost' as good as the Shannon-Fano code for $f$, no
matter what $f \in \{ f \}$ actually generates the sequence?
We now show that, provided $ \{ f \}$ is finite or countable, 
then (perhaps surprisingly), the answer is yes. To see this,
we need the notion of an {\em sequential information source},
Section~\ref{sec:preliminaries}. 

Suppose then that $\{ f \}$ represents a finite or countable set of
sequential information sources. Thus,
$\{ f \} = \{ f_1, f_2, \ldots \}$ and $f_k \equiv (f_k^{(1)},
f_k^{(2)}, \ldots)$ represents a sequential information source, abbreviated to
$f_k$. To each marginal distribution $f^{(n)}_k$, there corresponds a
unique Shannon-Fano code defined on the set $\{0,1\}^n$ with lengths
$L_{\langle n, k \rangle}(x) := \lceil \log 1/ f^{(n)}_k(x) \rceil$
and decoding function $D_{\langle n, k \rangle}$.

For given $f \in \{ f \}$, 
we define $H(f^{(n)}) := \sum_{x \in \{0,1\}^n} f^{(n)}(x) [  \log 1/
f^{(n)}(x)]$ as the entropy of the distribution of the first $n$
outcomes. 

Let $E$ be a prefix-code assigning codeword $E(x)$ to source word $x
\in \{0,1\}^n$.  The Noiseless Coding Theorem~\ref{thm:noiseless}
asserts that the minimal average codeword length
$\bar{L}(f^{(n)})
= \sum_{x \in \{0,1\}^n} f^{(n)}(x) l(E(x))$ among all such
prefix-codes $E$ satisfies
$$H(f^{(n)}) \leq L(f^{(n)}) \leq H(f^{(n)}) + 1.$$
The entropy $H(f^{(n)})$  
can therefore be interpreted as the expected code length of
encoding the first $n$ bits generated by the source $f$, when the
optimal (Shannon-Fano) code is used.

We look for a prefix code $\tilde{D}$ with length function $\tilde{L}$
that satisfies, for all fixed $f \in
\{ f \}$:
\begin{equation}
\label{eq:universalb}
\lim_{n \rightarrow \infty}
\frac{{\bf E}_f \tilde{L}(X_{[1:n]})}{H(f^{(n)})} \leq 1.
\end{equation}
where ${\bf E}_f \tilde{L}(X_{[1:n]}) = \sum_{x \in \{0,1\}^n}
f^{(n)}(x)L(x)$.
Define $\tilde{D}$ as the following two-part code: first, $n$ is
encoded using the standard prefix code for  natural numbers. Then, among
all codes $D_{\langle n, k \rangle}$, the $k$ that minimizes
$L_{\langle n, k \rangle}(x)$  is encoded (again using the standard
prefix code); finally, $x$ is encoded in $L_{\langle n, k \rangle}(x)$
bits. Then for all $n$, for all $k$, for {\em every\/} sequence
$x_{[1:n]}$,
\begin{equation}
\label{eq:probuni}
\tilde{L}(x_{[1:n]}) \leq L_{\langle n,k \rangle}(x_{[1:n]}) +L_{\cal
  N}(k) + L_{\cal N}(n)
\end{equation}
Since (\ref{eq:probuni}) holds for all strings of length $n$, it must
also hold in expectation for all possible distributions on strings
of length $n$. In particular, this gives, for all $k \in {\cal N}$,
$$
{\bf E}_{f_k} \tilde{L}(X_{[1:n]}) \leq {\bf E}_{f_k} L_{\langle n, k
  \rangle}(X_{[1:n]}) + O(\log n) = H(f^{(n)}_k) + O(\log n),
$$
from which (\ref{eq:universalb}) follows.

Historically, codes satisfying (\ref{eq:universalb}) have been called
{\em universal codes\/} relative to $\{ f \}$; codes satisfying
(\ref{eq:universal}) have been considered in the literature only much
more recently and are usually called `universal codes for individual
sequences' \cite{MerhavF98}.  The two-part code $\tilde{D}$ that we
just defined is universal both in an individual sequence and in an
average sense: $\tilde{D}$ achieves code lengths within a constant of
that achieved by $D_{\langle n,k \rangle}$ for {\em every individual
  sequence}, for {\em every\/} $k \in {\cal N}$; but $\tilde{D}$ also
achieves expected code lengths within a constant of the Shannon-Fano
code for $f$, for {\em every\/} $f \in \{ f \}$.  Note once again that
the $D^*$ based on Kolmogorov complexity does at least as well as
$\tilde{D}$.

\begin{example}
\rm
\label{ex:appy}
Suppose our sequence is generated by independent tosses of a coin with
bias $p$ of tossing ``head'' where $p \in (0,1)$. 
Identifying `heads' with $1$, the probability of $n-n_0$ outcomes
``1'' in an initial
segment $x_{[1:n]}$ is then $(1-p)^{n_0} p^{n- n_0}$. 
Let $\{ f \}$ be the set of corresponding information sources, 
containing one element for each $p \in (0,1)$. 
$\{ f \}$ is an uncountable set; nevertheless, a universal code for
$\{ f \}$ exists. In fact, it can be shown that 
the code $\tilde{D}$ with lengths (\ref{eq:stirling})
in Example~\ref{ex:universal} is universal for $\{ f \}$, i.e. it
satisfies (\ref{eq:universalb}). The reason for this is (roughly) as
follows: if data are generated by a coin with bias $p$, then with
probability $1$, the frequency $n_0/n$ converges to $p$, so that, by
(\ref{eq:stirling}),  $n^{-1} \tilde{L}(x_{[1:n]})$ tends to  
$n^{-1} H(f^{(n)}) = H(p,1-p)$.

If we are interested in practical data-compression, then the
assumption that the data are generated by a biased-coin source is very
restricted. But there are much richer classes of distributions 
$\{ f \}$ for which we can formulate universal codes. For example, we
can take $\{ f \}$ to be the class of all Markov sources of each
order; here the probability that $X_i = 1$ may depend on arbitrarily
many earlier
outcomes. Such ideas form the basis of most data compression schemes
used in practice. Codes which are universal for the class of
all Markov sources of each order and which encode and decode in real-time 
can easily be implemented. Thus, while we cannot find the
shortest program that generates a particular sequence, it is often
possible to effectively find the shortest encoding within a
quite sophisticated class of codes.
\end{example}
\bibliographystyle{plain}

\end{document}